# Artificial Intelligence and Arms Control

Paul Scharre and Megan Lamberth[1]

October 2022

**Abstract**
Potential advancements in artificial intelligence (AI) could have profound implications for how countries research and develop weapons systems, and how militaries deploy those systems on the battlefield. The idea of AI-enabled military systems has motivated some activists to call for restrictions or bans on some weapon systems, while others have argued that AI may be too diffuse to control. This paper argues that while a ban on all military applications of AI is likely infeasible, there may be specific cases where arms control is possible. Throughout history, the international community has attempted to ban or regulate weapons or military systems for a variety of reasons. This paper analyzes both successes and failures and offers several criteria that seem to influence why arms control works in some cases and not others. We argue that success or failure depends on the desirability (i.e., a weapon's military value versus its perceived horribleness) and feasibility (i.e., sociopolitical factors that influence its success) of arms control. Based on these criteria, and the historical record of past attempts at arms control, we analyze the potential for AI arms control in the future and offer recommendations for what policymakers can do today.

[1] Paul Scharre is the Vice President and Director of Studies at the Center for a New American Security (CNAS). Megan Lamberth is a former Associate Fellow for the Technology and National Security Program at CNAS. This report was made possible, in part, with the generous support of Open Philanthropy. Any errors are the responsibility of the authors. Please direct any communications to Paul Scharre at pscharre@cnas.org.

# Contents

## Introduction

Advances in artificial intelligence (AI) pose immense opportunity for militaries around the world. With this rising potential for AI-enabled military systems, some activists are sounding the alarm, calling for restrictions or outright bans on some AI-enabled weapon systems.[1] Conversely, skeptics of AI arms control argue that as a general-purpose technology developed in the civilian context, AI will be exceptionally hard to control.[2] AI is an enabling technology with countless nonmilitary applications; this factor differentiates it from many other military technologies, such as landmines or missiles.[3] Because of its widespread availability, an absolute ban on all military applications of AI is likely infeasible. There is, however, a potential for prohibiting or regulating specific use cases.

The international community has, at times, banned or regulated weapons with varying degrees of success. In some cases, such as the ban on permanently blinding lasers, arms control has worked remarkably well to date. In other cases, however, such as attempted limits on unrestricted submarine warfare or aerial bombardment of cities, states failed to achieve lasting restraint in war. States' motivations for controlling or regulating weapons vary. States may seek to limit the diffusion of a weapon that is particularly disruptive to political or social stability, contributes to excessive civilian casualties, or causes inhumane injury to combatants.

This paper examines the potential for arms control for military applications of AI by exploring historical cases of attempted arms control, analyzing both successes and failures. The first part of the paper explores existing academic literature related to why some arms control measures succeed while others fail. The paper then proposes several criteria that influence the success of arms control.[4] Finally, it analyzes the potential for AI arms control and suggests next steps for policymakers. Detailed historical cases of attempted arms control—from ancient prohibitions to modern agreements—can be found in appendix A. For a summary table of historical attempts at arms control, see appendix B.

History teaches us that policymakers, scholars, and members of civil society can take concrete steps today to improve the chances of successful AI arms control in the future. These include taking policy actions to shape the way the technology evolves and increasing dialogue at all levels to better understand how AI applications may be used in warfare. Any AI arms control will be challenging. There may be cases, however, where arms control is possible under the right conditions, and small steps today could help lay the groundwork for future successes.

## Understanding Arms Control

"Arms control" is a broad term that can encompass a variety of different actions. Generally, it refers to agreements that states make to control the research, development, production, fielding, or employment of certain weapons, features of weapons, applications of weapons, or weapons delivery systems.[5]

### Types of arms control

Arms control can occur at many stages in the development and use of a weapon (see figure 1). Nonproliferation regimes, such as the nuclear nonproliferation treaty (NPT), aim to prevent access to the underlying technology behind certain weapons. (See appendix C for a list of official treaty names and informal titles and acronyms). Bans, such as those on land mines and cluster munitions, allow access to the technology but prohibit developing, producing, or stockpiling the weapons. Arms-limitation treaties permit production; they simply limit the quantities of certain weapons that countries can have in peacetime.[6] Other agreements regulate the use of weapons in war, restricting their use in certain ways or prohibiting use entirely.

**Figure 1. Arms Control Measures Across the Life Cycle of Weapons Development and Use**

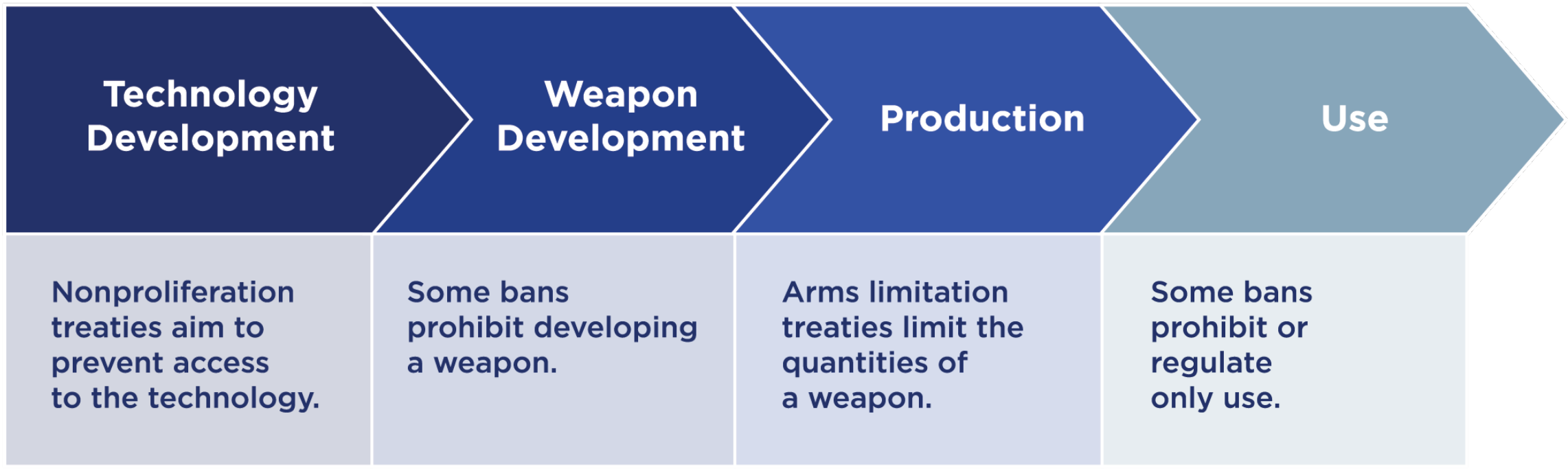

Arms control can be implemented in a variety of means, including legally binding treaties, customary international law that arises from state practice over time, or non-legally-binding instruments. Successful arms control can even be carried out through tacit agreements that are not explicitly stated between states but nevertheless result in mutual restraint.

Arms control among states is the exception rather than the rule.[7] Most of the time, states compete in military technologies without either formal or informal mechanisms of arms control to limit their competition. Several factors make arms control challenging. Arms control requires some measure of coordination and trust among states, and the circumstances in which arms control is most needed—intense militarized competition or war—are the ones in which coordination and trust are most difficult.[8] The kind of monitoring and verification that might enable trust is also a challenge, because the same transparency that might allay a competitor's fears about weapons development might also reveal vulnerabilities in one's own military forces, making states reluctant to adopt such measures.[9]

Despite these pressures, states have at times succeeded in restraining weapons development or use. Even at the height of total war, states have sought mutual restraint and refrained from using certain weapons, features of weapons, or tactics that would escalate fighting or unnecessarily increase suffering.[10] The key question for this paper is not why arms control is rare, but why it succeeds in some instances and not others.

**Factors that influence the success or failure of arms control**

Sean Watts and Rebecca Crootof analyzed historical cases of arms control to identify which social, legal, and technological factors influence whether arms control succeeds.

Watts identifies six criteria that he argues affect a weapon's tolerance or resistance to regulation: effectiveness, novelty, deployment, medical compatibility, disruptiveness, and notoriety.[11] An effective weapon that provides "unprecedented access" to enemy targets and has the capacity to ensure dominance is historically resistant to regulation. There is a mixed record for regulating novel weapons or military systems throughout history. Countries have pursued regulation of certain new weapons or weapons delivery systems (e.g., aerial bombardment) while also resisting regulation for other novel military systems (e.g., submarines). Weapons that are widely deployed—"integrated into States' military operations"—tend to be resistant to arms control. Weapons that cause "wounds compatible with existing medical protocols" in military and field hospitals are historically difficult to ban or regulate. Powerful nations have historically tried to regulate or ban weapons that are "socially and militarily disruptive" out of fear that such weapons could upend existing global or domestic power dynamics. Campaigns by civil society groups or widespread disapproval from the public can increase notoriety, making a weapon potentially more susceptible to arms control.[12]

Crootof's model overlaps with Watts's, but her focus is on weapons bans as opposed to arms control more generally. She identifies eight factors that influence the success of a weapons ban.[13] Weapons that cause superfluous injury or unnecessary suffering or that are inherently indiscriminate are more likely to be banned. Countries tend to resist regulating or banning a weapon that has demonstrated military or strategic utility. Weapons that are unique or provide a country with the "only means of accomplishing certain goals" are difficult to regulate or ban. A ban that is narrow and clearly defines what is and is not permitted is more likely to be effective. An existing treaty or regulation on a weapon may make future arms control more successful, unless technological developments increase the weapon's military utility. Advocacy groups and public opinion may influence countries' consideration of a weapons ban, although, as Crootof notes, "this factor is far from decisive." Finally, the success of a weapons ban is influenced by both the total number of countries willing to support the ban and which countries agree to sign on to it.[14]

Watts and Crootof agree that a weapon's effectiveness is arguably the most important factor that influences the success of arms control. Although their interpretations differ slightly, both argue that a weapon with uniquely valuable capabilities is difficult to regulate. Watts focuses on the social or military disruptiveness of a weapon—the capacity of the weapon to upset the existing balance of power. Although powerful countries may seek to restrain disruptive weapons, he argues that these efforts rarely succeed. Crootof argues that a weapon unique in its “ability to wreak a certain type of devastation” or accomplish certain military objectives is likely to be resistant to arms control.[15]

The next section will build upon their models to present a slightly revised approach toward understanding factors that affect the success or failure of arms control for different technologies.

## Desirability and Feasibility of Arms Control

Whether arms control succeeds or fails depends on both its desirability and its feasibility. The desirability of arms control encompasses states’ calculation of a weapon’s perceived military value versus its perceived horribleness (because it is inhumane, indiscriminate, or disruptive to the social or political order). Thus, desirability of arms control is a function of states’ desire to retain a weapon for their own purposes balanced against their desire to restrain its use by their adversaries.

The feasibility of arms control—the sociopolitical factors that influence its success—includes states’ ability to achieve clarity on the degree of restraint that is desired, states’ capacity to comply with an agreement to restrain use, states’ capacity to verify compliance, and the number of states needed to secure cooperation for an agreement to succeed. Arms control has the best chance of success when both desirability and feasibility are high.

Arms control is deemed successful when state behavior is restrained—in weapons development, quantity produced, deployment posture, or use. For the purposes of this paper, arms control agreements that fail to restrain state behavior are not considered successful. In rare instances, restraint occurs by tacit agreement, without any formal treaty or other mechanism. Generally, however, formal agreements are a useful coordination mechanism between states for reaching clarity on what is permitted and what is not. In many cases, success exists on a spectrum. Few arms control agreements are 100 percent successful, with zero violations. Some of the most successful agreements, such as modern bans on chemical and biological weapons or limits on the proliferation of nuclear weapons, have some exceptions and violators. Other agreements are successful only for a period of time, after which technology or the political environment changes in a way that causes them to collapse. Nevertheless, even partially successful agreements can be valuable in reducing harm by improving stability, reducing civilian casualties, or reducing combatant suffering.

### Desirability of arms control

A weapon that is effective, grants unique access or a capability, or provides a decisive battlefield advantage has high military value. Although relinquishment is not impossible, states will be reluctant to give up a weapon that provides a critical advantage or a unique capability even if the weapon arguably causes other significant harm. A weapon's military value is the most important factor that influences the desirability of arms control. Above all, states want to ensure their own security.

Weighed against a weapon's value is its perceived horribleness—meaning the type of injury it causes, its stability risks, its impact on the social or political order, or its indiscriminate nature. Although most successful bans are against weapons that are not particularly effective, it is oversimplified to suggest that bans are not feasible for any weapon with military value. War is horrible, and states have at times sought to temper its horror through arms control measures that restrain their actions or capabilities.

States have often sought to restrain weapons that increase combatant or civilian suffering in war beyond that required for battlefield effectiveness. States have at times restricted weapons that cause superfluous injury or unnecessary suffering to combatants, for example, if such weapons are not deemed to be uniquely effective. [16] A bullet that leaves glass shards in the body, for example, causes superfluous injury beyond that required to disable combatants and win on the battlefield, because glass shards are not detectable by x-rays, and are therefore more difficult to remove from wounded personnel. (Weapons that leave undetectable fragments in the body are prohibited under the Convention on Certain Conventional Weapons Protocol I.[17]) States have also attempted arms control for weapons or weapons delivery systems that are difficult to use in a discriminate manner to avoid civilian casualties. International humanitarian law already prohibits weapons that cause superfluous injury and indiscriminate attacks, yet states have sometimes coordinated on regulations that identify which specific weapons are worthy of special restraint.

Throughout history, those with political power have sought to control disruptive weapons, such as early firearms or the crossbow, that have threatened the existing political or social order. States have also tried to regulate weapons that could cause undue instability in crisis situations, such as intermediate-range ballistic missiles, anti-ballistic missile systems, or space-based weapons of mass destruction (WMD). Weapons that are perceived as destabilizing because they could provoke an arms race may also be susceptible to some form of regulation. For instance, a primary motivation for signatories to the 1922 Washington Naval Treaty was a desire to avoid a costly naval arms race.

A key factor to a state's continued desire for arms control is reciprocity. While there are myriad threats and inducements that compel states to comply with arms control agreements in times of peace, it is not international opprobrium that restrains militaries in the heat of war—it is the fear of enemy reprisal.

**Feasibility of arms control**

Whereas the desirability of arms control encompasses the criteria that incentivize or disincentive states to attempt some form of control, feasibility includes the factors that determine if long-term, successful arms control is possible.

An essential ingredient for effective arms control is clarity among states about the degree of restraint that is desired. Lines clearly delineating what is and is not permitted must exist for arms control to succeed; ambiguous agreements run the risk of a slippery slope to widespread use. Simplicity is key. Agreements with a clear focal point, akin to the “no gas” and permanently blinding lasers prohibitions, are more effective, because states have a clear understanding of the expectations of their behavior and that of their adversaries’.[18]

A closely related issue is the necessity of states’ being able to comply with an agreement to restrain use. In the early 20th century, states sought to limit the use of submarines and aerial bombardment, but practical realities in the ways submarines and aircraft were employed made it difficult for states to comply with the agreed-upon limits. States initially restrained their use in wartime, but restraint did not last once the practical difficulties of doing so in war were revealed.

Arms control’s feasibility is also affected by states’ ability to verify whether other parties are complying with an agreement. This ability can be accomplished through a formal verification regime, but it doesn’t necessarily have to be. The key to verification is ensuring sufficient transparency. For weapons that can be developed in secret—such as chemical or nuclear weapons—transparency may need to be assured through a verification regime. In other cases, countries may adopt less formal measures of verifying other states’ compliance, such as relying on national intelligence collection measures.

The overall number of countries needed for an agreement to succeed also influences the feasibility of arms control. Feasibility increases when fewer countries are necessary for arms control to succeed. If the polarity of the international system causes military power to be concentrated in a small number of states, getting those states to agree is crucial to success. Despite their mutual hostility, the Soviet Union (USSR) and the United States had a number of successful arms control treaties during the Cold War, some of which were bilateral agreements and some of which included many states but were led by the United States and USSR. Alternatively, in some cases, few states may be needed for an agreement to be successful simply because by virtue of technology, the weapons—such as nuclear weapons, long-range ballistic missiles, or space-based weapons—are accessible only to those few states. Diffuse weapons are more difficult to control, and more nations need to reach agreement for arms control on them to be lasting and successful. Which countries support an agreement is also important. As Rebecca Crootof explains, “If a treaty ban is ratified by the vast majority of states in the international community, but not by states that produce or use the weapon in question, it would be difficult to argue that the ban is successful.”[19]

Finally, arms control is often path-dependent, with successful regulations piggybacking on prior successful regulations of similar technologies. Modern bans on chemical and biological weapons build on long-standing ancient prohibitions on poison. The 2008 ban on cluster munitions was likely enabled by the successful 1997 ban on antipersonnel landmines. Cold War–era strategic arms control treaties likely had a snowballing effect, with successful agreements increasing the odds of future success.

The criteria within these two dimensions—desirability and feasibility—capture the most important factors that affect the success or failure of arms control. While not all-encompassing, these factors appear to be the most significant when examining the historical record of attempted arms control. If past historical experience turns out to be a useful guide for the future, then these factors are likely to influence the desirability and feasibility of arms control for new and emerging technologies, including military applications of AI.

## Why Some Arms Control Measures Succeed and Others Fail

The factors affecting the desirability and feasibility of arms control combine in ways that make arms control successful in some cases and not others. States desire arms control for some weapons over others because they are seen as more horrible and/or less useful. In some cases, states have sought arms control that was ultimately not feasible, and arms control failed.

A state's calculation of the desirability of arms control is best exemplified by the response to nuclear weapons versus chemical weapons. Nuclear weapons are undeniably more horrible—they cause greater suffering, more civilian casualties, and lasting environmental impact. Nuclear weapons are uniquely effective, however, giving states that wield them a decisive battlefield advantage. It's the military value of nuclear weapons that has prevented the nonproliferation community from achieving worldwide nuclear disarmament.

The result of this dynamic is that many examples of successful arms control are for weapons that are not particularly effective. There are instances, however, where states have chosen to place restrictions on effective weapons. If the military value of a weapon were the only factor, far more states would use chemical weapons on the battlefield. If nothing else, the threat of chemical weapons in war forces the enemy to fight in protective gear, slows down enemy troops, and reduces their effectiveness. Fighting in a gas mask is hard. It restricts situational awareness, makes it difficult to breathe, and diminishes firing accuracy. This alone is valuable. Despite these advantages, states have, for the most part, successfully controlled the use of chemical weapons in war. For most states, their military advantage is outweighed by the increased suffering they bring and the fear that using them would only cause adversaries to respond in kind.

There are many examples of states banning weapons seen as causing particularly problematic injuries to combatants, especially when these weapons have only marginal military value. For such weapons, the perceived horribleness outweighs its effectiveness, increasing desirability for arms control. Germany's sawback bayonet in World War I reportedly caused grievous injuries to combatants because of its serrated edge for sawing wood. Germany unilaterally withdrew the bayonet after reports that British and French troops would torture and kill German soldiers found with the weapon.[20]

A novel mechanism of injury can also increase the perception of a weapon's horribleness, increasing the desirability of its regulation. In the case of the ban on permanently blinding lasers, the type of injury (permanent blinding) is perceived to cause unnecessary suffering. It is not obvious why being blinded by a laser is worse than being killed, but the prohibition remains.[21] The permanently blinding laser ban also owes its success, however, to the fact that it is narrowly scoped enough that it does not inordinately constrain military effectiveness.[22] The ban permits laser "dazzlers" that temporarily blind an individual but do not cause lasting damage. Desirability for arms control is high in this case because militaries can use lasers to cause a similar battlefield effect, temporarily incapacitating the enemy, with lower levels of suffering and harm to combatants.

The process by which some weapons are deemed inhumane while others are allowable is path-dependent and not always logical. Long-standing prohibitions against poison date back to ancient times and likely influenced the success of modern-day bans against chemical and biological weapons. Ancient prohibitions on fire-tipped weapons also appear to have carried over to modern regulations on inflammable bullets and incendiary weapons. It's unclear why death by poison or a fire-tipped weapon is worse than many other means of death in war. These prohibitions, however, are enduring and cut across regions and cultures.

Path dependence has often enabled bans on weapons perceived to cause especially problematic injuries, even if those weapons are viewed as legitimate in other settings. Expanding bullets are regularly used for personal defense and by law enforcement, yet many states foreswear them because of the 1899 Hague Declaration ban, which itself built on the 1868 ban on exploding bullets.[23] Similarly, riot control agents are permissible for use against rioting civilians but, perversely, are banned for use against combatants because they fall under Chemical Weapons Convention.[24]

Countries have also regulated weapons that are seen as destabilizing or are difficult to use discriminately, and these are more likely to be successful when additional factors enhance the feasibility of regulation. Arms control measures on destabilizing weapons, such as the Seabed Treaty, Outer Space Treaty, 1972 Anti-Ballistic Missile (ABM) Treaty, and 1987 Intermediate-Range Nuclear Forces (INF) Treaty, have succeeded (at least temporarily), particularly in cases where the overall number of countries needed for cooperation was limited, making arms control more feasible. Prohibitions on expanding warfare into new domains, such as weapons on the moon or in Antarctica, have succeeded only when a clear focal point existed and the military value of deploying the weapons was low, making both the desirability and feasibility of arms control higher. Regulations on less-discriminate weapons—ones that are more difficult to use in a targeted fashion against combatants without also causing civilian harm—have succeeded in the past, but only when a weapon was banned entirely, thereby increasing the feasibility of control.

Clarity and simplicity of the agreement are essential for making arms control feasible. States need agreements with clear focal points to effectively coordinate with one another. Agreements that ban a weapon, such as poisonous gas or blinding lasers, are typically more successful than complex regulations that govern specific uses. Complete bans on weapons such as cluster munitions, antipersonnel land mines, exploding bullets, chemical and biological weapons, and blinding lasers have largely been successful because the bans were clearly defined and the weapons were prohibited entirely, not just in certain circumstances.[25] Conversely, arms control measures on weapons and delivery systems, such as air-delivered weapons and submarine warfare, that permitted their use in some circumstances but not others ultimately failed. In wartime, states expanded their use to prohibited targets.

Notable exceptions to this rule on simplicity are the bans on land mines and cluster munitions. Although the treaties seem simple enough on the surface—"never under any circumstances to use …"—the more complicated rules are concealed in the weapons' definitions.[26] The way these treaties were crafted suggest that the drafters understood the normative power of a complete prohibition to help stigmatize a weapon. Complex exceptions that were necessary for states to reach agreement were pushed to the fine print.

Not all treaties have simple rules, but successful treaties that have complex regulations often have other factors that favor success. Many of the bilateral arms control agreements between the United States and the Soviet Union/Russia, such as the INF Treaty, ABM Treaty, Strategic Arms Limitation Talks (SALT) I and II, Strategic Offensive Reductions Treaty (SORT), Strategic Arms Reduction Treaty (START), and New START, have complicated rules, but only two parties are needed to reach agreement. Additionally, these treaties apply to the production, stockpiling, or deployment of weapons in peacetime rather than wartime use, when the exigencies of war might increase pressures for defection. Complicated rules may be more viable in peacetime than wartime.[27]

Although states have often codified arms control agreements in treaties, an agreement's legal status seems to have little to no bearing on its success. Throughout history, countries have violated legally binding treaties, especially in wartime. Violations include the use of chemical weapons in World War I and the aerial bombardment of undefended cities in World War II. States have also complied with informal, non-legally-binding agreements, such as the 1985 Australia Group, which prevents the export of technologies used to produce chemical or biological weapons. There are even a few instances of tacit restraint among states without a formal agreement at all, such as the United States' and Soviet Union's decision to refrain from pursuing anti-satellite (ASAT) weapons and neutron bombs.

Integral to a state's continued adherence to an agreement is not the threat of legal consequences but the fear of reciprocity. Adolph Hitler refrained from ordering the bombing of British cities in the initial stages of World War II not because of legal prohibitions against doing so, but because of the fear that Britain would respond in kind (which it did after German bombers hit central London by mistake at night). Before the 1925 Geneva Gas Protocol was ratified, major powers, including the United Kingdom, France, and the USSR, declared that the protocol would cease to be binding if a nation failed to abide by it.[28] Even if the horribleness of a weapon far outweighs its utility, if the fear of reciprocity does not exist, states may use the weapon regardless. Syrian leader Bashar al-Assad used poisonous gas against his own people without fear of retribution. Germany used poisonous gas extensively in World War II, but never against powers that could retaliate in kind. When mutual restraint prevails, it is because state behavior is held in check either by internal norms of appropriateness or fear of how their adversary may respond.

When restraint depends upon reciprocity, states need some mechanism to verify that others are complying with an agreement. For some weapons, such as those that can be developed in secret, formal verification regimes may be necessary. Other cases may not require formal verification but do require some form of transparency. The Chemical Weapons Convention and the NPT have inspection measures in place to verify signatories' compliance. The Outer Space Treaty requires that states allow others to view launches and visit installations on the moon. While the prohibitions on land mine and cluster munitions do not have formal inspection regimes, they do require states to be transparent on their stockpile elimination.[29]

Arms control measures do not require formal or institutional verification to succeed, however. A host of arms control agreements—the 1899 ban concerning expanding bullets, the 1925 Geneva Gas Protocol, the Convention on Certain Conventional Weapons (CCW), and SORT—have no formal verification regimes in place. States will verify each other's compliance through their own observations in some cases. For the Environmental Modification Convention, Biological Weapons Convention, and Seabed Treaty, states can turn to the U.N. Security Council if they believe a signatory is violating the agreement.[30] The Strategic Arms Limitation Talks I and II agreements and ABM Treaty stated that the United States and Soviet Union would use their own means of verifying compliance, such as using satellite imagery. The Washington Naval Treaty had no verification provision, perhaps on the assumption that states could observe capital ship construction through their own means. The essential element is the ability of states to observe, through any number of means, whether or not a competitor is in compliance with the terms of the agreement.[31]

The one remaining factor that undergirds all the rest is time. Over time, the desirability and feasibility of arms control is subject to change. Technology advances and evolves, making some weapons or capabilities, such as air power, more valuable. Alternatively, a weapon—for example, chemical weapons—may be stigmatized over time if it is perceived to cause unnecessary suffering or does not provide a decisive battlefield advantage. It is very difficult to predict the developmental pathway of emerging technologies and their countermeasures. The 1899 Hague Declarations crafted regulations around a host of new weapons—balloon-delivered weapons, expanding bullets, and gas-filled projectiles—that were correctly anticipated to be problematic. Yet the regulations states crafted to restrain these technologies were built on assumptions that turned out to be false. For air-delivered weapons, Hague delegates failed to fully anticipate the futility in defending against air attacks. Expanded bullets were banned, even though their use became normalized in personal defense and law enforcement settings. And Hague delegates failed to ban poison gas in canisters, creating a loophole that Germany exploited in early gas use in World War I.[32]

The difficulty in anticipating how technologies will evolve is a challenge for regulating emerging technologies. The fact that a technology is new complicates the desirability of arms control in several ways. Some states may favor preemptively restricting a nascent technology or weapon, particularly if they fear a potential arms race. In other instances, however, states may be reluctant to give up a capability whose military value isn't fully known. States may also fail to comprehend the horror of a weapon until it is deployed in battle. Countries understood the potential harm of air-delivered weapons in civilian areas, but the horror of poisonous gas and nuclear weapons was not fully realized until their use.

Even if states desire arms control for emerging technologies, attempted regulations may not prove feasible if they misjudge the way the technology evolves. Complicated rules (in the fine print) are possible for bans on weapons that already exist, like cluster munitions and land mines. For preemptive bans on new weapons, however, states are unlikely to successfully predict the details of how the technology will evolve.[33] Preemptive regulations of emerging technologies are more likely to succeed when they focus on the intent of a weapon, such as the ban on lasers intended to cause permanent blinding, rather than technical details that may be subject to change.

Even when factors support the desirability and feasibility of arms control, success is not guaranteed. States may choose not to comply. Mutual restraint may collapse. A weapon may prove too valuable militarily, leading states to forgo arms control to retain a potentially war-winning weapon. These challenges have been faced for centuries, and they have concrete implications for future attempts at regulating emerging technologies, such as AI. Countries must keep them in mind as they reckon with how and when to regulate or restrict certain uses of military AI.

## Implications for Artificial Intelligence

AI technology poses challenges for arms control for a variety of reasons. AI technology is diffuse, and many of its applications are dual use. As an emerging technology, its full potential has yet to be realized—which may hinder efforts to control it. Verification of any AI arms control agreement would also be challenging; states would likely need to develop methods of ensuring that other states are in compliance to be comfortable with restraining their own capabilities. These hurdles, though significant, are not insurmountable in all instances. Under certain conditions, arms control may be feasible for some military AI applications. Even while states compete in military AI, they should seek opportunities to reduce its risks, including through arms control measures where feasible.

### AI as a general-purpose technology

AI is a general-purpose enabling technology akin to electricity or the internal combustion engine, rather than a discrete weapon such as the submarine, the expanding bullet, or the blinding laser. This aspect of the technology poses several challenges from an arms control standpoint.

First, AI technology is dual use, with both civilian and military applications, and thus is likely to be widely available. The diffuse nature of the technology makes arms control challenging in two ways. First, it makes a nonproliferation regime that would propose to "bottle up" AI and reduce its spread less likely to succeed. Additionally, the widespread availability of AI technology means that many actors would be needed to comply with an arms control regime for it to be effective. All things being equal, coordination is likely to be more challenging with a larger number of actors.

Second, the general-purpose nature of AI technology could make it more difficult to establish clear focal points for arms control. This is particularly true given that its very definition is fuzzy and open to many interpretations. "No AI" lacks the clarity of "no gas"; whether a technology qualifies as "AI" may be open to multiple interpretations. In practice, AI is a such a broad field of practice that declaring "no AI" would be analogous to states deciding "no industrialization" in the late 19th century. Although states attempted to regulate or ban many specific technologies that emerged from the industrial revolution (including submarines, aircraft, balloons, poison gas, and exploding or expanding bullets), a pledge by states to simply not use any industrial-era technologies in warfare would have been untenable. Nor, given the dual-use nature of civilian industrial infrastructure, is it at all clear where such lines would or could have been drawn, even if they had been desirable. Could civilian railways, merchant steamships, or civilian trucks have been used to transport troops? Could factories have been used to manufacture weapons? For AI technology today, many military applications are likely to be in non-weapons uses that improve business processes or operational efficiencies, such as predictive maintenance, image processing, or other forms of predictive analytics or data processing that may help streamline military operations. These AI applications could enhance battlefield effectiveness by improving operational readiness levels, accelerating deployment timelines, shortening decision cycles, improving situational awareness, or providing many other advances. Yet where the line should be drawn between acceptable military AI uses and unacceptable uses could be murky, and states would need clarity for any agreement to be effective.

States' experience with arms control for technologies that emerged during the industrial revolution is a useful historical guide because states did attempt to regulate (and succeeded in some cases) specific applications of general-purpose industrial technologies, including the internal combustion engine (submarines and airplanes) and chemistry (exploding bullets and poison gas). These efforts were not always successful, but not because states were unable to define what a submarine or airplane was, nor even because states could not limit their civilian use (which was not necessary for the bans to succeed). Rather, the reasons for failure had to do with the specific form of how those weapons were used in warfare. Had the offense-defense balance between bombers and air defenses, or submarines and merchant ships, evolved differently, arms control for those weapons might have been more successful. (Alternatively, had states attempted to ban these weapons entirely, rather than regulate their use in war, arms control for aircraft and submarines might have been successful.)

This analysis suggests that although banning all military AI applications may be impractical for many reasons, there is ample historical evidence to suggest that states may be able to agree to limit specific military applications of AI. The question then is which specific military AI applications may meet the necessary criteria for desirability and feasibility of arms control. Because AI could be used for many applications, there may be certain specific uses that are seen as particularly dangerous, destabilizing, or otherwise harmful. AI applications relating to nuclear stability, autonomous weapons, and cybersecurity have already been the focus of attention from scholars, and there may be other important AI applications that merit additional consideration.[34] Even within particular domains of interest, the desirability and feasibility of arms control for any specific applications may depend a great deal on the way the technology is applied. Bans or regulations could be crafted narrowly against specific instantiations of AI technology that are seen as particularly problematic, analogous to state restraint on bullets that are designed to explode inside the body, rather than all exploding projectiles.

**AI as an emerging technology**

One of the difficulties in anticipating which specific AI applications may merit further consideration for arms control is that, as is the case with other emerging technologies, it is not yet clear exactly how AI will be used in warfare. This problem is not new. States struggled in the late 19th and early 20th centuries to successfully control new industrial-age technologies precisely because they were continually evolving.

There are ways in which arms control is both easier and harder for emerging technologies. On the one hand, preemptive bans on new technologies can be easier in some respects, because states are not giving up a weapon that is already integrated into their militaries, upon which they depend for security (and for which there may be internal bureaucratic advocates). On the other hand, regulating emerging technologies can sometimes be more challenging. The cost-benefit tradeoff for militaries is unknown, because it may be unclear how militarily effective a weapon is. Similarly, its degree of horribleness may not be known until a weapon is used, as was the case for poison gas and nuclear weapons. States may be highly resistant to restraining development of a weapon that appears to be particularly valuable.

Militaries' perception of AI as a "game-changing" technology may be a hurdle in achieving state restraint. Militaries around the world are investing in AI and may be reluctant to place some applications off limits. The hype surrounding AI—much of which may not actually match militaries' investments in practice—may be an obstacle to achieving arms control. Additionally, the perception of AI systems as yielding superhuman capabilities, precision, reliability, or efficacy may reduce perceptions that some AI applications may be destabilizing or dangerous.

Perceptions of AI technology, even if they are unfounded, could have a significant impact on states' willingness to consider arms control for military AI applications. Over time, these perceptions are likely to become more aligned with reality as states field and use military AI systems. In some cases, though, even if some AI applications are eventually seen as worthy of arms control, it could be difficult to put the genie back into the bottle if they have already been integrated into states' military forces or used on the battlefield.[35]

### Challenges in verifying compliance

Even if states can agree on clear focal points for arms control and the cost-benefit tradeoff supports mutual restraint, verifying compliance with any arms control regime is critical to its success. One complication with AI technology is that, as is the case with other forms of software, the cognitive attributes that an AI system possesses are not easily externally observable. A "smart" bomb, missile, or car may look the same as a "dumb" system of the same type. The sensors that an autonomous vehicle uses to perceive its environment, particularly if it is engaged in self-navigation, may be visible, but the particular algorithm used may not be. This is a challenge when considering arms control for AI-enabled military systems. States may not be able to sustain mutual restraint if they cannot verify that others are complying with the agreement.

There are several potential approaches that could be considered in response to this problem: states could adopt intrusive inspections, restrict physical characteristics of AI-enabled systems, regulate observable behavior of AI systems, and restrict compute infrastructure.

***Adopt intrusive inspections.*** States could agree to intrusive inspection regimes that permit third-party observers access to facilities and to specific military systems to verify that their software complies with an AI arms control regime. AI inspection regimes would suffer from the same transparency problem that arises for other weapons: inspections risk exposing vulnerabilities in a weapon system to a competitor nation. Future progress in privacy-preserving software verification might help states overcome this challenge, however, by verifying the behavior of a piece of software without exposing private information.[36] Or states might simply accept that the benefits to verification outweigh the risks of increased transparency; there are precedents for intrusive inspection regimes. One challenge with inspections is that if the difference between the permitted and banned capability lay in software, a state could simply update its software after inspectors left. Software updates could be done relatively quickly and at scale, far more easily than building more missiles or nuclear enrichment facilities. In principle, states might be able to overcome this problem through the development of more advanced technical approaches in the future, such as continuous monitoring of software to detect changes or by embedding functionality into hardware.[37] Unless states can confidently overcome the challenge of fast and scalable post-inspection updates to AI systems, intrusive inspection regimes will remain a weak solution for verifying compliance, even if states were willing to agree to such inspections.

***Restrict externally observable physical characteristics of AI-enabled systems.*** States could focus not on the cognitive abilities of a system but on gross physical characteristics that are both easily observable and difficult to change, such as size, weight, power, endurance, payload, warhead, and so forth. Under this approach, states could adopt whatever cognitive characteristics (sensors, hardware, and software) they wanted for a system. Arms control limitations would apply only to the gross physical characteristics of a vehicle or munition, even if the actual concern were motivated by the military capabilities enabled by AI. For example, if states were concerned about swarms of antipersonnel small drones, rather than permitting only "dumb" small drones (which would be difficult to verify), states could simply prohibit all weaponized small drones, regardless of their cognitive abilities.[38] States have used similar approaches before, regulating the gross physical characteristics of systems (which could be observed), rather than their payloads (which were the states' actual concern but more difficult to verify). Multiple Cold War–era treaties limited or banned certain classes of ballistic and cruise missiles, rather than only prohibiting arming them with nuclear weapons.[39] An alternative approach, limiting only nuclear-armed missiles, would have permitted certain conventional missiles but would have been harder to verify.

***Regulate observable behavior of AI systems.*** States could choose to center regulations on the observable behavior of an AI system, such as how it operated under certain conditions. This would be analogous to the "no cities" concept of bombing restrictions, which prohibited not bombers but rather the way they were employed. This approach would be most effective when dealing with physical manifestations of AI systems in which the outward behavior of the system is observable by other states. For example, states might establish rules for how autonomous naval surface vessels ought to behave in proximity to other ships. States might even adopt rules for how armed autonomous systems might clearly signal escalation of force to avoid inadvertent escalation in peacetime or crises. The specific algorithm that a state used to program the behavior would be irrelevant; different states could use different approaches. The regulation would govern how the AI system behaved, not its internal logic. For some military AI applications that are not observable, however, this approach would not be effective. (For example, restrictions on the role of AI in nuclear command and control would likely not be observable by an adversary.) Another limitation to this approach is that, as is the case with intrusive inspections, the behavior of a system could potentially be modified quickly through a software update—which could undermine verifiability and trust.

***Restrict compute infrastructure.*** AI systems have physical infrastructure used for computation—chips—and one approach could be to focus restraint on elements of AI hardware that can be observed or controlled. This could be potentially done by restricting specialized AI chips, if these specialized chips could be controlled through a nonproliferation regime (and if these chips were essential for the prohibited military capability).[40] Another approach could conceivably focus on restricting large-scale compute, if compute resources were observable or could be tracked. Leading AI research labs have invested heavily in large-scale compute for machine learning in recent years, although it is unclear whether the value of this research outweighs its significant costs and for how long this trend can continue.[41] There are also countervailing trends in compute efficiency that may, over time, democratize AI capabilities by lowering compute costs for training machine learning systems.[42]

One important factor enabling arms control focused on AI hardware is the extent to which chip fabrication infrastructure is democratized globally versus concentrated in the hands of a few actors. Current semiconductor supply chains are highly globalized but have key chokepoints. These bottlenecks present opportunities for controlling access to AI hardware. For example, in 2020 the United States successfully cut off the Chinese telecommunications company Huawei from advanced chips needed for 5G wireless communications by restricting the use of U.S.-made equipment for chip manufacturing (even though the chips themselves were made in Taiwan).[43] Similar measures could conceivably be used in the future to control access to AI hardware if production of those chips were similarly limited to a few key actors.

The future evolution of semiconductor supply chains is highly uncertain. Supply chain shocks and geopolitical competition have accelerated state intervention in the global semiconductor market, causing significant uncertainties in how the market evolves. There are trends pointing toward greater concentration of hardware supply chains and other trends toward greater democratization. The high cost of semiconductor fabrication plants, or fabs, is one factor leading to greater concentration in the industry. On the other hand, geopolitical factors are leading China and the United States to accelerate indigenous fab capacity. There are powerful market and nonmarket forces affecting the global semiconductor industry, and the long-term effects of these forces on supply chains is unclear.

## The Way Ahead

The closest historical analogy to the current moment with artificial intelligence is the militarization of industrial-age technology around the turn of the 20th century and states' attempts at the time to control those dangerous new weapons. Following the St. Petersburg

Declaration in 1868, states engaged in a flurry of arms control activity, both in the run-up to World War I and in the interwar period before World War II. Leading military powers at the time met to discuss arms control in 1874, 1899, 1907, 1909, 1919, 1921, 1922, 1923, 1925, 1927, 1930, 1932, 1933, 1934, 1935, 1936, and 1938. Not all of these efforts reached agreements, and not all of the treaties that were ratified held in wartime, but the scale of diplomatic activity shows the effort and patience needed to achieve even modest results in arms control.

There are several steps that policymakers, scholars, and members of civil society can take today to explore the potential for AI arms control. These include meetings and dialogue at all levels to better understand the technology, how it may be used in warfare, and potential arms control measures. Academic conferences, Track II academic-to-academic exchanges, bilateral and multilateral dialogues, and discussions in various international forums are all valuable for helping advance dialogue and mutual understanding among international parties.[44] Analysis of potential arms control measures must be tightly linked to the technology itself and the conduct it enables, and these dialogues must include AI scientists and engineers to ensure that policy discussions are grounded in technical realities. Additionally, because AI technology remains fluid and rapidly evolving, those considering arms control must be prepared to be adaptive and to shift the focus of their attention to different aspects of AI technology or the military capabilities it enables as the technology matures. Metrics for tracking AI progress and proliferation will also help illuminate both possibilities for arms control and future challenges.[45]

Policymakers can take steps today that may make the technology more controllable in the long run by shaping its development, particularly in hardware. Enacting export controls on key choke points in the global supply chain may help to control the spread of underlying technologies that enable AI, concentrating supply chains and enhancing future controllability.[46] Export controls can have the effect of accelerating indigenization of technology, however, as actors who are cut off from a vital technology redouble their efforts to grow their national capacity. Policymakers should be judicious in applying various instruments of industrial policy to ensure that they are mindful of the long-term consequences of their actions and whether they are retaining centralized control over a technology, and thus the ability to restrict it in the future, or whether they are inadvertently accelerating its diffusion.

At the dawn of the AI revolution, it is unclear how militaries will adopt AI, how it will affect warfare, and what forms of arms control states may find desirable and feasible. Policymakers can take steps today, however, to lay the groundwork for potential arms control measures in the future, including not only shaping the technology's evolution but also the political climate. The history of arms control shows that it is highly path-dependent—and that arms control measures are often built on prior successful arms control agreements. Small steps now could yield larger successes down the road, and states should seek opportunities for mutual restraint to make war less terrible whenever possible.

# Appendices

## Appendix A.
## Case Studies of Historical Attempts at Arms Control

*This appendix includes a series of historical case studies of attempted arms control, from ancient prohibitions to modern treaties. These case studies illustrate how the success or failure of attempted arms control depends on its desirability—the weapon's effectiveness weighed against its perceived horribleness—and its feasibility.*

### Ancient prohibitions

Rules of war date back to antiquity and have existed across many civilizations. One of the oldest known texts, the Babylonian Code of Hammurabi, outlines rules governing what to do if a person is taken prisoner in war.[47] The Bible's book of Deuteronomy prohibits wanton environmental destruction in war.[48] Islamic texts similarly include instructions for proper conduct in war,[49] including prohibitions on unnecessary environmental destruction.[50] The Hindu Laws of Manu outline a number of rules governing conduct in war, such as delineating those who are out of combat and should not be attacked.[51] The Laws of Manu also specifically call out certain weapons as illegitimate: "When he fights with his foes in battle, let him not strike with weapons concealed (in wood), nor with (such as are) barbed, poisoned, or the points of which are blazing with fire."[52]

These prohibitions are mirrored in other ancient Hindu texts. The Hindu Dharmaśāstras and Mahābhārata similarly prohibit the use of poisoned or barbed arrows.[53] The Mahābhārata calls them "weapons of evil people."[54] The prohibitions in these texts mirror present-day bans on perfidy, poisoned weapons, and weapons designed to cause unnecessary suffering, reflecting millennia-old traditions about appropriate conduct in war.

The same section of the Mahābhārata also includes the curious admonition "One should not attack chariots with cavalry; chariot warriors should attack chariots."[55] It's possible this is tactical advice, but in context with the surrounding verses, it appears to be an ethical guideline on appropriate conduct in war. This is interesting because, unlike the prohibition on poisoned, barbed, or fire-tipped weapons, its motivation is presumably about fairness, rather than the specific type of injury.

These texts don't offer any clue as to whether these prohibitions were successful. These could have been rules that were scrupulously followed, or it's possible that they were rules that had to be written down precisely because they were routinely violated. Torkel Brekke, author of *The Ethics of War in Asian Civilizations,* put these bans in historical context: "When looking at history I always find it very hard to say something sensible about whether or not rules or norms were followed." The historical record just doesn't give us enough information. Brekke said that, in general, when there are "clear norms" about something, as in the case of these texts, "then these were most probably contested practices."[56] All we can know for certain is that the view that some types of weapons were illegitimate in war has ancient roots, dating back over 2,000 years.

**The diabolical crossbow**

One of the best known prohibitions on weapons—often held up as the archetype of futile weapons bans—is the popes' ban on the crossbow.[57] In the 1097 Lateran Synod, Pope Urban II banned the use of the crossbow (against Christians).[58] Forty years later, in the 1139 Second Lateran Council, Pope Innocent II restated the ban, issuing the decree: "We prohibit under anathema that murderous art of crossbowmen and archers, which is hateful to God, to be employed against Christians and Catholics from now on."[59]

The popes' decrees do not specify their rationale, but historians note that the weapon was seen as dishonorable and "despised as unchivalrous."[60] Today, there are different interpretations of what motivated this sentiment.

One theory is that the crossbow's ability to kill "beyond … range of hearing, vision, and retaliation" was "too remote and inhuman for contemporary opinion."[61] The problem with this theory is that traditional bows had long been used in medieval warfare with no objections. Traditional bows kill from greater distances than crossbows. They are also more psychologically remote. With regular bows, archers fire en masse into the air so that a hail of arrows falls on the enemy. Crossbows are held flat and aimed directly at the enemy. Crossbows are a shorter-range and more accurate weapon. A crossbow firer knows whom he or she is aiming at. In short, killing with crossbows is more intimate than regular bows, not less.

A more plausible theory suggests a realpolitik motivation for the ban. By allowing a relatively untrained commoner to kill an armored knight, the crossbow upset the existing political order. As military historian N.H. Mallett observed,

> Crossbows meant that no breast-plated nobleman, prince or king was safe on the battlefield. Any low-born peasant with just a bit of training could kill a lord or sovereign with [*sic*] simple squeeze of a trigger—a platoon of crossbowmen could wipe out a kingdom's aristocracy with just a few volleys. And that was something Medieval elites feared might shatter the natural order of society. … [A]ny technology that could put the power to instantly kill a chivalric knight, a nobleman, or even a king into the hands of a rank amateur was seen as an abomination. Crossbows weren't just weapons that could quickly win battles, [*sic*] to the ruling class they were a great equalizer.[62]

Crossbows were hated by those invested in the existing social and political order. Historians have noted that the crossbow was seen as "immoral," "devilish," "diabolical," "inherently evil," and “a bow of the barbarians.”[63] In paintings and sculptures at the time, demons were often depicted holding crossbows.[64] One can understand why knights despised the crossbow. It was the 12th-century equivalent of the modern handgun, whose equalizing power is captured in the 19th-century adage, “God created men equal. Colonel Colt made them equal.”[65] Knights saw the crossbow as dishonorable and unchivalrous because it neutralized their advantages. To the knight, who had trained hard and was superior at hand-to-hand fighting, this shift to a different style of fighting must have seemed unfair, unsporting, barbaric, and immoral.[66] To the weaker fighter, though, the equalizing power of the crossbow must have been welcome.

Whatever moral qualms Europeans had about the crossbow, they fell to the exigencies of military necessity. The ban was a total failure. The crossbow was effective, and that was all that mattered. One historian observed, “As with most measures of moral condemnation of highly useful and richly rewarding things, these grand proscriptions had no effect.”[67] If the ban caused any pause at all in the crossbow’s diffusion into armies, it was fleeting. By the end of the 12th century, only a few decades after the second papal degree, the crossbow was in widespread use.[68] Medieval rulers may not have liked the development of the crossbow, but they hastened to add crossbowmen to their armies.[69] Historians have noted that the crossbow was the “standard archery weapon in France” for the next several centuries, until it was gradually replaced by firearms.[70]

**The way of the gun**

The dramatic failure of the crossbow ban stands in stark contrast to one of the most successful weapons restrictions ever: the 250-year Japanese relinquishment of firearms.

Firearms came to Japan at roughly the same time that they were being incorporated into armies elsewhere around the world, in the mid-1500s. Early matchlock firearms were used by Japanese feudal lords, sometimes in large numbers. During the Japanese invasion of Korea in 1592, tens of thousands of Japanese soldiers carried guns.[71]

This situation changed when Tokugawa Ieyasu consolidated his hold over Japan in 1603, ending a bloody era of feudal wars and ushering in the Tokugawa Shogunate, which ruled Japan for more than 250 years. The new government moved swiftly to consolidate firearm production. In 1607, it issued a decree that all gunsmiths should relocate to the city of Nagahama.[72] All firearm production had to be authorized by the government and, according to scholars, it authorized “almost no orders” except the government’s.[73] In 1609, the government salaried all gunsmiths, paying them whether they manufactured guns or not. The government kept a “a dribble of orders” open, paying “outrageously high prices” for each gun to keep the gunsmiths gainfully employed.[74] Even still, business was so scarce that some gunsmiths switched to sword making, a business that was alive and well in Japan.[75]

Firearms were never officially outlawed in Japan.[76] They simply ceased to be relevant. The government kept producing firearms throughout the 1600s and 1700s, but at insignificant numbers: a few dozen large guns in even years and 250 to 300 small guns in odd years.[77] At this level, even a century's worth of production could arm but a tiny fraction of the half a million samurai in Japan.[78] Gun technology development similarly fell by the wayside, as did cannon development.

Japan so thoroughly ignored gunpowder weapons that when U.S. Navy Commodore Matthew Perry entered Tokyo Bay in 1853 to compel Japan to open to international trade, Japan had no effective defenses.[79] Its coastal defense cannons were over 200 years old and fired only six- to eight-pound shot, compared with Perry's 64-pound cannons.[80] Perry's visit changed everything. The Tokugawa Shogunate fell in 1867, and the new Meiji government set out to rapidly modernize its military forces. By the end of the century, Japan was a global military power on par with European great powers.[81]

In his book *Giving up the Gun: Japan's Reversion to the Sword, 1543–1879,* Noel Perrin explains the unique circumstances that led to Japan's 250-year relinquishment of firearms while Europe was embracing them wholesale. There was clearly a cultural resistance to firearms within Japanese samurai culture. Swords were a central part of samurai culture, "the soul of the samurai."[82] The samurai occupied an outsize role in Japanese society, making up 7 percent to 10 percent of the Japanese population.[83] By comparison, the warrior class of European countries was less than 1 percent of the population. Firearms were also foreign-originated weapons. Even though they had already been used in Japan, this factor may have contributed to their being viewed in a negative light. The Tokugawa Shogunate had a foreign policy of *sakoku*, or "national isolation," that heavily restricted engagement with foreigners.[84] All of these factors made firearms culturally less appealing.

It was the unique political circumstances of Japan at the time, however, that were decisive in allowing it to effectively abandon firearms as a weapon of war. Japan wasn't the only nation to attempt to restrict firearms. In 1523, King Henry VIII of England forbade anyone earning less than 100 pounds a year from owning a gun, essentially restricting gun ownership to the upper class.[85] The problem was that England had external threats that Japan didn't face. In 1543, when England went to war with France, Henry VIII reversed course, authorizing guns for any male aged 16 or older. Japan faced no such external threats. Its geography combined with weak neighbors meant that Japan had essentially no threat to its sovereignty until Perry's arrival in 1853. Internally, the Tokugawa Shogunate had consolidated power within Japan. There were no major internal threats, either, except for a brief rebellion by a small Christian minority in 1637.[86] For 250 years in Japan, there simply were no battles in which firearms would have been useful.

Outright bans of firearms proved untenable in the long run. The technology was simply too diffuse, and over time even nations that desired prohibitions on firearms were forced to adopt them to compete with rivals.

Prohibitions on types of bullets, however, did prove possible. In 1675, in the midst of the Franco-Dutch War, France and the Holy Roman Empire signed the Strasbourg Agreement—the first known international agreement banning the use of poisoned weapons.[87] Influenced by a long-standing taboo against poison, the Strasbourg Agreement prohibited the use of poisoned bullets for the duration of the war.[88] Comprehensive agreements between states on practices in war did not materialize until the late 19th century, however.[89]

**Modern weapons bans**

The late 19th and early 20th centuries saw a wave of international treaties that set out to regulate war and ban certain weapons. These efforts had mixed track records.

Modern laws of war date back to the Lieber Code of 1863, a set of regulations for conduct during war issued by Abraham Lincoln for the Union Army during the American Civil War.[90] The rules were written by Franz Lieber, who had fought for Prussia during the Napoleonic Wars, and codified traditions that had previously been customary. The Lieber Code forbid perfidy, torture, cruelty, wanton destruction, the killing of wounded or disabled combatants, and the murder and enslavement of civilians. It called on commanders, “whenever admissible,” to give advance notice of bombardment so that noncombatants could leave—an early form of today’s rules on taking feasible precautions in attack.[91] These rules laid the foundations for the early-20th-century Geneva Conventions. The only prohibition on a specific weapon was against poison, in accordance with a millennia-old aversion.[92] On the other side of the globe, however, Europeans were designing frightening new weapons.

In 1863, the Russian military developed a bullet that would explode when it hit a hard surface, originally envisioned as a weapon for exploding ammunition stores.[93] In 1867, the bullet was modified so that it could explode even upon hitting soft targets, such as people.[94] The wounds caused by such a bullet would be grievous, far worse than those caused by a non-exploding bullet. In response to this development, Russia convened a conference of European powers in 1868 to ban these weapons. The result was the 1868 St. Petersburg Declaration, which banned explosive or inflammable projectiles below 400 grams (roughly equivalent to a 30 mm shell).[95] The declaration clearly stated the signatories’ reasons for doing so. In the text, they agreed that “[t]he only legitimate object … during war is to weaken the military forces of the enemy.” Weapons that “uselessly aggravate the sufferings of disabled men,” they declared, were “contrary to the laws of humanity.”[96] This ban is an early articulation of the principle of prohibiting weapons that cause unnecessary suffering.

The St. Petersburg Declaration is an interesting ban, because states have adhered to the spirit, though not the letter, of the law. The specific regulations laid out in the declaration have been rendered obsolete because of changing technology. Militaries now regularly use tracer ammunition, anti-materiel exploding bullets, and grenade projectiles, all of which are below 400 grams.[97] These are technically prohibited under the St. Petersburg Declaration, which bans "any projectile of a weight below 400 grammes, which is either explosive or charged with fulminating or inflammable substances."[98] None of these types of ammunition are intended, however, to explode within the body in order to cause more harmful wounds.[99] There have been notable instances of exploding bullets being used, such as the 1981 assassination attempt on President Ronald Reagan, but militaries have not commonly used bullets intended to explode within the body.[100] Moreover, the underlying principle of prohibiting weapons that were designed solely to cause unnecessary suffering, or superfluous injury, has been repeated in numerous subsequent treaties.

Following the St. Petersburg Declaration, European nations embarked on a project of crafting treaties that established the modern-day laws of war. In 1874, they met in Brussels to write down laws of war that had long been customary on the battlefield. The 1874 Brussels Declaration was never ratified and so never went into effect, but it contained provisions against poison and weapons intended to cause unnecessary suffering that would later become law.[101]

The industrial revolution had more novel weapons in store for European powers, which struggled to contain these technological demons. In 1899, Europeans came together again, this time in the Hague, to pass a series of declarations banning expanding bullets, asphyxiating gases, and balloon-delivered projectiles.[102] States also codified into law long-standing customary prohibitions on poison and weapons intended to cause superfluous injury, bans that were reaffirmed at a second Hague conference in 1907.[103]

The 1907 Hague conference also renewed the ban on balloon-delivered projectiles, which had expired a few years earlier.[104] Unlike the bans on expanding bullets and asphyxiating gases, the balloon-delivered projectile ban was not motivated by concern about unnecessary suffering. There was no reason to think that death by a projectile dropped from a balloon would cause more suffering than by a projectile fired from a cannon. Rather, the ban was motivated by uncertainty surrounding the effects of this new vehicle, the hot air balloon. Concerned about the possibility of weapons that could fly over battle lines and indiscriminately bombard undefended cities, countries had agreed in the 1899 Hague convention to ban projectiles delivered from balloons for a period of five years.[105] In 1907, they reaffirmed the ban. They additionally adopted a regulation prohibiting "attack or bombardment, by whatever means, of towns, villages, dwellings, or buildings which are undefended."[106] The prohibition was aimed at air-delivered projectiles of any kind, whether from balloons or aircraft, which had first flown at Kitty Hawk four years earlier.[107]

Europeans didn’t have long before these bans were put to the test. World War I broke out less than a decade later, in 1914, and with it the awful power of industrial-era technology was unleashed on the battlefield.

**Poison gas**

World War I began on July 28, 1914, and the first use of gas came only a few weeks later. Before the war, France had openly developed tear gas grenades. French police had even used them to catch a gang of bank robbers. The French wasted no time in trying them in war, deploying them in August 1914. They were not effective. French troops threw the grenades in open areas where the gas quickly diffused, and the French abandoned them as not useful.[108]

Despite the grenades’ futility, Germany rushed to experiment with poison gas, lest it miss out on a valuable weapon. In October 1914, Germany fired 3,000 projectiles filled with a chemical irritant on British troops in Neuve-Chapelle, France. The chemicals had no effect, but the Germans persisted. On January 31, 1915, Germany fired 18,000 shells filled with xylyl bromide tear gas on Russian positions during the Battle of Bolimów, but the gas liquefied in the cold and did nothing.[109]

The first large-scale successful use of poison gas was the German attack during the Second Battle of Ypres on April 22, 1915. Germany released 170 tons of chlorine from 5,730 canisters that had been carried to the front lines and opened by hand.[110] A gray-green cloud formed and, picked up by a breeze, began to float toward French and British lines. The chlorine reacted with water to form hydrochloric acid, burning the lungs and blinding the eyes of soldiers.

A British soldier described what happened next:

> Plainly something terrible was happening. What was it? Officers, and Staff officers too, stood gazing at the scene, awestruck and dumbfounded; for in the northerly breeze there came a pungent nauseating smell that tickled the throat and made our eyes smart. The horses and men were still pouring down the road, two or three men on a horse, I saw, while over the fields streamed mobs of infantry, the dusky warriors of French Africa; away went their rifles, equipment, even their tunics that they might run the faster.
>
> One man came stumbling through our lines. An officer of ours held him up with leveled revolver, "What's the matter, you bloody lot of cowards?" says he. The Zouave [French Algerian] was frothing at the mouth, his eyes started from their sockets, and he fell writhing at the officer's feet.[111]

Six thousand French troops in the path of the cloud were injured or killed. Because of the manual method of dispersal, many Germans were also injured or killed. This new and horrible weapon created mass panic among the French troops, who fled the advancing cloud of gas. Germany failed to effectively exploit the gap in lines, however, in part because German troops were hesitant to advance into the gas themselves.[112]

The French and British governments protested the attack, declaring it a violation of international law. Germany responded that the Hague declaration banned only projectile-delivered gas; they had released the gas manually by canister.[113] This was technically correct, although clearly in violation of the spirit of the ban. (Their argument is also undermined by the fact that Germany had twice before used gas-filled projectiles; they simply had been ineffective.)

Soldiers were clearly horrified by gas. One soldier at Ypres described the gas victims dying from "the slow poison of suffocation" as "a slow and lingering death of agony unspeakable."[114] The British and French response, though, was gas of their own. A few months later, the British deployed chlorine at the Battle of Loos using the same technique of opening gas canisters by hand. The wind was less favorable this time. The gas hung in the no man's land between the two trenches for a period of time, before a breeze blew it back into British lines. British troops who were gassed staggered around vomiting and gasping for air, while the Germans were largely unharmed.[115] Before long, both sides were using artillery and mortar-delivered gas to overcome this problem, abandoning all pretense of complying with the Hague rules.[116]

All sides expanded their arsenals of poison gas in World War I, developing new chemicals including phosgene, which was 18 times more toxic than chlorine, and mustard gas.[117] Militaries evolved their tactics for using gas, coupling gas attacks with traditional artillery, and countermeasures also evolved. By the end of the war, the warring powers had developed 21 different toxic agents fired by 66 million artillery shells. Poison gas had a profound psychological effect on combatants, inspiring poems that captured the horror of the war. In his poem "Dulce et Decorum Est," Wilfred Owen described seeing a fellow soldier die after failing to get his gas mask on in time:

> Dim through the misty panes and thick green light,
> As under a green sea, I saw him drowning.
> In all my dreams before my helpless sight,
> He plunges at me, guttering, choking, drowning.[118]

Despite its horrors, gas had little effect on the outcome of the war. When used against troops who had protective gear, gas caused few fatalities. In total, poison gas killed an estimated 90,000 people in World War I, less than 1 percent of the 17 million killed in the war.[119]

Following the war, there were differing perspectives on how to view gas. General Amos Fries, who headed the U.S. Chemical Warfare Service, argued it was a weapon that "civilized nations should not hesitate to use." He argued it was "just as sportsman-like to fight with chemical warfare materials as it is to fight with machine guns."[120] General Fries was in the minority, though. In 1922, the five victors of World War I (France, the United Kingdom, Italy, the United States, and Japan) signed the Treaty relating to the Use of Submarines and Noxious Gases in Warfare. It banned "asphyxiating, poisonous or other gases … having been justly condemned by the general opinion of the civilized world."[121] The treaty was never ratified by the French and so did not take effect, but the same prohibition was included in the 1925 Geneva Gas and Bacteriological Protocol.[122] This time the treaty took hold. The 1925 Geneva Protocol banned chemical and, for the first time, bacteriological weapons.[123] Many states declared upon ratification, however, that the prohibitions would cease to hold if an enemy used gas against them.[124]

When World War II broke out, all the major powers had chemical weapons in their inventory. The British, Americans, Germans, and Russians had tens of thousands of tons of stockpiled mustard gas each.[125] Incredibly, though, during a total war that ravaged cities across Europe and Asia, chemical weapons remained largely unused on the battlefield. Japan used them in small amounts against China, which did not have chemical weapons, and there were a few isolated incidents of their use by German and Polish troops in Poland, which may have been unauthorized or accidental.[126] Germany continued to experiment with gas, developing a number of nerve agents, and used poison gas in the Holocaust. Germany killed millions of people during the Holocaust, in part in gas chambers using Zyklon B and carbon monoxide.[127]

Nations did consider using them on the battlefield. Britain planned to use chemical weapons if Germany invaded the United Kingdom, and the United States contemplated using them as part of a planned invasion of Japan.[128] But none of the parties actually used gas in any meaningful way in combat. Instead, it was held in reserve as a deterrent against the other side's use.

The mutual restraint of all parties during World War II from using chemical weapons is astonishing, particularly given the ferocity of the war, which included direct attacks on cities. Why was gas used in World War I and not World War II? In both cases, there were treaties banning the weapons. The 1925 treaty was a helpful focal point for coordination, but not the decisive factor. The key difference seems to have been that in World War II, countries knew that gas would not dramatically change the outcome on the battlefield.

In the event of a German invasion, Britain didn't plan to use gas on the beaches against German landing troops; explosive bombs were thought to be more effective. Instead, Britain planned to use gas against enemy-occupied ports, where unprotected dockworkers would be more vulnerable.[129] This would have been a desperate measure, though. Both sides knew that the chief consequence of unleashing gas on the battlefield would simply be enemy retaliation in kind. Gas would bring the war to a new level of horror, without adding any significant military advantage. Gas was held at bay by its own form of deterrence: mutual assured suffering.

In World War I, on the other hand, countries did not yet know that gas would prove to not be very useful. Gas was an unknown commodity. European nations had gone into World War I expecting it to be over quickly, and by the spring of 1915 all sides were desperate for a method to break the stalemate. Gas seemed like a potentially good option. Furthermore, Germany feared the French might use gas first. Germany was aware that France had experimented with gas grenades early in the war. Despite the grenades' ineffectiveness, Germany's fear persisted. It may have been exacerbated by reports in American newspapers about a purported French artillery shell that released poison gas. These reports later turned out to be erroneous, but the myth may have been enough to spur Germany to develop gas of its own, before it lost the race to field a potentially war-ending weapon.[130]

Following World War II, views against chemical and biological weapons solidified. If the great powers had not used them during a total, global war, then perhaps they truly were weapons that civilized nations did not use. Great Britain, the United States, and the Soviet Union all experimented with chemical weapons during the Cold War but eventually agreed to relinquish them. In 1972, states signed the Biological Weapons Convention (BWC); in 1993, the Chemical Weapons Convention (CWC).[131] The BWC and CWC go beyond the 1925 Geneva Protocols, by prohibiting the development, production, and stockpiling of biological and chemical weapons. The CWC includes an obligation to destroy existing stockpiles; as of 2022, 99 percent of all declared chemical weapon stockpiles have been destroyed under verification by the Organisation for the Prohibition of Chemical Weapons, or OPCW.[132]

The bans on chemical and biological weapons are two of the most successful weapons bans ever, but their track records aren't perfect. Even as the Soviet Union was signing on to the BWC, it was simultaneously building its second-generation biological weapons program.[133] The weapons program was a massive, highly secretive undertaking that continued into the 1990s.[134] Egypt, though not a CWC treaty member, used chemical weapons during the 1960s North Yemen Civil War.[135] Chemical weapons were used extensively throughout the Iran-Iraq War by both sides.[136] Though neither country was party to the CWC then, both countries later joined the treaty.[137] Chemical weapons have been used most recently by Bashar al-Assad in Syria, both before and after Syria joined the CWC.[138] Many of these uses were against civilian populations, who did not have access to the same protective gear that advanced militaries have to defend themselves from gas. Even the strongest taboo against a weapon can be overcome by those who do not care about international opprobrium if they see a benefit in doing so.

**Air-delivered weapons**

While a norm against poison gas solidified over time, in part because of the horrors seen by their use in World War I, sentiment shifted in the other direction on air-delivered weapons. Despite causing vastly more suffering, with whole cities leveled and hundreds of thousands of civilians killed, aerial bombardment became normalized in war over time.

Despite the 1899 and 1907 Hague prohibitions, Germany showed no hesitation in World War I in launching balloon attacks. Only a month after the start of the war, Germany launched an attack on the Belgian city of Liège using a zeppelin (a type of airship). The bombs missed their intended target but killed nine civilians, a telling sign of what was to come.[139] Soon all sides were using airships and airplanes to bomb each other's cities. Germany carried out the most extensive strategic bombing campaign, launching 51 airship raids and 27 airplane bombing raids against Britain. The raids were largely ineffective militarily. The bombs were hopelessly inaccurate. They were useful, however, as a psychological weapon for terrorizing the citizens of London. "Zeppelinitis" afflicted London, with residents referring to the airships as "baby-killers."[140] The air attacks caused minimal damage, killing approximately 1,400 people total—a tiny number compared with the 700,000 soldiers from the British Isles killed in the war.[141] Nevertheless, they "left a lasting impression on the British population and its government," one commentator observed.[142]

In the interwar period, all of the great powers further developed air power to take advantage of this rapidly developing technology. Two main schools of thought emerged on how to best employ aircraft in war. One was to use aircraft in support of ground forces, a philosophy embraced by the German *Luftwaffe* (air force). Germany developed tight integration between aircraft and ground forces, and the result was the *blitzkrieg*, a "lightning war" of rapidly maneuvering forces that allowed Germany to swiftly conquer much of Europe. Britain focused its efforts in a different direction, emphasizing strategic bombing. Promoted by Italian airpower strategist Giulio Douhet, strategic bombing aimed to swiftly bring a country to its knees and end the war by directly attacking cities. Douhet argued that air attacks would bring "a complete breakdown in social order" and "the people themselves, driven by the instinct of self-preservation, would rise up and demand an end to the war."[143]

Douhet's ideas found fertile ground among airpower advocates in both the U.S. and British air corps. One underlying idea motivating the philosophy of strategic bombing was the assumption that the offense-defense balance between air defenses and bombers favored bombers. This was reflected in the maxim "[T]he bomber will always get through," a phrase that captured the idea that trying to defend cities from enemy bombers was largely fruitless.[144] If true, then the best strategy was to hit the enemy's cities first and hit hard with strategic bombing attacks of one's own.

Douhet and his acolytes had wildly unrealistic expectations about what strategic bombing could accomplish. They understood that air attacks were not precise enough to cripple enemy infrastructure, but they vastly overestimated their effect on morale. This was particularly true in Britain, whose experience with "zeppelinitis" on the receiving end of German bombing in World War I led the British to believe that a scaled-up version could crush civilian morale. Royal Air Force chief Hugh Trenchard argued that the "moral effect" of air attacks on populations was 20 times greater than the material effect on the enemy (apparently based on no evidence).[145] Britain's Royal Air Force heavily invested in long-range bombers, and when World War II began, it was ready to go on the offensive in the air against Germany.

Before World War II began, there were strong appeals to show restraint in attacking cities. In 1923, nations came together to negotiate the Hague Rules of Air Warfare. They were never formally adopted, but the principles show the concern about civilian casualties. The rules prohibited "[a]erial bombardment for the purpose of terrorizing the civilian population" and permitted bombing only military targets.[146] They acknowledged the difficulty of separating civilian and military targets in practice, though, given that they might be co-located and that bombs at the time were hopelessly inaccurate. The contorted guidelines the rules laid out on how to address this problem presaged what was to come. The rules stated that if military targets were located such that "they cannot be bombarded without the indiscriminate bombardment of the civilian population, the aircraft must abstain from bombardment." The Hague Rules had different criteria for bombing near front lines, however, stating that civilian targets "in the immediate neighborhood of the operations of land forces" could be bombed "provided that there exists a reasonable presumption that the military concentration is sufficiently important to justify such bombardment, having regard to the danger thus caused to the civilian population."[147] This language shows the lengths to which negotiators went to try to balance avoiding civilian casualties with military necessity, but the result was a complicated set of rules that lacked the simplicity needed to ensure compliance in wartime. The rules, which were never adopted, highlighted the challenges nations faced in trying to craft guidelines that would minimize attacks on civilians while allowing for attacks against military targets, given the fact that vital military-industrial targets such as refineries, factories, railway stations, ports, warehouses, and other facilities were inevitably located in and near cities.

Calls for restraint on attacks against cities continued right up until the outbreak of World War II. In 1938, just a year before Germany invaded Poland, the League of Nations unanimously agreed on a resolution condemning aerial bombing against civilians. The resolution declared that "on numerous occasions public opinion has expressed through the most authoritative channels its horror of the bombing of civilian populations."[148] It stated that bombing of civilians had "no military necessity," "only causes needless suffering," and "is condemned under the recognised principles of international law." Because aerial bombing of military targets was considered legitimate and lawful, the declaration outlined three principles for aerial warfare: intentional bombing of civilians was illegal; bombs must be directed at identifiable, legitimate military targets; and attacks on military targets must be carried out in such a way as to not hit nearby civilian populations through negligence.[149] (These can be seen as essentially mirroring the broader international humanitarian law concepts of distinction and precautions in attack.)

It was far from clear, though, that this resolution would have the desired effect. When war broke out in September 1939, American President Franklin Roosevelt appealed to European governments to avoid attacks on civilians. He wrote:

> The ruthless bombing from the air of civilians in unfortified centers of population during the course of the hostilities which have raged in various quarters of the earth during the past few years, which has resulted in the maiming and in the death of thousands of

> defenseless men, women, and children, has sickened the hearts of every civilized man and woman, and has profoundly shocked the conscience of humanity.
>
> If resort is had to this form of inhuman barbarism during the period of the tragic conflagration with which the world is now confronted, hundreds of thousands of innocent human beings who have no responsibility for, and who are not even remotely participating in, the hostilities which have now broken out, will lose their lives. I am therefore addressing this urgent appeal to every government which may be engaged in hostilities publicly to affirm its determination that its armed forces shall in no event, and under no circumstances, undertake the bombardment from the air of civilian populations or of unfortified cities, upon the understanding that these same rules of warfare will be scrupulously observed by all of their opponents. I request an immediate reply.[150]

Britain, France, and Germany all agreed to abide by Roosevelt's call for restraint. Then, Germany immediately carried out a massive bombing campaign in Poland as part of its invasion.[151] Germany stated this was legal because the 1907 Hague rules prohibited attacks against only "undefended cities" and Warsaw had defenses. Like Germany's justification for its use of gas at Ypres in World War I, this was technically true, although its actions were clearly in violation of the spirit of the law. Air defenses against bombers were so weak that in practice any city was effectively defenseless against air attack, even such cities as London, which had invested heavily in air defense. An unstated, but probably even more important, reason having to do with Germany's bombardment of Poland was that Poland had little ability to retaliate in kind. Poland carried out only a single air attack in Germany, hitting a factory in Ohlau.[152]

Against Britain, though, Germany showed far more restraint. Hitler prohibited air attacks on British naval forces unless the British attacked first, stating, "The guiding principle must be not to provoke the initiation of aerial warfare on the part of Germany."[153] Britain was more aggressive, permitting air attacks on German ships, but held back from strategic bombing of land targets because of different risk to civilians. The German bombing of Rotterdam in the Netherlands changed the British calculus. One thousand Dutch civilians were killed, although allied newspapers reported 30,000 dead.[154] Following the attack, Britain expanded bombing to military targets on land in Germany. Britain hit oil, rail, and other industrial targets in Gelsenkirchen, Hamburg, Bremen, Cologne, Essen, Duisburg, Dusseldorf, Hanover, Dortmund, Mannheim, Frankfurt, Bochum, and Hamm.[155] While ostensibly aimed only at military targets, in practice the British bombers were so inaccurate that they were indiscriminately bombing cities. Germany responded by small-scale bombing raids against Britain, but still the objective was to stick to only military targets. Hermann Göring, head of the Luftwaffe, directed: "The war against England is to be restricted to destructive attacks against industry and air force targets which have weak defensive forces. ... It is also stressed that every effort should be made to avoid unnecessary loss of life amongst the civilian population."[156]

Hitler's Directive 17, which gave guidance "for the conduct of air and sea war against England," gave explicit instructions to bomb only military targets: "The attacks are to be directed primarily against flying units, their ground installations, and their supply organizations, but also against the aircraft industry, including that manufacturing antiaircraft equipment." Hitler additionally instructed that he alone held the right to "decide on terror attacks as measures of reprisal."[157] Even as the aerial war heated up between the two nations and Germany planned for an invasion of England, both sides sought to prevent the war from spilling over into attacks on each other's vulnerable civilian populations. But it was not to be.

Restraint collapsed when, on August 24, 1940, several German bombers lost their way at night and mistakenly bombed central London.[158] The British retaliated by bombing Berlin. Hitler was enraged, and in a public speech, declared: "If the British Air Force throws 2,000 or 3,000 or 4,000 kilograms of bombs, then we will throw 150,000, 180,000, 230,000, 300,000, 400,000, one million in one night. … If they declare that they will attack our cities on a large scale—we will eradicate their cities."[159]

Germany launched the London Blitz, and both sides stopped holding back. Britain and Germany both pursued "terror bombing" campaigns against the civilian populations to break the other's will to fight. The bombing campaigns that ensued devastated Germany and Britain. The Royal Air Force estimated that more than 20 German cities were 40 percent destroyed or more.[160] Tens of thousands of civilians were killed. Yet civilian morale did not crumble. Douhet was wrong. The war did not end quickly. People simply suffered more.

Why did restraint succeed with chemical weapons but fail with bombing attacks on civilians? Legal restrictions were not the decisive factor. Nor was it a lack of a stigma against air attacks. Both nations feared "terror bombing" by the other and sought mutual restraint. Neither side decided, suddenly, that it would be in its interest to escalate to this new level of conflict. Rather, accident, miscalculation, and emotion caused the conflict to escalate. Each nation simply believed that it was retaliating for the other's actions. Soon, it had spiraled out of control, and the two sides could not de-escalate.

This dynamic is not surprising. That countries can restrain their behavior at all in the midst of total war is remarkable. What made poison gas different and allowed restraint? Or, to put it another way, since Germany and Britain were already bombing each other's cities, why didn't they escalate to poison gas attacks on cities? Without the kind of protective gear that militaries had, civilian populations would have been highly vulnerable to gas attacks. In fact, Douhet had advocated precisely that approach, using both explosives and poison gas against unprotected civilians in cities.[161]

In his books *Strategy of Conflict* and *Arms and Influence,* Thomas Schelling explained the kinds of dynamics that allow cooperation even amid conflict. A key factor, he said, was the existence of "focal points" upon which adversaries could coordinate their behavior.[162] He explained that "the most powerful limitations, the most appealing ones, are those that have a conspicuousness and simplicity, that are qualitative and not a matter of degree, that provide recognizable boundaries."[163] "No gas" provided a clear focal point for coordination among the parties in World War II. Schelling observed:

> Gas was not used in World War II. The agreement, though not without antecedents, was largely a tacit one. It is interesting to speculate on whether any alternative agreement concerning poison gas could have been arrived at without formal communication (or even, for that matter, with communication). "Some gas" raises complicated questions of how much, where, under what circumstances: "no gas" is simple and unambiguous. Gas only on military personnel; gas used only by defending forces; gas only when carried by vehicle or projectile; no gas without warning—a variety of limits is conceivable; some may make sense, and many might have been more impartial to the outcome of the war. But there is a simplicity to "no gas" that makes it almost uniquely a focus of agreement when each side can only conjecture at what rules the other side would propose and when failure to coordinate on the first try may spoil the chances for acquiescence in any limits at all.[164]

Attempts to avoid bombing attacks on civilian targets were of a different nature, not necessarily because bombing attacks were more effective than gas or less horrible. They were, in fact, largely ineffective and universally reviled. The main purpose of Britain's and Germany's launching them seemed to be retaliation for the other's having done so. Although air power fanatics may have believed Douhet's absurd predictions that aerial bombardment would end a war in days, there is no evidence that the political leadership of Britain or Germany thought strategic bombing would lead to surrender so quickly. Over time, both sides hoped that strategic bombing would weaken the other's resolve (it didn't), but they knew they were embarking on a long-term war of mutual assured suffering.

Aerial bombardment differed from gas because restraint against civilian targets lacked the simplicity and unambiguity of the "no gas" prohibition. Rules on aerial bombardment were more complicated and a matter of degrees, unlike the binary distinction of "no gas." Aerial bombardment was allowed in some circumstances and not others, whereas gas was prohibited entirely. Regulating the use of a technology in war proved far more difficult than simply banning a technology entirely. The use of aerial bombardment expanded over time, from ships to land-based military targets to, finally, cities. Further complicating matters was the possibility of accidental escalation, as was the case when German bombers hit central London by mistake. National leaders have imperfect control over their own military forces on the battlefield, and directives to ban targets near cities, but not populated areas, proved impractical in reality given the imprecision of bombing technology at the time.[165] The total prohibition on gas, on the other hand, made restraint easier. Using gas crossed a clear threshold, demonstrating a decision to escalate. The distinction between restraint and unrestricted use was sharper in the case of gas and could be clearly communicated between adversaries, with any violations easily observed. It was also easier for national leaders to control their own forces' use (or nonuse) with gas, because they could ensure that the weapon remained stockpiled under central control, rather than deployed on the battlefield, where discretion about its use lay in the hands of lower-level subordinates. If gas had been used against troops, its use could have easily spread to attacks on cities over time.[166]

World War II's strategic bombing campaigns culminated in the atomic bombs dropped on Hiroshima and Nagasaki, ending the war and ushering in the atomic age. Nuclear weapons were the logical endpoint of strategic bombing—they were so destructive that a single bomb could bring a city to ruin. During the Cold War, "counter-value" became the term for deliberately targeting the opponent's cities in order to deter the opponent from attacking. The resulting "delicate balance of terror" between the United States and the Soviet Union held entire societies at risk.[167] In 1962, then–Secretary of Defense Robert McNamara gave his "no cities" speech, in which he outlined a shift away from targeting the enemy's civilian population to targeting only its nuclear forces, a strategy that came to be known as "counter-force" targeting.[168] The old debates about restraint in air attacks were back. We don't know—and hopefully will never know—whether a nuclear war could have been fought in a limited fashion, but accepting collateral damage from aerial bombardment became normalized in war.[169] Even a counter-force nuclear attack to wipe out the enemy's nuclear arsenal would have killed tens of millions of people: "collateral damage."[170]

**Submarines**

Just as war expanded into the air domain, it expanded undersea as well. Submarines were first introduced in the American Civil War, but only in a marginal way.[171] Submarine technology dramatically improved around the turn of the century. At the 1899 Hague convention, Russia proposed banning submarines, suggesting that states prohibit "the use in naval battles of submarine or diving torpedo-boats or of other engines of destruction of the same nature."[172] Delegates were divided on the prohibition and overall had mixed views on the utility of submarines. At the time, many believed that submarines would be relevant only for coastal defenses because of their limited range and would primarily benefit smaller nations that couldn't afford large navies. No one had an inkling of the central role submarines would play in naval warfare only a few decades later.[173] The proposed submarine ban was voted down.[174]

In 1907, nations met to codify laws of maritime warfare, laying out a series of provisions relating to the treatment of hospital ships, merchant ships, and prisoners. The submarine ban was not revisited, except for a passing reference by the Belgian delegate, who observed:

> A torpedo-boat or a submarine can annihilate in a few moments a magnificent vessel representing an enormous outlay and a thousand lives. In 1899 Russia proposed that the employment of such engines of destruction be given up, just as the poisoning of arms and of springs had been prohibited, and most of the Powers seemed ready to adhere to the proposal provided it were accepted unanimously. But unfortunately I do not now see any indication among us of such an idea.[175]

There was no further discussion of a ban. The 1907 convention did, however, establish maritime laws that would prove problematic for submarines.[176] The 1909 Declaration of London, which was never ratified, similarly laid out rules for maritime war that, as it turned out, would be essentially impossible for submarines to comply with.[177] Neither convention specifically mentioned submarines.

The essential problem with submarines lay in their ability to comply with maritime law in relation to attacks on merchant ships. Much of maritime warfare was aimed not only at enemy warships but also at stopping civilian merchant vessels from supplying goods for the opposing nation. Under customary international law, warships could fire without warning on enemy warships, but merchant vessels had to be given a warning and the opportunity to surrender. Under a maritime concept called "prize law," if the vessel was found to be carrying contraband cargo, the cargo and even the ship itself could be seized as a "prize."[178] Under certain circumstances the ship could be sunk, but the attacking ship had to provide for safe passage for the crew. In effect, maritime law permitted attacks against materiel supplying an opposing nation but still treated the merchant sailors themselves as noncombatants.[179]

These rules regulating the conduct of attacks on enemy shipping posed a fundamental problem for submarines. Submarines relied on stealth for their effectiveness. Once surfaced, they were extremely vulnerable. If submarines were required to give warning before firing, they would in effect be giving away their only advantage. Taking on merchant sailors after sinking a ship was even more problematic, because there simply was not space aboard a submarine.

Initially in World War I, Germany submarines (U-boats) complied with these rules. German submarines primarily attacked only enemy warships. In the few instances where they did attack merchant vessels, they fired a warning shot first, searched the ship for contraband, and ensured the crew's safety before sinking the ship.[180] This restraint didn't last.

In November 1914, only a few months into the war, Britain declared the entirety of the North Sea a military area, essentially blockading Germany. The blockade was effective, and Germany retaliated in February 1915, declaring a "war zone" around the British Isles.[181] Germany declared that within this zone, merchant ships could be sunk without warning. The United States, concerned about the escalating hostilities, called on both sides to restrict their submarine attacks against merchant ships. The United States proposed that each side should agree "that neither will use submarines to attack merchant vessels of any nationality except to enforce the right of visit and search."[182] Germany agreed, but Britain refused, stating that Germany's declaration of a war zone around the British Isles already amounted to, "in effect, a claim to torpedo on sight." Additionally, Britain pointed out that submarines could not effectively comply with the traditional laws to give warning before attacking merchant ships and provide for the safety of the crew.[183]

Part of this problem was of Britain's own making. Before the war even began, in 1913 Britain began arming its merchant sailors. Then–First Lord of the Admiralty Winston Churchill informed Parliament that the weapons were only defensive and merchant ships would retain their noncombatant status. Churchill declared that these defenses were aimed only at other hostile merchant ships, not enemy warships. A year later, in 1914, he told Parliament: "They are armed solely for defensive purposes. … They are not allowed to fight any ship of war. ... They are, however, thoroughly capable of self-defence against an enemy's armed merchantmen."[184] The British ambassador further informed the United States that these armed merchant ships "will never fire unless first fired upon, and that they will never under any circumstances attack any vessel."[185] The British government gave its merchant sailors different instructions, though, telling them:

> If a submarine is obviously pursuing a ship by day and it is evident to the master that she has hostile intentions, the ship pursued should open fire in self-defence, notwithstanding the submarines [*sic*] may not have committed a definite hostile act, such as firing a gun or torpedo.
>
> Any submarine approaching a merchant vessel may be treated as hostile.[186]

Britain had effectively made its merchant ships combatants by arming them and authorizing them to attack submarines, but continued to claim noncombatant status for them.

With the British rejection of the American proposal for restraint in submarine attacks against merchant ships, the last chance for cooperation ended. Germany began a campaign of sinking merchant ships headed for Great Britain without warning. On May 7, 1915, the German U-boat *U-20* torpedoed and sank the *Lusitania*, killing more than 1,000 passengers, including 128 Americans.[187] The American public was outraged. The U.S. government protested, noting

> the practical impossibility of employing submarines in the destruction of commerce without disregarding those rules of fairness, reason, justice, and humanity which all modern opinion regards as imperative. It is practically impossible for the officers of a submarine to visit a merchantman at sea and examine her papers and cargo. It is practically impossible for them to make a prize of her; and, if they cannot put a prize crew on board of her, they cannot sink her without leaving her crew and all on board of her to the mercy of the sea in her small boats.[188]

In August 1915, after another protest by the United States following another U-boat sinking, Germany sent orders to its U-boat commanders not to sink passenger ships without warning and to provide for safe passage for the crew.[189] The remainder of 1915 saw a general lull in submarine attacks on merchant ships, with intermittent periods of attack and restraint throughout 1916.[190]

The underlying tensions remained, though. Britain had not only armed its merchant vessels but had also given them instructions to ram surfaced submarines.[191] The United States was aware of what Britain was doing and in January 1916 sent a letter to all parties to World War I, stating that

> the placing of guns on merchantmen at the present day of submarine warfare can be explained only on the ground of a purpose to render a merchantman superior in force to submarines and to prevent warning and visit and search by them. An armament, therefore, on a merchant vessel would seem to have the character of an offensive armament.[192]

The United States again called for restraint in both submarine attacks and arming merchant vessels, but to no avail. In January 1917, hoping to break the war's stalemate, Germany declared unrestricted submarine warfare against Great Britain.[193] The United States severed diplomatic relations a few days later and, in April 1917, declared war on Germany.[194]

Following the war, nations continued to attempt to reconcile the submarine with traditional maritime law. In the 1921–1922 Washington Naval Conference, Britain proposed banning the submarine. France and other nations disagreed, saying the problem was not the weapon itself but the way it was being used.[195] The 1922 Treaty relating to the Use of Submarines and Noxious Gases in Warfare (which was never ratified), 1930 London Naval Treaty, and 1936 London Protocol all reaffirmed the position that submarines should give warning before attacks on merchant vessels and provide for the safety of the crew.[196]

These statements were effectively pointless. World War I had demonstrated that submarines had fundamental challenges in complying with these rules, which had been designed for surface warships. Submarines were so vulnerable when surfaced that it was too tempting for merchant ships to resist ramming them or firing on them, measures that would have been suicidal against a warship. While Britain claimed its merchant vessels were armed only for "defensive" purposes, in practice the line between defense and offense became blurred as ships interacted on the high seas. There were many ways nations could have cooperated to restrain submarine warfare. They could have agreed to use submarines only for attacks on enemy warships, for example. But short of banning submarines entirely, there was no clear focal point on par with "no gas." The incentives for one or both actors to violate the rules of submarine conduct were simply too great. Despite the numerous treaties that had tried to bend and twist submarines to fit existing maritime law, when World War II began, nations had not found a satisfactory solution.

On September 3, 1939, the day Britain declared war on Germany, Hitler issued a directive that all German ships, including submarines, would follow maritime law with regard to giving warning to merchant ships.[197] Britain used intensive anti-submarine measures that made following these rules effectively impossible, though. It again armed merchant ships. Furthermore, with the advent of radio, even unarmed merchant ships could call in the position of German U-boats to British warships, effectively acting as naval scouts. Within a matter of weeks, Germany changed its policies, permitting attacks without warning on "merchantmen and troopships recognized beyond doubt as hostile," ships sailing without lights near the British Isles, and merchant ships that used their radio transmitters when stopped.[198] It would be unfair to characterize these rules as permitting attacks on civilian vessels, as merchant ships engaging in these behaviors effectively made themselves combatants.[199] Attacks on passenger ships were still prohibited. As the war progressed, all sides resorted to unrestricted submarine warfare against other nations' commerce. By the time the United States entered the war, in 1941, there was no attempt at restraint. Mere hours after the attack at Pearl Harbor, the U.S. Navy ordered its fleet to "execute against Japan unrestricted air and submarine warfare." [200]

**Expanding bullets**

Of the three emerging technologies banned in 1899—asphyxiating gases, balloon-delivered projectiles, and expanding bullets—only one, expanding bullets, was successfully kept off the battlefield. Motivated by the same concerns about superfluous injury that drove the 1868 St. Petersburg Declaration on exploding bullets, nations agreed at the 1899 Hague Convention to ban "the use of bullets which expand or flatten easily in the human body."[201] Expanding bullets, also called "dumdum bullets," have a soft point or a hollow point that causes them to expand when they enter the body.[202] This expansion, sometimes called "mushrooming," increases the size of the wound as the bullet travels through the body, causing more injury.

The prohibition on using expanding bullets in war is strange, because expanding bullets are regularly used in civilian firearms applications such as hunting, law enforcement, and self-defense. It is entirely legal, for example, to buy expanding bullets in the United States for one's personal firearm. In fact, expanding bullets are preferred for these applications for two key reasons. First, they are significantly more effective in accomplishing the purpose of a bullet: to take down the intended target. Additionally, expanding bullets are less likely to pass through the victim's body, potentially injuring bystanders. Non-expanding bullets can often pass directly through the body, leaving the intended target still standing and potentially harming other people nearby. Expanding bullets unquestionably cause greater injury, but the injury is not "unnecessary" or "superfluous." The greater injury has a purpose: to incapacitate the intended target. As the U.S. Department of Defense pointed out in its law of war manual, the fact that "expanding bullets are widely used by law enforcement agencies today ... supports the conclusion that States do not regard such bullets are [*sic*] inherently inhumane or needlessly cruel."[203]

Nevertheless, militaries have generally abided by the prohibition on expanding bullets.[204] Most countries treat expanding bullets as prohibited, and the International Committee of the Red

Cross considers the prohibition to be customary international law.[205] The United States, which is not a signatory to the 1899 Hague Convention, disagrees that the ban on expanding bullets is customary international law and has taken the position that expanding bullets are prohibited only to the extent that their design is intended to cause unnecessary suffering.[206] The U.S. military has used expanding bullets in some narrow situations, such as hostage rescue and counterterrorism operations, but they are generally not used.[207] Overall, the ban on expanding bullets has largely been successful.

**Naval arms limitations**

The period between World War I and World War II saw one other dramatic attempt at arms control. In 1922, the victors of World War I (Great Britain, France, Italy, the United States, and Japan) signed the Washington Naval Treaty (also known as the Five-Power Treaty), which placed limits on the sizes of their navies. The Washington Naval Treaty was a different kind of treaty than earlier prohibitions on weapons. It did not prohibit any type of weapon, but limited the size and quantity of warships that countries could develop. It was not motivated by concern about unnecessary suffering or civilian casualties. Rather, the intent was to avoid a costly arms race.

The treaty specified a ratio of 5 : 5 : 3 : 1.75 : 1.75 for allowable tonnage of naval warships among Great Britain, the United States, Japan, Italy, and France, respectively.[208] That is, for every five ships Great Britain and the United States had, Japan could have three and France and Italy could have 1.75 each. The rules applied to aircraft carriers, battleships, and battle cruisers, specifying the allowable tonnage as well as number and size of guns for each ship. With this fixed ratio, the idea was that the five nations could freeze the existing balance of power and avoid exhausting themselves in an arms race that depleted their national treasure and gained them no long-term lasting advantage. (Germany was not included because its navy was already restricted by the Treaty of Versailles, signed at the conclusion of World War I.)[209]

The five nations met again in Geneva in 1927, with the hope of extending the limitations down to smaller ships—cruisers, destroyers, and submarines—which had been left out of the Washington Treaty. This round of negotiations failed, but countries reached agreement three years later in the 1930 London Naval Treaty. In 1932, 31 nations came together in the Second Geneva Conference to attempt to establish broader arms limitations, including limits on the size of armies, but these negotiations failed. In 1935, the five naval powers came together yet again to renew the Washington and London treaties, which were set to expire in 1936. Japan, which had long chafed at the ratio it was allocated relative to America, withdrew from the conference and announced its intention to let the treaty expire.[210] While unsuccessful in the long term, the agreements achieved a 14-year period of naval arms limitations. Great Britain, France, and the United States negotiated a Second London Naval Treaty in 1936, setting limits among themselves, as well as an “escalator clause” that would raise those limits if Japan did not sign the treaty. A few years later, in 1939, World War II broke out and all attempts to limit an arms race ended.

**Cold War–era weapons bans**

The Cold War brought a suite of new, even more terrible weapons for nations to contend with. Chief among these were nuclear weapons. Despite the enmity between them, the United States and the Soviet Union embarked on a series of unilateral and cooperative measures to avoid instability and inadvertent war by placing limits on military competition. These treaties are examples of states' willingness to regulate weapons they are not necessarily willing to ban outright.

A number of treaties placed geographic or other limits on Cold War military competition. The Limited Test Ban Treaty (1963) banned nuclear weapons tests in the atmosphere, in outer space, and under water.[211] The Outer Space Treaty (1967) forbade placing weapons of mass destruction in space or any kind of weapon on the moon, and the Seabed Treaty (1971) banned placing weapons of mass destruction (WMD) on the seabed beyond a 12-mile coastal zone.[212] The Antarctic Treaty (1959) and Outer Space Treaty further declared the Antarctic and the moon (and other celestial bodies) entirely off-limits for military use of any kind.[213] The Treaty of Tlatelolco (1967) declared Latin American and the Caribbean a nuclear-free zone. (Cuba did not sign until 1995.) This was followed by the Treaty of Rarotonga (1985), Treaty of Bangkok (1995), and Treaty of Pelindaba (1996), which created nuclear-free zones in the South Pacific, Southeast Asia, and Africa, respectively. The Comprehensive Nuclear Test Ban Treaty, which bans all nuclear tests, was signed in 1996.[214]

Negotiations on these treaties were not always straightforward. The United States and the Soviet Union had differing positions that needed to be reconciled. For example, the Soviet Union initially sought to link space disarmament in the Outer Space Treaty to more general disarmament of short- and medium-range missiles.[215] The Seabed Treaty was more contentious. The Soviet Union wanted complete demilitarization of the seabed, while the United States wanted to restrict only nuclear weapons.[216] The Soviet proposal would have prohibited submarine surveillance sensors on the ocean floor that the United States saw as essential for self-defense. The Soviet Union also wanted a more intensive verification regime, with all installations on the seabed open for inspection. (The Outer Space Treaty had a similar provision, with any installations on the moon or other celestial bodies open for inspection.) The United States disagreed, saying that a nuclear weapons installation on the seabed would be large and elaborate, making it difficult to conceal. (Inspections would have also permitted the Soviets access to nonnuclear U.S. military installations on the seabed, an example of the transparency problem of intrusive verification regimes.) Eventually, agreement was reached with a provision for parties to verify treaty compliance through their own observation.[217]

Many of these treaties were large, multinational ones with many countries participating, although in practice the chief negotiating parties were the U.S. and USSR. The United States and the Soviet Union also negotiated bilateral arms limitations treaties. The 1972 Anti-Ballistic Missile Treaty (ABM Treaty) restrained the U.S. and USSR's deployment of missile defense shields by placing restrictions on anti-ballistic missiles.[218] The treaty was in place until 2002, when the United States withdrew. The United States and USSR also placed limitations on intermediate-range missiles—potentially destabilizing nuclear delivery vehicles that gave little warning time before hitting their targets. After seven years of negotiations, the 1987 Intermediate-Range Nuclear Forces Treaty (INF Treaty) was signed, banning both ground-launched ballistic missiles and cruise missiles with a range of 500 to 5,500 kilometers.[219]

These treaties prohibited certain types of weapons or delivery vehicles, but the United States and the Soviet Union also cooperated to restrict nuclear weapons quantities through arms limitations treaties. The two countries began the Strategic Arms Limitation Talks (SALT) in 1969, which led to the ABM Treaty and an interim SALT I agreement in 1972. SALT I froze the construction of new intercontinental ballistic missile (ICBM) launchers and permitted new submarine-launched ballistic missile (SLBM) launchers only as replacements for older ICBM and SLBM launchers.[220] SALT II, signed in 1979, set more comprehensive limits on nuclear weapons delivery systems.[221] The Strategic Arms Reduction Treaty (START), signed in 1991, led to not just arms limitations but actual arms reductions.[222]

The Soviet Union and United States pursued conventional arms reductions, as well. Throughout the 1970s and 1980s, NATO and the Warsaw Pact had held negotiations on mutual conventional arms reductions but were unable to reach agreement. The talks finally bore fruit at the very end of the Cold War. The Treaty on Conventional Armed Forces in Europe (CFE), which placed limits on conventional ground and air forces in Europe, was negotiated in 1989 and ratified in 1991, just as the Soviet Union was collapsing.

All of these treaties were successful in restraining nuclear weapons and other armaments during the Cold War. In some cases, restraint was even possible without formal agreements. The Soviet Union and the United States engaged in tacit cooperation in a few instances in limiting the development or deployment of weapons perceived as particularly destabilizing.

The United States and USSR refrained from arms races in anti-satellite (ASAT) weapons and neutron bombs without any formal agreement in place. Both the Soviet Union and the United States demonstrated anti-satellite capabilities but canceled their programs in the 1980s. Several countries have carried out destructive ASAT tests, although nations have stopped short of the widespread deployment of ASAT weapons. Destructive ASAT demonstrations have come under criticism because they can generate thousands of pieces of space debris that can stay in orbit for years.[223] In 2022, the United States unilaterally announced that it was committing "not to conduct destructive, direct-ascent anti-satellite (ASAT) missile testing" and that it "seeks to establish this as a new international norm for responsible behavior in space."[224]

Neutron bombs—"cleaner" bombs that could kill people but leave buildings largely undamaged—are perceived as destabilizing because their use could allow for an attacker to annihilate a city's population while preserving its infrastructure. The United States planned to deploy neutron bombs to Europe in the late 1970s, but upon receiving considerable public pushback, halted its deployment. Although nuclear powers have the ability to create neutron bombs, no country has ever openly pursued the development of large numbers of them.[225] France and China tested the weapon in the 1980s, Israel is suspected to have tested one, and India admitted to having the capacity to develop them.[226]

**Restraint in the post–Cold War world**

A major factor behind the success of mutual restraint during the Cold War was undoubtedly the fact that there were only two great powers. This factor makes reaching agreement significantly easier. After the Cold War, some bilateral agreements began to unravel precisely because of concerns about other nations.

In 2001, the United States announced its withdrawal from the ABM Treaty because of concerns about long-range WMD-capable missiles from North Korea and Iran.[227] In 2007, likely in response to U.S. plans to base missile defense sites in Eastern European NATO nations, Russia suspended its participation in CFE. Russia announced it was completely halting its participation in CFE in 2015.[228]

In 2007, Russia also began to voice concerns about the INF Treaty and the fact that it did not restrict China, which is not a signatory. U.S. defense strategists have also raised concerns about Chinese missiles.[229] In 2012, it came to light that Russia had tested certain missiles that were prohibited under the INF Treaty, and in 2014 the United States formally accused it of violating the treaty.[230] In response, Russia indicated it was considering leaving the INF Treaty entirely.[231] By 2018, continued Russian violations had made sustaining the treaty untenable, and the United States threatened to withdraw from the treaty unless Russia returned to compliance.[232] Both the United States and Russia suspended their treaty obligations in early 2019, and the United States formally withdrew later that year.[233]

The multilateral treaties, however, have held. The United States and Russia also have continued to cooperate on nuclear arms reductions, which require only bilateral agreement, because both nations have been so far ahead of others in nuclear weapons quantities. The Strategic Offensive Reductions Treaty (SORT) of 2002, also known as the Moscow Treaty, and 2011's New START Treaty further reduced both nations' nuclear arsenals.[234] China's recent nuclear weapons buildup could change this dynamic, however, complicating further reductions.[235]

**New weapons, new treaties**

The Cold War also saw new kinds of weapons that were regulated or banned, sometimes preemptively. These bans have had mixed records of success.

In the early 1970s, the United States became concerned that future technology might permit militaries to modify the environment, such as through geo-engineering, as a method of warfare. In 1972, the U.S. government unilaterally renounced the use of the environment for hostile purposes, and the U.S. Senate adopted a resolution calling for an international treaty.[236] The 1977 Environmental Modification Convention banned "military or any other hostile use of environmental modification techniques having widespread, long-lasting or severe effects," prohibiting using the environment as a weapon of war.[237]

The Convention on Certain Conventional Weapons (CCW), signed in 1980, sought to limit several weapons that were perceived to be "excessively injurious" or having "indiscriminate effects."[238] Protocol I of the CCW prohibits the use of weapons that create fragments in the body that cannot be detected by x-rays. For example, a bullet that left glass shards in the body would be prohibited. This is an excellent example of an injury that would be superfluous. Leaving shards in the body that could not be found by x-ray, so they could not be removed by a surgeon, would have no additional military benefit in incapacitating a combatant and would cause needless suffering. Although in theory, weapons designed to cause such an effect would already be prohibited under international humanitarian law by the general prohibition against weapons intended to cause unnecessary suffering, the treaty negotiation process allows for the clarification of norms and expectations, and the treaty itself provides a valuable focal point for coordination. This is especially important given the inherently subjective concept of "unnecessary suffering."

CCW Protocol II lays out a series of rules governing the use of mines to keep them away from civilians. The rules prohibit placing mines near towns or cities unless they are near an enemy military objective or the minefield is marked by signs or fences to warn civilians.[239] Remotely delivered mines, such as those dropped from aircraft or launched via artillery, would not be feasible to mark with fences or signs. Therefore, remotely delivered mines can be used only if the location can be accurately recorded or if the mine has a self-deactivating mechanism that will disable it after a period of time.[240]

Protocol III, similarly, regulates incendiary weapons, in an attempt to reduce their potential for civilian harm. It prohibits directly attacking the civilian population or civilian objects. It also prohibits attacking military targets located within populated areas with air-delivered incendiary weapons. For example, the aerial firebombing tactics used during World War II would be prohibited. Ground-launched incendiary weapons are permitted only if the military target "is clearly separated from the concentration of civilians and all feasible precautions are taken" to avoid harm to civilians and civilian objects.[241]

In the early 1990s, there was concern that laser technology was maturing to the point where it might be possible to field battlefield weapons that could permanently blind soldiers.[242] In 1995, parties to the CCW adopted Protocol IV, which bans blinding lasers. The protocol states: “It is prohibited to employ laser weapons specifically designed, as their sole combat function or as one of their combat functions, to cause permanent blindness to unenhanced vision.”[243] This was another preemptive ban on a weapon that was perceived to cause unnecessary suffering, in the same vein as the prior bans on explosive and expanding bullets, poison gas, and non-x-ray-detectable fragments. Like the ban on expanding bullets, it has a strange rationale; blinding not only would have military value but certainly would lead to less suffering than killing, which is permitted.[244]

These bans, like those before them, have had a mixed track record. The bans on environmental modification and on non-x-ray-detectable fragments have been successful. Militaries have not developed weapons to create such effects nor incorporated them into their armed forces. This has been the case even though some of them would undoubtedly have military value (environmental weapons and blinding lasers). Blinding lasers are also clearly within the technological capacity of many militaries. Yet none of these weapons have been fielded, partly because the prohibitions are so narrow.[245]

The CCW regulations on land mines and incendiary weapons are of a different nature. They don’t seek to ban a weapon outright, but rather to regulate its use. Like prior rules on air-delivered weapons and submarines that sought to separate military targets from civilians, these rules have generally been a miserable failure.

Many countries did not follow the rules on using marked minefields, in some cases deliberating using mines as a tool to inflict harm on the civilian population. Because mines linger after a conflict, the harm was cumulative. By the mid-1990s, there were more than 110 million mines in 68 countries around the globe. Because they were not in marked minefields, civilians often encountered them, dying by the thousands and being injured by tens of thousands.[246]

Incendiary weapons also continue to be used in populated areas.[247] The Assad regime and Russia have used them in Syria against civilians.[248] Israel also used white phosphorus in populated areas in Gaza in 2008 and 2009.[249] (White phosphorus is not technically considered an incendiary weapon, because its primary purpose is as an obscurant, but it also has incendiary effects.)

**Humanitarian disarmament: Land mines and cluster munitions**

The failure of the CCW protocol on land mines led to a novel development in the history of disarmament: a weapons ban pushed by nongovernmental organizations (NGOs) rather than states. In the early 1990s, frustration over civilian harm from landmines reached a boiling point. In 1991, the Vietnam Veterans of America Foundation and the German NGO Medico agreed to jointly launch an advocacy campaign to bring together NGOs to create a coordinated effort to ban land mines. The following year, the International Campaign to Ban Land Mines was launched, with six NGOs on board. In response to growing concern about the humanitarian consequences of mines, U.S. President George H.W. Bush signed a one-year moratorium on antipersonnel landmines.[250] By 1993, the issue had begun to gain traction internationally. The Campaign to Ban Land Mines held an international conference of 40 NGOs, and France called for the CCW to take up the issue.

The CCW held its first conference on land mines in 1995, but only 14 nations supported a ban. They adjourned without agreement, and the following year states adopted an amended protocol that still permitted antipersonnel land mines. One of the challenges of the CCW is that it is a consensus-based organization. All 125 states in the CCW need to agree in order for it to adopt a measure. This makes it an extremely weak body, with any agreement being invariably watered down to the barest minimum that all can agree on.

On the sidelines of the CCW conference, the Campaign to Ban Land Mines met with representatives of the 14 nations that had spoken out in support of a ban to plan their next move. Many were frustrated and wanted to pass an immediate ban—a more aggressive stance than the CCW's position of working toward mines' "eventual elimination." In October 1996, only a few months after the adoption of the CCW's amended protocol, 75 nations met in Ottawa to discuss a way forward. At the meeting's conclusion, Canada challenged the nations to meet again the following year to negotiate a treaty. In 1997, after six years of vigorous public and private lobbying by NGOs, 122 nations signed the Mine Ban Treaty (Ottawa Treaty) banning the production, stockpiling, transfer, and use of antipersonnel land mines.[251] That fall, the International Campaign to Ban Land Mines and Jody Williams, who led the campaign, were awarded the Nobel Peace Prize.[252]

A few years later, many of the same NGOs spearheaded an effort to ban another weapon causing civilian harm: cluster munitions. Cluster munitions are bombs that release small submunitions, or "bomblets," which then spread out over an area. Cluster munitions are used as an area weapon, for example to render an enemy air base unusable. If the bomblets do not explode, they can remain unexploded on the ground, and months or possibly years later can maim or kill civilians who step on them or pick them up. The problem with cluster munitions is not the releasing of submunitions over a wide area, per se, but rather the "dud rate" of the submunitions. In theory, submunitions with a very low dud rate (a very high percentage of them exploding on contact with the ground) would be safe. In practice, the dud rate for many commonly used cluster munitions ranges anywhere from 5 percent to 15 percent or higher.[253] Because a single cluster bomb may have hundreds of submunitions, even a short war could result in tens of thousands of unexploded bomblets scattered around a country. During the Kosovo air campaign, the United States and Great Britain reportedly dropped 2,000 cluster bombs, releasing 380,000 submunitions. Even assuming the bombs performed at their specified 5 percent dud rate, they would have left nearly 20,000 unexploded bomblets across Serbia.[254]

Longer wars, such as the U.S. bombing of Laos from 1964 to 1973, can leave tens of millions of unexploded bomblets. The resulting "cluster contamination" can leave large swaths of areas no-go zones for civilians. Laos has reported that 37 percent of its farmable land has been rendered unsafe because of unexploded bombs.[255] In 2008, 94 nations came together in Oslo to sign the Convention on Cluster Munitions, which banned the production, stockpiling, transfer, and use of cluster munitions.[256]

Unlike early weapons bans, which prohibited only a weapon's use in war, the bans on land mines and cluster munitions follow in the footsteps of the chemical and biological weapons conventions in also banning production and stockpiling. For many nations, this provision has meant that they had to retire their existing stockpiles of land mines and cluster munitions, as they have had to do with chemical weapons. By taking prohibited weapons out of the hands of states entirely, these treaties adopted a more proactive approach to compliance. This is certainly a decision justified by history. In the heat of warfare, perceived military necessity overcame a desire to refrain from using poison gas, unrestricted submarine warfare, and air attacks against cities. Mutual restraint might have been easier in these circumstances had militaries not had these weapons in their inventories at the outset of war.

One interesting aspect to the bans on land mines and cluster munitions is the elegant way in which they solve the Schelling focal point problem—the importance of having clear rules to help states coordinate their behavior. The prohibitions in both treaties' text are clear and unambiguous. States that sign the land mine ban pledge "never under any circumstances to use anti-personnel mines" and those that join the cluster munitions ban agree "never under any circumstances to use cluster munitions." A ban does not get much more straightforward than that. When you dig into the definitions, however, the details are more complicated. The land mine treaty defines an antipersonnel mine as

> a mine designed to be exploded by the presence, proximity or contact of a person and that will incapacitate, injure or kill one or more persons. Mines designed to be detonated by the presence, proximity or contact of a vehicle as opposed to a person, that are equipped with anti-handling devices, are not considered anti-personnel mines as a result of being so equipped.[257]

Anti-vehicle mines are permitted under this definition, including those with anti-handling devices (that are lethal to people).[258] The ban on cluster munitions has an even more complicated definition:

> **"Cluster munition"** means a conventional munition that is designed to disperse or release explosive submunitions each weighing less than 20 kilograms, and includes those explosive submunitions. It does not mean the following:
> (a) A munition or submunition designed to dispense flares, smoke, pyrotechnics or chaff; or a munition designed exclusively for an air defence role;

(b) A munition or submunition designed to produce electrical or electronic effects;
(c) A munition that, in order to avoid indiscriminate area effects and the risks posed by unexploded submunitions, has all of the following characteristics:
(i) Each munition contains fewer than ten explosive submunitions;
(ii) Each explosive submunition weighs more than four kilograms;
(iii) Each explosive submunition is designed to detect and engage a single target object;
(iv) Each explosive submunition is equipped with an electronic self-destruction mechanism;
(v) Each explosive submunition is equipped with an electronic self-deactivating feature;[259]

The result of the fine print is that many weapons that appear to be cluster munitions are permitted under this definition. This is not an oversight. Some states did not want to give up existing inventories of weapons that could be deemed cluster munitions, and they ensured that the final definition allowed them to retain these munitions. Australia, for instance, made certain that the final agreement would not prohibit its SMArt 115 artillery shells, which dispense two anti-tank submunitions.[260] With the more complex details obscured in the definition, the cluster munitions ban has the appearance of simplicity, which makes it stronger. “No cluster munitions” is much clearer and more straightforward than “some cluster munitions, but not all,” although in reality, this is exactly what the prohibition articulates.

The NGO campaigns to ban land mines and cluster munitions were undoubtedly influential. As of 2022, 164 nations have joined the Mine Ban Treaty banning anti-personnel land mines and 123 nations have joined the Convention on Cluster Munitions.[261] The extent to which these treaties have led to restraint is mixed. The Mine Ban Treaty has unquestionably reduced the number of antipersonnel land mines in the world. Since the treaty was signed, states have removed from their arsenals or destroyed over 53 million antipersonnel mines.[262] As of 2016, 158 states no longer have any stockpiled mines at all.[263]

Still, a number of major military powers—the United States, Russia, and China—have not signed the Mine Ban Treaty. In 2014, Ukraine and Finland suggested that they might withdraw from the treaty.[264] With a revanchist Russia attempting to seize territory in Europe by force, one can understand why neighboring countries might suddenly feel differently. Antipersonnel land mines have clear military value, and their effect on civilians can be mitigated by only using self-deactivating mines that will not persist after the war ends, as the United States does. In the afterglow of the Cold War’s demise, many nations might have felt that they were giving up weapons that they weren’t using anyway. Yet in today’s international security environment, those weapons may seem more relevant.

Antipersonnel mines have played a role in Russia’s invasions of Ukraine, during both its illegal annexation of Crimea in 2014 and its larger invasion in 2022. According to a June 2022 Human Rights Watch report, “Russian forces have used at least seven types of antipersonnel mines in at least four regions of Ukraine: Donetsk, Kharkiv, Kyiv, and Sumy."[265] The report goes on to say, "There is no credible information that Ukrainian government forces have used antipersonnel mines in violation of the Mine Ban Treaty since 2014 and into 2022."[266] Other nations may look at Ukraine and wonder whether the fleeting benefits of international goodwill for signing a treaty are worth the lasting military disadvantage of giving up a valuable weapon for defending their borders.

Still, the normative power of these bans in stigmatizing certain weapons is undeniable and can put tremendous pressure on states. In 2014, the United States pledged to refrain from using antipersonnel mines outside of the Korean Peninsula.[267] The cluster munitions ban has been less successful. Fewer nations have signed it, and there has been continued use of cluster munitions by Syria, Saudi Arabia, and, most recently, Russia, during its 2022 invasion of Ukraine.[268] The treaty has had some effect, though. No treaty signatories have used cluster munitions, and 29 countries have completely destroyed their stockpiles.[269] Even some countries that have not signed, such as the United States, have changed their cluster munitions policies as a result.[270]

**Nonproliferation agreements: Controlling the spread of dangerous technology**

In addition to treaties that ban weapons, regulate use, or limit arms, there is another class of arms control treaty: nonproliferation regimes. These regimes, sometimes in the form of non-legally-binding agreements, seek to limit the spread of a harmful technology to reduce its availability.

The most significant and well-known nonproliferation regime is the Treaty on the Non-Proliferation of Nuclear Weapons (NPT). Established in 1970, the treaty was designed to control the spread of nuclear weapons. At the time, many believed that nuclear weapons would rapidly proliferate, with as many as 25 to 30 nations possessing them within a few decades.[271] The NPT was designed to halt this process. The NPT recognizes only five nuclear weapons states: the United States, Soviet Union (now Russia), Britain, France, and China. (They are also the five permanent members of the U.N. Security Council.) The NPT prohibits any other signatories from acquiring nuclear weapons. In exchange, the treaty permits all nations access to nuclear power for peaceful purposes. It also commits the five preexisting nuclear powers to "pursue negotiations in good faith on effective measures relating to cessation of the nuclear arms race at an early date and to nuclear disarmament."[272]

By the standards of expected proliferation at the time, the agreement is a tremendous success. Today only nine nations have nuclear weapons, instead of the 25 to 30 that many predicted. In addition to the five nations listed in the NPT, India, Pakistan, and North Korea have tested nuclear weapons, and Israel is widely believed to have nuclear weapons as well (but has never publicly confirmed that it does).[273] As a treaty, the NPT has held up well. North Korea is the only country ever to withdraw from the NPT. India, Pakistan, and Israel never signed the treaty. The NPT is credited with rolling back or containing the nuclear ambitions of a number of countries, including Libya, Syria, South Korea, Iran, and South Africa.

Several other nonproliferation agreements followed the NPT. The Australia Group (1985) limits the spread of technologies that could assist in developing chemical or biological weapons.[274] The Missile Technology Control Regime (MTCR) (1987) is intended to slow the spread of technologies that could be used in nuclear-capable missiles and unmanned vehicles.[275] The Wassenaar Arrangement (1996) is an agreement among 42 countries to restrict export of a variety of conventional weapons and dual-use technologies, from tanks and artillery to lasers and small arms.[276] The Hague Code of Conduct (2002) has a similar goal to the MTCR's, preventing the spread of WMD-capable ballistic missiles, but has a larger membership, with 138 member states to the MTCR's 35.[277] Unlike the NPT, which is a legally binding treaty, none of these other agreements are legally binding. They do not commit states to giving up these technologies. States simply agree not to export them.

Nonproliferation regimes fall into two categories. Those in the first category, which the MTCR, Wassenaar Arrangement, and Hague Code of Conduct fall into, are intended to prevent the spread of weapons that many signatories themselves have. This effectively makes them security cartels. Countries with these technologies have a military advantage over others that do not, and they are attempting to retain that military advantage by collectively agreeing to restrict exports. The Australia Group is different, in that it is aimed at restricting access to chemical and biological weapons, which states have already agreed to give up. The NPT straddles the gap, with the stated intention of moving to a world without any nuclear weapons, but for now permitting nuclear arms by some states.

Because these agreements do not ban possessing weapons, except for the NPT, noncompliance is not generally a major concern for nonproliferation regimes. States that join have their own incentives to slow the spread of these technologies to others. That doesn't mean these regimes are perfect. At best, they can be seen as slowing down a natural process of technology diffusion across the international system. How successful they are in doing so depends heavily on how hard it is for states outside the regime to develop these weapons on their own or through dual-use commercial applications. The NPT has had tremendous success because building nuclear weapons is difficult. The MTCR, on the other hand, has been significantly challenged by the rapid proliferation of uninhabited aircraft.[278]

## Appendix B.
## Summary of Historical Attempts at Arms Control

| Era | Weapon | Year | Regulation or Treaty | Legally binding? | Type of Regulation | Successful? | Motivation |
|---|---|---|---|---|---|---|---|
| Pre-Modern Era | poisoned or barbed arrows | Dates vary—1500 to 200 BC | Laws of Manu; Dharmaśāstras; Mahābhārata | legally binding | banned use | success unknown | avoid unnecessary suffering |
| | concealed weapons | Dates vary—1500 to 200 BC | Laws of Manu | legally binding | banned use | success unknown | inhibit perfidy |
| | fire-tipped weapons | Dates vary—1500 to 200 BC | Laws of Manu | legally binding | banned use | success unknown | avoid unnecessary suffering |
| | crossbow | 1097; 1139 | 1097 Lateran Synod; 1139 Second Lateran Council | legally binding | banned use | failed | retain political control |
| | firearms | 1607–1867 | Order by Tokugawa Shogunate, Japan | legally binding | effectively prohibited production | successful (lasted ~250 years) | retain political control |
| | firearms | 1523–1543 | Order by King Henry VIII (by Act of Parliament) | legally binding | limited ownership among civilian population | short lived | retain political control |
| | poisoned bullets | 1675 | Strasbourg Agreement | legally binding | banned use | successful | avoid unnecessary suffering |

| Era | Weapon | Year | Regulation or Treaty | Legally binding? | Type of Regulation | Successful? | Motivation |
|---|---|---|---|---|---|---|---|
| Turn of the Century | explosive or inflammable projectiles below 400 grams | 1868 | 1868 St. Petersburg Declaration | legally binding | banned use | superseded by technology, but adhered to in spirit | avoid unnecessary suffering |
| | expanding bullets | 1899 | 1899 Hague Declaration | legally binding | banned use | successful in limiting battlefield use, although lawful in civilian applications | avoid unnecessary suffering |
| | asphyxiating gases (from projectiles) | 1899 | 1899 Hague Declaration | legally binding | banned use | failed—used in WWI | avoid unnecessary suffering |
| | poison | 1899; 1907 | 1899 and 1907 Hague Declarations | legally binding | banned use | successful | avoid unnecessary suffering |
| | weapons that cause superfluous injury | 1899; 1907 | 1899 and 1907 Hague Declarations | legally binding | banned use | mixed, but generally successful | avoid unnecessary suffering |
| | balloon-delivered projectiles or explosives | 1899; 1907 | 1899 and 1907 Hague Declarations | legally binding | banned use | short lived | reduce civilian casualties |
| | aerial bombardment against undefended cities | 1907 | 1907 Hague Declaration | legally binding | banned use | failed | reduce civilian casualties |

| Era | Weapon | Year | Regulation or Treaty | Legally binding? | Type of Regulation | Successful? | Motivation |
|---|---|---|---|---|---|---|---|
| World War I to World War II | sawback bayonets | World War I | tacit cooperation on the battlefield | no explicit agreement | norm against possession | successful | avoid unnecessary suffering |
| | chemical and bacteriological weapons | 1925 | 1925 Geneva Gas and Bacteriological Protocol | legally binding | banned use | largely successful in restraining battlefield use in WWII | avoid unnecessary suffering |
| | submarines | 1899; 1921–1922 | 1899 Hague convention; 1921–1922 Washington Naval Conference | never ratified | attempted bans—never ratified | failed—treaty never ratified | reduce civilian casualties |
| | submarines | 1907; 1930; 1936 | 1907 Hague Declaration; 1930 London Naval Treaty; 1936 London Protocol | legally binding | regulated use | failed—compliance collapsed in war | reduce civilian casualties |
| | size of navies | 1922; 1930; 1936 | 1922 Washington Naval Treaty; 1930 London Naval Treaty; 1936 Second London Naval Treaty | legally binding | limited quantities and size of ships | short lived | limit arms races |

| Era | Weapon | Year | Regulation or Treaty | Legally binding? | Type of Regulation | Successful? | Motivation |
|---|---|---|---|---|---|---|---|
| Cold War | nuclear tests | 1963; 1967; 1985; 1995; 1996 | Limited Test Ban Treaty; Treaty of Tlatelolco; Treaty of Raro tonga; Treaty of Bangkok; Treaty of Pelindaba; Comprehensive Nuclear Test Ban Treaty | legally binding | restricted testing | generally successful, with some exceptions | reduce effects on civilians; limit arms races |
| | weapons in Antarctica | 1959 | Antarctic Treaty | legally binding | banned deployment | successful | limit arms races |
| | weapons of mass destruction in space | 1967 | Outer Space Treaty | legally binding | banned deployment | successful | maintain strategic stability |
| | weapons on the moon | 1967 | Outer Space Treaty | legally binding | banned deployment | successful | limit arms races |
| | nuclear-free zones | 1967; 1985; 1995; 1996 | Treaty of Tlatelolco; Treaty of Rarotonga; Treaty of Bangkok; Treaty of Pelindaba | legally binding | banned developing, manufacturing, possessing, or stationing | successful | limit arms races |
| | nuclear weapons | 1970 | Nuclear Non-Proliferation Treaty | legally binding | banned proliferation | generally successful, with some exceptions | maintain strategic stability |
| | nuclear weapons on the seabed | 1971 | Seabed Treaty | legally binding | banned deployment | successful | maintain strategic stability |

| Era | Weapon | Year | Regulation or Treaty | Legally binding? | Type of Regulation | Successful? | Motivation |
|---|---|---|---|---|---|---|---|
| Cold War | ballistic missile defenses | 1972 | Anti-Ballistic Missile Treaty | legally binding | limited deployment | successful during Cold War; collapsed in multipolar world | maintain strategic stability |
| | biological weapons | 1972 | Biological Weapons Convention | legally binding | banned development, production, stockpiling, and use | generally successful, with some exceptions | avoid unnecessary suffering; reduce civilian casualties; prevent arms race |
| | using the environment as a weapon | 1976 | Environmental Modification Convention | legally binding | banned use | successful | reduce civilian casualties; prevent arms race |
| | anti-satellite weapons | 1970s & 1980s | tacit cooperation between U.S. and USSR | no explicit agreement | norm against deployment | successful, but currently threatened in multipolar world | maintain strategic stability |
| | neutron bombs | 1970s | tacit cooperation between U.S. and USSR | no explicit agreement | norm against deployment | successful | maintain strategic stability |
| | non-x-ray detectable fragments | 1980 | Convention on Certain Conventional Weapons (CCW) Protocol I | legally binding | banned use | successful | avoid unnecessary suffering |

| Era | Weapon | Year | Regulation or Treaty | Legally binding? | Type of Regulation | Successful? | Motivation |
|---|---|---|---|---|---|---|---|
| Cold War | land mines | 1980 | CCW Protocol II | legally binding | regulated use | unsuccessful | reduce civilian casualties |
| | incendiary weapons | 1980 | CCW Protocol III | legally binding | regulated use | mixed success | reduce civilian casualties |
| | chemical and biological weapons | 1985 | Australia Group | not legally binding | banned proliferation | generally successful, with some exceptions | avoid unnecessary suffering; reduce civilian casualties |
| | ballistic and cruise missiles | 1987; 2002 | Missile Technology Control Regime; Hague Code of Conduct | not legally binding | limited proliferation | has had some success | maintain strategic stability |
| | intermediate-range missiles | 1987 | Intermediate-Range Nuclear Forces (INF) Treaty | legally binding | banned possession | successful during Cold War; collapsed in multipolar world | maintain strategic stability |
| | nuclear weapons and launcher quantities | 1972; 1979; 1991; 2002; 2011 | SALT I; SALT II; START; SORT; New START | legally binding | limited quantities | successful | limit arms races |
| | conventional air and ground forces | 1991 | Conventional Forces in Europe | legally binding | limited quantities | collapsed in multipolar world | limit arms races |

| Era | Weapon | Year | Regulation or Treaty | Legally binding? | Type of Regulation | Successful? | Motivation |
|---|---|---|---|---|---|---|---|
| Post–Cold War | chemical weapons | 1993 | Chemical Weapons Convention | legally binding | banned development, production, stockpiling, and use | generally successful, with some exceptions | avoid unnecessary suffering; reduce civilian casualties |
| | blinding lasers | 1995 | CCW Protocol IV | legally binding | banned use | successful | avoid unnecessary suffering |
| | conventional weapons | 1996 | Wassenaar Arrangement | not legally binding | limited proliferation | has had some success | retain political control |
| | land mines | 1997 | Mine Ban Treaty (Ottawa Treaty) | legally binding | banned development, production, stockpiling, and use | generally successful, with some exceptions | reduce civilian casualties |
| | cluster munitions | 2008 | Convention on Cluster Munitions | legally binding | banned development, production, stockpiling, and use | generally successful, with some exceptions | reduce civilian casualties |

Source: Used with permission. Scharre, “Autonomous weapons and stability.”

## Appendix C.
## List of International Agreements

| Official Title | Informal Title or Acronym (referenced in report) |
|---|---|
| Strasbourg Agreement (1675) | N/A |
| Declaration Renouncing the Use, in Time of War, of Explosive Projectiles Under 400 Grammes Weight (1868) | 1868 St. Petersburg Declaration |
| | 1868 ban on exploding bullets |
| The First Hague Convention (1899) | 1899 Hague convention |
| | 1899 ban on expanding bullets |
| | 1899 ban concerning expanding bullets |
| | 1899 Hague declarations on balloon-delivered weapons, expanding bullets, and gas-filled projectiles |
| | 1899 Hague prohibitions |
| The Second Hague Convention (1907) | 1907 Hague conference |
| | 1907 Hague prohibitions |
| Protocol for the Prohibition of the Use in War of Asphyxiating, Poisonous or Other Gases, and of Bacteriological Methods of Warfare (1925) | Geneva Gas Protocol |
| | Geneva Gas and Bacteriological Protocol |
| Washington Naval Treaty (1922) | Five-Power Treaty |
| Treaty for the Limitation and Reduction of Naval Armament (1930) | London Naval Treaty |
| Second London Naval Treaty (1936) | N/A |
| The Antarctic Treaty (1959) | N/A |
| Treaty Banning Nuclear Weapons Tests in the Atmosphere, in Outer Space, and Under Water (1963) | Limited Test Ban Treaty |
| Treaty for the Prohibition of Nuclear Weapons in Latin America and the Caribbean (1967) | The Treaty of Tlatelolco |
| Treaty on Principles Governing the Activities of States in the Exploration and Use of Outer Space, Including the Moon and Other Celestial Bodies (1967) | Outer Space Treaty |
| Treaty on the Non-Proliferation of Nuclear Weapons (1970) | Nuclear Nonproliferation Treaty |
| | NPT |
| Treaty on the Prohibition of the Emplacement of Nuclear Weapons and Other Weapons of Mass Destruction on the Seabed and the Ocean Floor and in the Subsoil Thereof (1971) | Seabed Treaty |
| Treaty Between the United States of America and The Union of Soviet Socialist Republics on The Limitation of Anti-Ballistic Missile Systems (1972) | Anti-Ballistic Missile Treaty |
| | ABM Treaty |
| Strategic Arms Limitation Talks I (1972) and II agreements (1979) | SALT I and SALT II |
| The Convention on the Prohibition of the Development, Production and Stockpiling of Bacteriological (Biological) and Toxin Weapons and on their Destruction (1972) | Biological Weapons Convention |
| | BWC |
| Convention on the Prohibition of Military or Any Other Hostile Use of Environmental Modification Techniques (1976) | Environmental Modification Convention |
| Convention on Certain Conventional Weapons (1980) | CCW |
| South Pacific Nuclear Free Zone Treaty (1985) | The Treaty of Rarotonga |
| Australia Group (1985) | N/A |
| Treaty Between the United States of America And The Union Of Soviet Socialist Republics On The Elimination Of Their Intermediate-Range And Shorter-Range Missiles (1987) | Intermediate-Range Nuclear Forces Treaty |
| | INF Treaty |
| Missile Technology Control Regime (1987) | MTCR |
| Strategic Arms Reduction Treaty (1991) | START |
| Treaty on Conventional Armed Forces in Europe (1991) | CFE |
| Convention on the Prohibition of the Development, Production, Stockpiling and Use of Chemical Weapons and on their Destruction (1993) | Chemical Weapons Convention |
| | CWC |
| Treaty on the Southeast Asia Nuclear Weapon-Free Zone (1995) | Treaty of Bangkok |
| African Nuclear-Weapon-Free Zone Treaty (1996) | Treaty of Pelindaba |
| Comprehensive Nuclear Test Ban Treaty (1996) | N/A |
| Wassenaar Arrangement (1996) | N/A |
| Convention on the Prohibition of the Use, Stockpiling, Production and Transfer of Anti-Personnel Mines and on their Destruction (1997) | Mine Ban Treaty |
| | Ottawa Treaty |
| | 1997 ban on antipersonnel landmines |
| Treaty Between the United States of America and the Russian Federation on Strategic Offensive Reductions (2002) | Strategic Offensive Reductions Treaty |
| | The Moscow Treaty |
| | SORT |
| Hague Code of Conduct (2002) | N/A |
| Convention on Cluster Munitions (2008) | 2008 ban on cluster munitions |
| | Convention on Cluster Munitions |
| Measures for the Further Reduction and Limitation of Strategic Offensive Arms (2011) | New START |

## Notes

[1] "Autonomous Weapons: An Open Letter from AI & Robotics Researchers," Future of Life Institute, July 28, 2015, https://futureoflife.org/open-letter-autonomous-weapons/; "Lethal Autonomous Weapons Pledge," Future of Life Institute, https://futureoflife.org/lethal-autonomous-weapons-pledge/; Adam Satariano, "Will There Be a Ban on Killer Robots?" *The New York Times,* October 19, 2018, https://www.nytimes.com/2018/10/19/technology/artificial-intelligence-weapons.html; "Less Autonomy, More Humanity," Stop Killer Robots, https://www.stopkillerrobots.org/; Matt McFarland, "Leading AI researchers vow to not develop autonomous weapons," CNNMoney, July 18, 2018, https://money.cnn.com/2018/07/18/technology/ai-autonomous-weapons/index.html; Tsuya Hisashi, "Can the use of AI weapons be banned?" NHK, April 18, 2019, https://www3.nhk.or.jp/nhkworld/en/news/backstories/441/; and Mary Wareham, "Stopping Killer Robots: Country Positions on Banning Fully Autonomous Weapons and Retaining Human Control" (Human Rights Watch, August 2020), https://www.hrw.org/sites/default/files/media_2020/08/arms0820_web_0.pdf.

[2] Will Knight, "AI arms control may not be possible, warns Henry Kissinger," *MIT Technology Review*, March 1, 2019, https://www.technologyreview.com/f/613059/ai-arms-control-may-not-be-possible-warns-henry-kissinger/; Vincent Boulanin, "Regulating military AI will be difficult. Here's a way forward," *Bulletin of the Atomic Scientists*, March 3, 2021, https://thebulletin.org/2021/03/regulating-military-ai-will-be-difficult-heres-a-way-forward/; Forrest E. Morgan, Benjamin Boudreaux, Andrew J. Lohn, Mark Ashby, Christian Curriden, Kelly Klima, and Derek Grossman, "Military Applications of Artificial Intelligence: Ethical Concerns in an Uncertain World" (RAND Corporation, 2020), https://www.rand.org/pubs/research_reports/RR3139-1.html; National Security Commission on Artificial Intelligence, *Final Report* (March 2021), 96, https://www.nscai.gov/wp-content/uploads/2021/03/Full-Report-Digital-1.pdf; and Evan Ackerman, "We Should Not Ban 'Killer Robots' and Here's Why: What we really need is a way of making autonomous armed robots ethical, because we're not going to be able to prevent them from existing," IEEE Spectrum, July 28, 2015, https://spectrum.ieee.org/we-should-not-ban-killer-robots.

[3] Michael C. Horowitz, "Artificial Intelligence, International Competition, and the Balance of Power," *Texas National Security Review,* 1 no. 3 (May 2018), https://tnsr.org/2018/05/artificial-intelligence-international-competition-and-the-balance-of-power/.

[4] Sean Watts, "Regulation-Tolerant Weapons, Regulation-Resistant Weapons and the Law of War," *International Law Studies,* 91 (August 2015), https://digital-commons.usnwc.edu/cgi/viewcontent.cgi?article=1411&context=ils; Sean Watts, "Autonomous Weapons: Regulation Tolerant or Regulation Resistant?"*Temple International and Comparative Law Journal,* 30 no. 1 (2016), https://sites.temple.edu/ticlj/files/2017/02/30.1.Watts-TICLJ.pdf; Rebecca Crootof, "The Killer Robots Are Here: Legal and Policy Implications," *Cardozo Law Review,* 36 no. 5 (December 2014), http://cardozolawreview.com/wp-content/uploads/2018/08/CROOTOF.36.5.pdf; and Rebecca Crootof, "Why the Prohibition on Permanently Blinding Lasers is Poor Precedent for a Ban on Autonomous Weapon Systems," Lawfare, November 24, 2015, https://www.lawfareblog.com/why-prohibition-permanently-blinding-lasers-poor-precedent-ban-autonomous-weapon-systems.

[5] Michael C. Horowitz and Paul Scharre, "AI and International Stability: Risks and Confidence-Building Measures" (Center for a New American Security, January 2021), https://www.cnas.org/publications/reports/ai-and-international-stability-risks-and-confidence-building-measures. Some definitions of arms control include post-conflict disarmament imposed by the victors on losing states, such as the Treaty of Versailles. For alternative definitions of arms control, see "Arms control, disarmament and non-proliferation in NATO," NATO, April 6, 2022, https://www.nato.int/cps/en/natohq/topics_48895.htm; Thomas C. Schelling and Morton H. Halperin, *Strategy and Arms Control* (Washington, DC: Pergamon-Brassey's, 1985), 2; Robert R. Bowie, "Basic Requirements of Arms Control," *Daedalus* 89 no. 4 (Fall 1960), 708, http://www.jstor.org/stable/20026612; Hedley Bull, "Arms Control and World Order," *International Security* 1, no. 1 (Summer 1976), 3, https://www.jstor.org/stable/2538573; Julian Schofield, "Arms Control Failure and the Balance of Power," *Canadian Journal of Political Science / Revue Canadienne de Science Politique*, 33 no. 4 (December 2000), 748, http://www.jstor.org/stable/3232662; Coit D. Blacker and Gloria Duffy, *International Arms Control: Issues and Agreements* (Stanford, CA: Stanford University Press, 1984), 3;

Lionel P. Fatton, “The impotence of conventional arms control: why do international regimes fail when they are most needed?" *Contemporary Security Policy*, 37 no. 2 (June 2016), 201, https://doi.org/10.1080/13523260.2016.1187952; and Henry A. Kissinger, “Arms Control, Inspection and Surprise Attack,” *Foreign Affairs*, 38 no. 4 (July 1960), 559, https://www.foreignaffairs.com/articles/1960-07-01/arms-control-inspection-and-surprise-attack.

[6] Used with permission. Paul Scharre, “Autonomous weapons and stability” (PhD diss., King’s College London, March 2020), https://kclpure.kcl.ac.uk/portal/files/129451536/2020_Scharre_Paul_1575997_ethesis.pdf.

[7] Andrew J. Coe and Jane Vaynman, “Why Arms Control Is So Rare,” *American Political Science Review*, 114 no. 2 (May 2020), 342–55, https://www.cambridge.org/core/journals/american-political-science-review/article/abs/why-arms-control-is-so-rare/BAC79354627F72CDDDB102FE82889B8A: John D. Maurer, “The Purposes of Arms Control,” *Texas National Security Review* 2, no. 1 (November 2018), https://tnsr.org/2018/11/the-purposes-of-arms-control/; Charles H. Anderton and John R. Carter, “Arms Rivalry, Proliferation, and Arms Control,” in *Principles of Conflict Economics: A Primer for Social Scientists*, eds. Charles H. Anderton and John R. Carter (Cambridge: Cambridge University Press, 2009), 185–221, https://doi.org/10.1017/CBO9780511813474.011; Andrew Webster, “From Versailles to Geneva: The many forms of interwar disarmament,” *Journal of Strategic Studies*, 29 no. 2 (2006), 225–246, https://www.tandfonline.com/doi/abs/10.1080/01402390600585050; Charles L. Glaser, “When Are Arms Races Dangerous? Rational versus Suboptimal Arming,” *International Security*, 28 no. 4 (Spring 2004), 44–84, http://www.jstor.org/stable/4137449; Robert Jervis, "Arms Control, Stability, and Causes of War,” *Political Science Quarterly*, 108 no. 2 (Summer 1993), 239–253, https://doi.org/10.2307/2152010; and Marc Trachtenberg, “The Past and Future of Arms Control,” *Daedalus*, 120 no. 1 (Winter 1991), 203–216, http://www.jstor.org/stable/20025364.

[8] Lionel P. Fatton, “The impotence of conventional arms control: why do international regimes fail when they are most needed?”, *Contemporary Security Policy*, 37 no. 2 (2016), 200–222, https://doi.org/10.1080/13523260.2016.1187952; Andrew Kydd, “Arms Races and Arms Control: Modeling the Hawk Perspective,” *American Journal of Political Science*, 44 no. 2 (2000), 228–229, https://doi.org/10.2307/2669307; Colin Gray, *House of Cards: Why Arms Control Must Fail* (Cornell University Publishing, 1992), 5, 27; Stuart Croft, *Strategies of arms control* (Manchester University Press: 1996), 5.

[9] Coe and Vaynman, “Why Arms Control Is So Rare,” 353; Jane Vaynman, “Enemies in Agreement: Domestic Politics, Uncertainty, and Cooperation Between Adversaries” (PhD diss., Harvard University, 2014), 12–16.

[10] Used with permission. Paul Scharre, “Autonomous weapons and stability.”

[11] Watts, “Regulation-Tolerant Weapons, Regulation-Resistant Weapons, and the Law of War”; Watts, “Autonomous Weapons: Regulation Tolerant or Regulation Resistant?”

[12] Watts, “Regulation-Tolerant Weapons, Regulation-Resistant Weapons, and the Law of War,” 609–618.

[13] Rebecca Crootof defines a “successful” weapons ban as "both enacted and effective at limiting the usage of the banned weapon.” Crootof, “The Killer Robots Are Here,” 1910; Crootof, “Why the Prohibition on Permanently Blinding Lasers is Poor Precedent for a Ban on Autonomous Weapon Systems.”

[14] Crootof, “The Killer Robots Are Here,” 1884–1890.

[15] Crootof, “The Killer Robots Are Here,” 1888.

[16] The use of “means and methods of warfare which are of a nature to cause superfluous injury or unnecessary suffering” is barred under customary international humanitarian law: “Rule 70. Weapons of a Nature to Cause Superfluous Injury or Unnecessary Suffering,” IHL database, International Committee of the Red Cross, https://ihl-databases.icrc.org/customary-ihl/eng/docs/v1_rul_rule70.

[17] “Protocol of Non-Detectable Fragments (Protocol I). Geneva, 10 October 1980,” International Committee of the Red Cross, https://ihl-databases.icrc.org/applic/ihl/ihl.nsf/Article.xsp?action=openDocument&documentId=1AF77FFE8082AE07C12563CD0051EDF5.

[18] Thomas C. Schelling, *The Strategy of Conflict* (Cambridge, MA: Harvard University Press, 1980), 75.

[19] Crootof, "The Killer Robots Are Here," 1890.

[20] "The German Sawback Blade Bayonet," Armourgeddon Blog, January 22, 2015, https://www.armourgeddon.co.uk/the-german-sawback-blade-bayonet.html; and Used with permission. Scharre, "Autonomous weapons and stability."

[21] Charles J. Dunlap Jr., "Is it Really Better to be Dead than Blind?" Just Security, January 13, 2015, https://www.justsecurity.org/19078/dead-blind/.

[22] Crootof, "Why the Prohibition on Permanently Blinding Lasers is Poor Precedent for a Ban on Autonomous Weapon Systems."

[23] Used with permission. Scharre, "Autonomous weapons and stability."

[24] "Each State Party undertakes not to use riot control agents as a method of warfare." Convention on the Prohibition of the Development, Production, Stockpiling and Use of Chemical Weapons and on Their Destruction, Article I.5, August 31, 1994, 2, https://www.opcw.org/fileadmin/OPCW/CWC/CWC_en.pdf; Executive Order No. 11850, 3 C.F.R. 980 (1975), https://www.archives.gov/federal-register/codification/executive-order/11850.html; Michael Nguyen, "Senate Struggles with Riot Control Agent Policy," *Arms Control Today*, 36 no. 1 (January–February 2006), https://www.armscontrol.org/act/2006_01-02/JANFEB-RiotControl; and Used with permission. Scharre, "Autonomous weapons and stability."

[25] This dynamic seems to suggest that if lasers were used in future wars for non-blinding purposes and ended up causing incidental blinding, then their use would quickly evolve to include intentional blinding.

[26] "Convention Text," Convention on Cluster Munitions, https://www.clusterconvention.org/convention-text/; International Campaign to Ban Landmines, "Treaty in Detail," http://www.icbl.org/en-gb/the-treaty/treaty-in-detail/treaty-text.aspx.

[27] Used with permission. Scharre, "Autonomous weapons and stability."

[28] Protocol for the Prohibition of the Use in War of Asphyxiating, Poisonous or Other Gases, and of Bacteriological Methods of Warfare (Geneva Protocol), June 17, 1925, https://2009-2017.state.gov/t/isn/4784.htm.

[29] Used with permission. Scharre, "Autonomous weapons and stability."

[30] "Any State Party to this Convention which has reason to believe that any other State Party is acting in breach of obligations deriving from the provisions of the Convention may lodge a complaint with the Security Council of the United Nations. Such a complaint should include all relevant information as well as all possible evidence supporting its validity." Convention on the Prohibition of Military or Any Other Hostile Use of Environmental Modification Techniques, October 5, 1978, https://www.state.gov/t/isn/4783.htm#treaty.

[31] Used with permission. Scharre, "Autonomous weapons and stability."

[32] Used with permission. Scharre, "Autonomous weapons and stability."

[33] Used with permission. Scharre, "Autonomous weapons and stability."

[34] Vincent Boulanin, Lora Saalman, Petr Topychkanov, Fei Su, and Moa Peldán Carlsson, "Artificial Intelligence, Strategic Stability and Nuclear Risk" (Stockholm International Peace Research Institute, June 2020), https://www.sipri.org/publications/2020/other-publications/artificial-intelligence-strategic-stability-and-nuclear-risk; Technology for Global Security, "AI and the Military: Forever Altering Strategic Stability" (T4GS , February 13, 2019), https://securityandtechnology.org/wp-content/uploads/2020/07/ai_and_the_military_forever_altering_strategic_stability__IST_research_paper.pdf; Forrest E. Morgan, Benjamin Boudreaux, Andrew J. Lohn, Mark Ashby, Christian Curriden, Kelly Klima, and Derek Grossman, "Military Applications of Artificial Intelligence: Ethical Concerns in an Uncertain World" (RAND Corporation, 2020), https://www.rand.org/pubs/research_reports/RR3139-1.html; Michael C. Horowitz,

Paul Scharre, and Alexander Velez-Green, "A Stable Nuclear Future? The Impact of Autonomous Systems and Artificial Intelligence," 2019, https://arxiv.org/abs/1912.05291; Edward Geist and Andrew J. Lohn, "How Might Artificial Intelligence Affect the Risk of Nuclear War?" (RAND Corporation, 2018), https://www.rand.org/pubs/perspectives/PE296.html; Ben Buchanan, "A National Security Research Agenda for Cybersecurity and Artificial Intelligence," CSET Issue Brief (Center for Security and Emerging Technology, May 2020), https://cset.georgetown.edu/research/a-national-security-research-agenda-for-cybersecurity-and-artificial-intelligence/; Michael C. Horowitz, Lauren Kahn, Christian, Ruhl, Mary Cummings, Erik Lin-Greenberg, Paul Scharre, and Rebecca Slayton, "Policy Roundtable: Artificial Intelligence and International Security," *Texas National Security Review*, June 2, 2020, https://tnsr.org/roundtable/policy-roundtable-artificial-intelligence-and-international-security/; Melanie Sisson, Jennifer Spindel, Paul Scharre, and Vadim Kozyulin, "The Militarization of Artificial Intelligence" (Stanley Center for Peace and Security, August 2019), https://stanleycenter.org/publications/militarization-of-artificial-intelligence/; Giacomo Persi Paoli, Kerstin Vignard, David Danks, and Paul Meyer, "Modernizing Arms Control: Exploring responses to the use of AI in military decision-making" (United Nations Institute for Disarmament Research, 2020), https://www.unidir.org/publication/modernizing-arms-control; Andrew Imbrie and Elsa B. Kania, "AI Safety, Security, and Stability Among Great Powers: Options, Challenges, and Lessons Learned for Pragmatic Engagement," CSET Policy Brief (Center for Security and Emerging Technology, December 2019), https://cset.georgetown.edu/research/ai-safety-security-and-stability-among-great-powers-options-challenges-and-lessons-learned-for-pragmatic-engagement/; Michael C. Horowitz, Lauren Kahn, and Casey Mahoney, "The Future of Military Applications of Artificial Intelligence: A Role for Confidence-Building Measures?" *Orbis*, 64 no. 4 (Fall 2020), 528–543; and Horowitz and Scharre, "AI and International Stability."

[35] Rebecca Crootof, "Regulating New Weapons Technology," in *The Impact of Emerging Technologies on the Law of Armed Conflict,* Eric Talbot Jensen and Ronald T.P. Alcala, eds. (New York: Oxford University Press, 2019), https://papers.ssrn.com/sol3/papers.cfm?abstract_id=3195980; and Rebecca Crootof and BJ Ard, "Structuring Techlaw," *Harvard Journal of Law and Technology,* 34 no. 2 (Spring 2021), https://papers.ssrn.com/sol3/papers.cfm?abstract_id=3664124.

[36] For more on privacy-preserving approaches for sharing information and verifying algorithms' behavior, see Andrew Trask, Emma Bluemke, Ben Garfinkel, Claudia Ghezzou Cuervas-Mons, and Allan Dafoe,"Beyond Privacy Trade-offs with Structured Transparency," December 15, 2020, https://arxiv.org/pdf/2012.08347.pdf; Joshua A. Kroll , Joanna Huey , Solon Barocas , Edward W. Felten , Joel R. Reidenberg , David G. Robinson, and Harlan Yu, "Accountable Algorithms," *University of Pennsylvania Law Review*, 165 no. 3 (2017), https://scholarship.law.upenn.edu/penn_law_review/vol165/iss3/3/; and Matthew Mittelsteadt, "AI Verification: Mechanisms to Ensure AI Arms Control Compliance," CSET Issue Brief (Center for Security and Emerging Technology, February 2021), https://cset.georgetown.edu/publication/ai-verification/.

[37] One possible solution to the problem of post-inspection software updates could be installing continuous monitoring devices that would alert inspectors to any changes in software. Adopting such an approach requires further technological advancements, as well as states' commitment to continuous intrusive monitoring, rather than periodic inspections. It is also possible that such an approach, if implemented, could have unforeseen destabilizing effects in certain scenarios. For example, a software update to improve functionality on the eve of a conflict could trigger an alert that would lead other states to assume arms control noncompliance. Alternatively, regime-compliant code that should not be altered could be embedded into physical hardware, for example through read-only memory (ROM) or application-specific integrated circuits (ASICs). See Mittelsteadt, "AI Verification," 18–24.

[38] For an example of how such an approach might be implemented, see Ronald C. Arkin, Leslie Kaelbling, Stuart Russell, Dorsa Sadigh, Paul Scharre, Bart Selman, and Toby Walsh, "A Path Towards Reasonable Autonomous Weapons Regulation: Experts representing a diversity of views on autonomous weapons systems collaborate on a realistic policy roadmap," *IEEE Spectrum*, October 21, 2019, https://spectrum.ieee.org/a-path-towards-reasonable-autonomous-weapons-regulation.

[39] For example, see the INF, SALT I, SALT II, START, SORT, and New START Treaties.

[40] Saif M. Khan, "U.S. Semiconductor Exports to China: Current Policies and Trends," CSET Issue Brief (Center for Security and Emerging Technology, October 2020), https://cset.georgetown.edu/publication/u-s-semiconductor-exports-to-china-current-policies-and-trends/.

[41] “AI and Compute,” OpenAI blog, May 16, 2018, https://openai.com/blog/ai-and-compute/; Jaime Sevilla et al., *Compute Trends Across Three Eras of Machine Learning* (arXiv.org, March 9, 2022), https://arxiv.org/pdf/2202.05924.pdf; Ryan Carey, “Interpreting AI compute trends,” AI Impacts blog, July 2018, https://aiimpacts.org/interpreting-ai-compute-trends/; and Ben Dickson, “DeepMind’s big losses, and the questions around running an AI lab,” VentureBeat, December 27, 2020, https://venturebeat.com/2020/12/27/deepminds-big-losses-and-the-questions-around-running-an-ai-lab/.

[42] “AI and Efficiency,” OpenAI blog, May 5, 2020, https://openai.com/blog/ai-and-efficiency/.

[43] Rule No. 2020-18213, 85 Fed. Reg. 51596 (August 20, 2020), https://www.federalregister.gov/documents/2020/08/20/2020-18213/addition-of-huawei-non-us-affiliates-to-the-entity-list-the-removal-of-temporary-general-license-and; Jeanne Whalen and Ellen Nakashima, “U.S. tightens restrictions on Huawei yet again, underscoring the difficulty of closing trade routes,” *Washington Post*, August 17, 2020, https://www.washingtonpost.com/business/2020/08/17/us-cracks-down-huawei-again/; David Shepardson, “U.S. tightening restrictions on Huawei access to technology, chips,” Reuters, August 17, 2020, https://www.reuters.com/article/us-usa-huawei-tech-exclusive/exclusive-u-s-to-tighten-restrictions-on-huawei-access-to-technology-chips-sources-say-idUSKCN25D1CC; “4Q20 Quarterly Management Report,” press release, January 14, 2021, Taiwan Semiconductor Manufacturing Company, https://investor.tsmc.com/english/encrypt/files/encrypt_file/reports/2021-01/4Q20ManagementReport.pdf; “The struggle over chips enters a new phase,” *The Economist*, January 23, 2021, https://www.economist.com/leaders/2021/01/23/the-struggle-over-chips-enters-a-new-phase.

[44] For more on the potential for multilateral dialogues, see Horowitz and Scharre, “AI and International Stability.”

[45] “The AI Index,” https://aiindex.stanford.edu/; Jess Whittlestone and Jack Clark, “Why and How Governments Should Monitor AI Development,” August 28, 2021, https://arxiv.org/pdf/2108.12427.pdf.

[46] Martijn Rasser, Megan Lamberth, Ainikki Riikonen, Chelsea Guo, Michael Horowitz, and Paul Scharre, “The American AI Century: A Blueprint for Action” (Center for a New American Security, December 17, 2019), https://www.cnas.org/publications/reports/the-american-ai-century-a-blueprint-for-action; and Saif M. Khan, “Securing Semiconductor Supply Chains,” CSET Issue Brief (Center for Security and Emerging Technology, January 2021), https://cset.georgetown.edu/publication/securing-semiconductor-supply-chains/.

[47]“If a chieftain or a man is captured on the ‘Way of the King’ (in war), and a merchant buy him free, and bring him back to his place; if he have the means in his house to buy his freedom, he shall buy himself free: if he have nothing in his house with which to buy himself free, he shall be bought free by the temple of his community; if there be nothing in the temple with which to buy him free, the court shall buy his freedom. His field, garden, and house shall not be given for the purchase of his freedom.” In context, the rule appears to apply to the payment of debts for one’s own soldiers who are taken prisoner and then later freed by a merchant paying ransom to release the soldier, rather than treatment of enemy prisoners. Fordham University, “Ancient History Sourcebook: Code of Hammurabi, c. 1780 BCE,” as translated by L.W. King, https://sourcebooks.fordham.edu/ancient/hamcode.asp.

[48] “When you lay siege to a city for a long time, fighting against it to capture it, do not destroy its trees by putting an ax to them, because you can eat their fruit. Do not cut them down. Are the trees people, that you should besiege them?"[19] "However, you may cut down trees that you know are not fruit trees and use them to build siege works until the city at war with you falls.” Deuteronomy 20:19–20 (New International Version), https://biblehub.com/deuteronomy/20-19.htm.

[49] Ahmed Zaki Yamani, “Humanitarian International Law in Islam: A General Outlook,” *Michigan Journal of International Law* 7 no. 1 (1985), 189–215, https://repository.law.umich.edu/cgi/viewcontent.cgi?article=1784&context=mjil.

[50] Ali Raza Naqvi, "Laws of War in Islam," *Islamic Studies*, 13 no. 1 (1974), 31, http://www.jstor.org/stable/20846902; and Youssef H. Aboul-Enein and Sherifa Zuhur, “Islamic Rulings on Warfare” (Strategic Studies institute, U.S. Army War College, October 2004), 22, https://publications.armywarcollege.edu/pubs/1693.pdf.

[51] Fordham University, "Indian History Sourcebook: The Laws of Manu, c. 1500 BCE," Chapter VII: 87–93, as translated by G. Buhler, https://sourcebooks.fordham.edu/india/manu-full.asp.

[52] Fordham University, "Indian History Sourcebook: The Laws of Manu, c. 1500 BCE," Chapter VII: 90, as translated by G. Buhler, https://sourcebooks.fordham.edu/india/manu-full.asp.

[53] "Let [the warrior] act according to his instructions. 9. Let him not turn back in battle. 10. Let him not strike with barbed or poisoned (weapons). 11. Let him not fight with those who are in fear, intoxicated, insane or out of their minds, (nor with those) who have lost their armour, (nor with) women, infants, aged men, and Brâhmanas, 12. Excepting assassins (âtatâyin)." Dharmaśāstras 1.10.18.8, quoted in A. Walter Dorn, "The Justifications for War and Peace in World Religions Part III: Comparison of Scriptures from Seven World Religions," (Defence R&D Canada–Toronto, March 2010), 20, http://www.dtic.mil/dtic/tr/fulltext/u2/a535552.pdf.

[54] "There should be no arrows smeared with poison, nor any barbed arrows—these are the weapons of evil people." "Law, Force, and War" in *The Mahabhbarata. Vol. 7, Book 11, The Book of the Women/ Book 12, The Book of Peace*, Chapter 841, verse 96.10, as translated by J.L. Fitzgerald (Chicago and London: University of Chicago Press, 2003), 411.

[55] "Law, Force, and War," 411.

[56] Torkel Brekke, email to Paul Scharre, January 20, 2017.

[57] The crossbow was also called the *arbalète.* For example, see Kelly DeVries, "Crossbow," in William W. Kibler, Grover A. Zinn, John Bell Henneman Jr., and Lawrence Earp, eds., *Medieval France: An Encyclopedia* (Routledge, 1995), 521, https://books.google.com/books/about/Medieval_France.html?id=TFxI-rhpEvAC; and Matthew Hipple, "Autonomy whether you like it or not," *War on the Rocks*, April 29, 2015, https://warontherocks.com/2015/04/autonomy-whether-you-like-it-or-not/.

[58] Cathal J. Nolan, *The Age of Wars of Religion, 1000–1650: An Encyclopedia of Global Warfare and Civilization, Volume 1* (Westport, CT: Greenwood, 2006), 200; and DeVries, "Crossbow," 521.

[59] It is worth acknowledging that there are dissenting views to the standard interpretation of the ban. In one version, the original Latin is translated to English as "We forbid under penalty of anathema that that deadly and God-detested art of stingers and archers be in the future exercised against Christians and Catholics." Some interpret this as referring to a tournament or wager involving arrows. "The reference seems to be to a sort of tournament, the nature of which was the shooting of arrows and other projectiles on a wager. The practice had already been condemned by Urban II in canon 7 of the Lateran Synod of 1097, no doubt because of the danger it involved." Second Lateran Council, Canon 29, https://www.papalencyclicals.net/councils/ecum10.htm; and H. J. Schroeder, *Disciplinary Decrees of the General Councils: Text, Translation and Commentary*, (St. Louis: B. Herder, 1937), 113, https://archive.org/details/DisciplinaryCouncils.

[60] There are other theories for the crossbow ban's motivation. One theory is that the ban was based on crossbows' "inability to discriminate" because they fired projectile weapons and had the potential to kill innocents. This theory doesn't stand up to scrutiny. For one thing, the crossbow was a fairly accurate weapon. Additionally, this theory presumes that there would be innocent civilians mixed in among soldiers on the battlefield—which strains credulity. Another theory is that the objection was not the crossbow per se, but the fact that it was employed by mercenaries. This theory, similarly, doesn't match the historical record. The crossbow was clearly reviled as an immoral weapon, depicted as such in paintings and sculptures. Additionally, although mercenaries did use the crossbow, mercenaries were also used widely throughout the Middle Ages in other roles, and non-mercenaries also fired the crossbow. It seems clear that the objection is to the weapon itself, for moral or political reasons—or both. Both theories are presented in Colm McKeogh, *Innocent Civilians: The Morality of Killing in War* (Palgrave Macmillian, 2002), 67–68. A related theory is that the ban applied to the employment of crossbowmen in armies (rather than the weapons themselves), but the motivation was to harm Pope Innocent II's chief rival, King Roger II of Sicily. See Monte S. Turner, *The Not So Diabolical Crossbow: A Re-Examination of Innocent II's Supposed Ban of the Crossbow at the Second Lateran Council* (Lulu.com, 2004), Robert L. O'Connell, *Of Men and Arms: A History of War, Weapons, and Aggression* (Oxford: Oxford University Press, 1990), 96, and Richard Arthur Preston and Sydney F. Wise, *Men in Arms: A History of Warfare and its Interrelationships with Western Society* (Holt, Rinehart, and Winston, 1979), 75.

[61] "[E]nemies could be killed from beyond their range of hearing, vision, and retaliation. This aspect of the mechanical device may have made killing in war too remote and inhuman for contemporary opinion." McKeogh, *Innocent Civilians,* 67.

[62] N.H. Mallett, "The Crossbow—A Medieval Doomsday Device?" MilitaryHistoryNow.com, May 23, 2012, captured by the Internet Archive on October 20, 2017, https://web.archive.org/web/20171020213110/https://militaryhistorynow.com/2012/05/23/the-crossbow-a-medieval-wmd/.

[63] McKeogh, *Innocent Civilians,* 67; Nolan, *The Age of Wars of Religion, 1000–1650*, 200.

[64] "Christian Europe at first viewed the weapon as morally ambiguous. A widely perceived diabolical nature was illustrated by placing crossbows in daemons' hands in illuminated manuscripts. At Toulouse cathedral daemon gargoyles were sculpted as having trouble drawing crossbows, which at least got part of the tale right. This early sense that the crossbow was inherently evil led to its condemnation in 1096 by Pope Urban II." Nolan, *The Age of Wars of Religion, 1000–1650*, 200.

[65] Kat Eschner, "On This Day in 1847, a Texas Ranger Walked Into Samuel Colt's Shop and Said, Make Me a Six-Shooter," *Smithsonian Magazine*, January 4, 2017, https://www.smithsonianmag.com/smart-news/day-1847-texas-ranger-walked-samuel-colts-shop-and-said-make-me-six-shooter-180961621/.

[66] Herbert Kikoy, "Super Weapons That Ended the Reign of Knights," Warhistoryonline.com, September 25, 2018, https://www.warhistoryonline.com/instant-articles/weapons-medieval-warriors.html?chrome=1.

[67] Nolan, *The Age of Wars of Religion, 1000–1650*, 200.

[68] "[B]y the end of the 12th century the crossbow was in wide use as both an offensive and defensive weapon…" Nolan, *The Age of Wars of Religion, 1000–1650*, 200. "[T]his condemnation was rarely heeded as the *arbalète* became increasingly popular in Europe. This was especially the case in France, where most kings and nobles used crossbowmen in their armies between the 12th and 15th centuries … " DeVries, "Crossbow"*,* 521.

[69] "[K]ings and nobles used crossbowmen in their armies between the 12th and 15th centuries, frequently employing mercenary crossbowmen, principally Italians, when they failed to recruit sufficient numbers of these troops from among their own subjects." DeVries, "Crossbow," 521.

[70] "By the mid-14th century, crossbows were being replaced in defense of castles by small-caliber cannons … [U]ltimately the *arbalète* could not survive the late 15th-century influx of handguns. By 1550, the weapon had disappeared from the battlefield." DeVries, "Crossbow," 521

[71] Noel Perrin, *Giving Up the Gun: Japan's Reversion to the Sword, 1543–1879* (Boston: David R. Godine, 1988), 27.

[72] Alexander Astroth, "The Decline of Japanese Firearm Manufacturing and Proliferation in the Seventeenth Century," *Emory Endeavors in History*, 5 (2013), 141, http://history.emory.edu/home/documents/endeavors/volume5/gunpowder-age-v-astroth.pdf.

[73] Perrin, *Giving Up the Gun*, 62.

[74] Perrin, *Giving Up the Gun*, 62.

[75] Perrin, *Giving Up the Gun*, 62–63.

[76] Perrin, *Giving Up the Gun*, 45.

[77] Astroth, "The Decline of Japanese Firearm Manufacturing and Proliferation in the Seventeenth Century."

[78] Perrin, *Giving Up the Gun*, 63.

[79] "The United States and the Opening to Japan, 1853," U.S. Department of State, Office of the Historian, https://history.state.gov/milestones/1830-1860/opening-to-japan.

[80] Perrin, *Giving Up the Gun*, 72.

[81] Japan defeated Russia in the Russo-Japanese War, 1904–1905, cementing Japan's place as a great power. Perrin, *Giving Up the Gun*, 76.

[82] Kieron Monks, "Blade runners: The powerful mystique of the samurai sword," CNN, July 15, 2015, https://www.cnn.com/style/article/samurai-swords/index.html.

[83] "Samurai: Japanese warrior," Britannica, https://www.britannica.com/topic/samurai.

[84] "Sakoku: National Isolation," Britannica, https://www.britannica.com/topic/sakoku.

[85] Perrin, *Giving Up the Gun*, 58.

[86] Perrin, *Giving Up the Gun,* 65.

[87] Catherine Jefferson, "Origins of the norm against chemical weapons," *International Affairs*, 90 no. 3 (May 2014), 648.

[88] Jean Pascal Zanders, "International Norms Against Chemical and Biological Warfare: An Ambiguous Legacy," *Journal of Conflict & Security Law*, 8 no. 2, (2003), 394, https://www.the-trench.org/wp-content/uploads/2018/07/200312-JCSL-International-norms.pdf.

[89] Jefferson, "Origins of the norm against chemical weapons."

[90] Jenny Gesley, "The 'Lieber Code'—the First Modern Codification of the Laws of War," The Library of Congress blog, April 24, 2018, https://blogs.loc.gov/law/2018/04/the-lieber-code-the-first-modern-codification-of-the-laws-of-war/.

[91] Adjutant General's Office, "General Orders No. 100: The Lieber Code: Instructions for the Government of Armies of the United States in the Field," The Avalon Project, http://avalon.law.yale.edu/19th_century/lieber.asp.

[92] The Lieber Code prohibits poison twice. Article 16 states, "Military necessity does not admit of cruelty—that is, the infliction of suffering for the sake of suffering or for revenge, nor of maiming or wounding except in fight, nor of torture to extort confessions. It does not admit of the use of poison in any way, nor of the wanton devastation of a district. It admits of deception, but disclaims acts of perfidy; and, in general, military necessity does not include any act of hostility which makes the return to peace unnecessarily difficult." Article 70 states, "The use of poison in any manner, be it to poison wells, or food, or arms, is wholly excluded from modern warfare. He that uses it puts himself out of the pale of the law and usages of war." Adjutant General's Office, "General Orders No. 100."

[93] "Declaration Renouncing the Use, in Time of War, of Explosive Projectiles Under 400 Grammes Weight, Saint Petersburg. 29 November / 11 December 1868," International Committee of the Red Cross , https://ihl-databases.icrc.org/ihl/full/declaration1868.

[94] "1868 Saint Petersburg Declaration," Weapons Law Encyclopedia, http://www.weaponslaw.org/instruments/1968-Saint-Petersburg-Declaration.

[95] "30x173MM Ammunition Suite for MK44 Cannon," General Dynamics Ordnance and Tactical Systems,, March 9, 2016, https://www.gd-ots.com/wp-content/uploads/2017/11/30x173mm-Ammunition-Suite-MK44-Cannon-Version-3.pdf.

[96] "Declaration Renouncing the Use, in Time of War, of Explosive Projectiles Under 400 Grammes Weight."

[97] U.S. Department of Defense, Office of General Counsel, *Department of Defense Law of War Manual* (December 2016), 346–347, 1157–1158, https://dod.defense.gov/Portals/1/Documents/pubs/DoD%20Law%20of%20War%20Manual%20-%20June%202015%20Updated%20Dec%202016.pdf?ver=2016-12-13-172036-190.

[98] "Declaration Renouncing the Use, in Time of War, of Explosive Projectiles Under 400 Grammes Weight."

[99] The United States is not a signatory to the St. Petersburg Declaration and does not view the prohibition on antipersonnel exploding bullets as customary international law. Department of Defense, *Law of War Manual,* 1157–1158.

[100] B. Swift and G. N. Ruffy, "The exploding bullet," *Journal of Clinical Pathology,* 57 no. 1 (January 2004), https://www.ncbi.nlm.nih.gov/pmc/articles/PMC1770159/; Bernard Knight, "Explosive bullets: a new hazard for doctors," *British Medical Journal,* 284 (March 13, 1982), https://www.ncbi.nlm.nih.gov/pmc/articles/PMC1496427/pdf/bmjcred00597-0008.pdf.

[101] "Project of an International Declaration concerning the Laws and Customs of War. Brussels, 27 August 1874," Introduction, International Committee of the Red Cross, https://ihl-databases.icrc.org/ihl/INTRO/135; "Project of an International Declaration concerning the Laws and Customs of War. Brussels, 27 August 1874," Article 13, International Committee of the Red Cross, https://ihl-databases.icrc.org/applic/ihl/ihl.nsf/Article.xsp?action=openDocument&documentId=31364F80ED69E269C12563CD00515549.

[102] "Declaration (IV,3) concerning Expanding Bullets. The Hague, 29 July 1899," International Committee of the Red Cross, https://ihl-databases.icrc.org/applic/ihl/ihl.nsf/385ec082b509e76c41256739003e636d/f1f1fb8410212aebc125641e0036317c; "Declaration of the Declaration (IV,2) concerning Asphyxiating Gases. The Hague, 29 July 1899," International Committee of the Red Cross, https://ihl-databases.icrc.org/applic/ihl/ihl.nsf/Article.xsp?action=openDocument&documentId=2531E92D282B5436C12563CD00516149; and "Declaration (IV,1), to Prohibit, for the Term of Five Years, the Launching of Projectiles and Explosives from Balloons, and Other Methods of Similar Nature. The Hague, 29 July 1899," International Committee of the Red Cross, https://ihl-databases.icrc.org/applic/ihl/ihl.nsf/385ec082b509e76c41256739003e636d/53024c9c9b216ff2c125641e0035be1a?OpenDocument.

[103] "Convention (II) with Respect to the Laws and Customs of War on Land and its Annex: Regulations Concerning the Laws and Customs of War on Land. The Hague, 29 July 1899," Annex to Section II, Chapter 1, Article 23, International Committee of the Red Cross, https://ihl-databases.icrc.org/applic/ihl/ihl.nsf/Article.xsp?action=openDocument&documentId=14BF8E8D6537838EC12563CD00515E22; "Convention (IV) Respecting the Laws and Customs of War on Land and its Annex: Regulations Concerning the Laws and Customs of War on Land. The Hague, 18 October 1907," Annex to Section II, Chapter 1, Article 23, International Committee of the Red Cross, https://ihl-databases.icrc.org/applic/ihl/ihl.nsf/Article.xsp?action=openDocument&documentId=61CDD9E446504870C12563CD00516768.

[104] "Declaration (XIV) Prohibiting the Discharge of Projectiles and Explosives from Balloons. The Hague, 18 October 1907," International Committee of the Red Cross, https://ihl-databases.icrc.org/ihl/INTRO/245?OpenDocument.

[105] "Declaration (XIV) Prohibiting the Discharge of Projectiles and Explosives from Balloons."

[106] "Convention (IV) Respecting the Laws and Customs of War on Land and Its Annex: Regulations Concerning the Laws and Customs of War on Land. The Hague, 18 October 1907," Article 25, International Committee of the Red Cross, https://ihl-databases.icrc.org/customary-ihl/eng/docs/v2_rul_rule37_sectionc.

[107] "Declaration (XIV) Prohibiting the Discharge of Projectiles and Explosives from Balloons."

[108] Charles E. Heller, "Chemical Warfare in World War I: The American Experience, 1917–1918," Combined Arms Research Library, Leavenworth Papers No. 10 (Combat Studies Institute, Command & General Staff College, September 1984), https://www.webharvest.gov/peth04/20041017045619/http://www.cgsc.army.mil/carl/resources/csi/Heller/HELLER.asp.

[109] The shells were a "105-mm shrapnel shell with dianisidine chlorosulphonate, an agent known to cause irritation of the mucous membrane." Heller, "Chemical Warfare in World War I." See also "Germans introduce

poison gas," History.com, February 9, 2010, https://www.history.com/this-day-in-history/germans-introduce-poison-gas.

[110] Sarah Everts, "A Brief Histroy of Chemical War," Science History Institute, May 11, 2015, https://www.sciencehistory.org/distillations/a-brief-history-of-chemical-war.

[111] Anthony R. Hossack, "The First Gas Attack," Firstworldwar.com, August 22, 2009, https://firstworldwar.com/diaries/firstgasattack.htm.

[112] James E. Edmonds and G.C. Wynne, *Military Operations, France and Belgium, 1915: Winter 1914–1915: Battle of Neuve Chapelle: Battles of Ypres.* History of the Great War Based on Official Documents by Direction of the Historical Section of the Committee of Imperial Defence. Imperial War Museum and Battery Press ed., vol 1. (London: Macmillan, 1995), 220–225.

[113] Jonathan B. Tucker, *War of Nerves: Chemical Warfare from World War I to Al-Qaeda* (Pantheon Books, 2006).

[114] A.O. Pollard, *Fire-Eater: The Memoirs of a VC* (Naval & Military Press, 2005).

[115] Philip Warner, *Battle of Loos* (Pen and Sword Military Classics, 2009), 37–38.

[116] Tucker, *War of Nerves.*

[117] Tucker, *War of Nerves.*

[118] Wilfred Owen, "Dulce et Decorum Est," *Poems* (Viking Press, 1921), https://www.poetryfoundation.org/poems-and-poets/poems/detail/46560.

[119] Tucker, *War of Nerves.*

[120] Amos Fries, as quoted in Tucker, *War of Nerves.*

[121] "Treaty relating to the Use of Submarines and Noxious Gases in Warfare. Washington, 6 February 1922," Article 5, International Committee of the Red Cross, https://ihl-databases.icrc.org/applic/ihl/ihl.nsf/Article.xsp?action=openDocument&documentId=3063E6C738AFAA65C12563CD00518323.

[122] "Treaty relating to the Use of Submarines and Noxious Gases in Warfare," Introduction, International Committee of the Red Cross, https://ihl-databases.icrc.org/ihl/INTRO/270; "Protocol for the Prohibition of the Use of Asphyxiating, Poisonous or Other Gases, and of Bacteriological Methods of Warfare. Geneva, 17 June 1925," introduction, International Committee of the Red Cross, https://ihl-databases.icrc.org/ihl/INTRO/280?OpenDocument.

[123] "Protocol for the Prohibition of the Use of Asphyxiating, Poisonous or Other Gases, and of Bacteriological Methods of Warfare. Geneva, 17 June 1925," Protocol, International Committee of the Red Cross, https://ihl-databases.icrc.org/applic/ihl/ihl.nsf/Article.xsp?action=openDocument&documentId=58A096110540867AC12563CD005187B9.

[124] The United States, the United Kingdom, France, the Soviet Union, Belgium, and Canada included this reservation. Department of Defense, *Law of War Manual,* 1149.

[125] Gerard J. Fitzgerald, "Chemical Warfare and Medical Response during World War I," *American Journal of Public Health*, 98 no. 4 (2008), 611–625, https://www.ncbi.nlm.nih.gov/pmc/articles/PMC2376985/; Sarah Pruitt, "The Nazis Developed Sarin Gas During WWII, But Hitler Was Afraid to Use It," History.com, April 12, 2017, https://www.history.com/news/the-nazis-developed-sarin-gas-but-hitler-was-afraid-to-use-it; "Chemical Weapons—Russia," Federation of American Scientists, https://fas.org/nuke/guide/russia/cbw/cw.htm; and "Chemical Weapons—United States," Federation of American Scientists, https://fas.org/nuke/guide/usa/cbw/cw.htm.

[126] "Chemical Weapons—United States"; Brendan Koerner, "Who Used Mustard Gas in WII?" Slate, October 1, 2003, https://slate.com/news-and-politics/2003/10/who-used-mustard-gas-in-wwii.html; Office of Chemical and Biological Weapons, "Blister Agent: Sulfur Mustard (H, HD, HS)," U.S. Department of State, archived, https://web.archive.org/web/20070724063602/http://www.cbwinfo.com/Chemical/Blister/HD.shtml; Joseph F. Bunnett and Marian Mikołajczyk, eds. *Arsenic and Old Mustard: Chemical Problems in the Destruction of Old Arsenical and 'Mustard' Munitions*," NATO ASI Series 19 (Springer, 1998), 92; *The Problem of Chemical and Biological Warfare* (Stockholm International Peace Research Institute, 2000), 153–154, https://www.sipri.org/sites/default/files/CBW_VOL1.PDF.

[127] "Documenting Numbers of Victims of the Holocaust and Nazi Persecution," United States Holocaust Memorial Museum, https://www.ushmm.org/wlc/en/article.php?ModuleId=10008193; "Gassing Operations," and United States Holocaust Memorial Museum, https://www.ushmm.org/wlc/en/article.php?ModuleId=10005220.

[128] Paul Iddon, "Why Didn't World War II Become a Chemical Weapons War?" National Interest, January 2, 2021, https://nationalinterest.org/blog/reboot/why-didnt-world-war-ii-become-chemical-weapons-war-175475; Richard B. Frank, "The Horribles: American Strategic Options Against Japan in 1945," The National WWII Museum, August 13, 2020, https://www.nationalww2museum.org/war/articles/american-strategic-options-against-japan-1945; Tucker, *War of Nerves.*

[129] "Chemical Warfare—Suffolk Anti-invasion defences," Pillboxes-suffolk, https://web.archive.org/web/20200807085739/http://pillboxes-suffolk.webeden.co.uk/chemical-warfare/4543908803; Iddon, "Why Didn't World War II Become a Chemical Weapons War?"

[130] Heller, "Chemical Warfare in World War I."

[131] Convention on the Prohibition of the Development, Production and Stockpiling of Bacteriological (Biological) and Toxin Weapons and on their Destruction, April 10, 1972, ttps://front.un-arm.org/wp-content/uploads/2020/12/BWC-text-English-1.pdf; "Chemical Weapons Convention," Organisation for the Prohibition of Chemical Weapons, https://www.opcw.org/chemical-weapons-convention.

[132] The process of dismantling existing stockpiles has not been without its challenges. See "OPCW by the Numbers," Organisation for the Prohibition of Chemical Weapons, https://www.opcw.org/news-publications/publications/facts-and-figures/#c1920; "OPCW Executive Council briefed on chemical weapons destruction progress in the United States," Organisation for the Prohibition of Chemical Weapons, March 25, 2022, https://www.opcw.org/media-centre/news/2022/03/opcw-executive-council-briefed-chemical-weapons-destruction-progress; and David Koplow, "Train Wreck: The U.S. Violation of the Chemical Weapons Convention," Georgetown Public Law and Legal Theory Research Paper No. 12-128 (Georgetown University Law Center, 2012), https://scholarship.law.georgetown.edu/facpub/1042/.

[133] Lukas Trakimavičius, "Is Russia Violating the Biological Weapons Convention?" New Atlanticist blog, May 23, 2018, https://www.atlanticcouncil.org/blogs/new-atlanticist/is-russia-violating-the-biological-weapons-convention/.

[134] Steven Aftergood, "The History of the Soviet Biological Weapons Program," Secrecy News blog, July 18, 2012, https://fas.org/blogs/secrecy/2012/07/soviet_bw/.

[135] Asher Orkaby, "Forgotten Gas Attacks in Yemen Haunt Syria Crisis," Bloomberg, September 15, 2013, https://www.bloomberg.com/opinion/articles/2013-09-15/forgotten-gas-attacks-in-yemen-haunt-syria-crisis.

[136] Javed Ali, "Chemical Weapons and the Iran-Iraq War: A Case Study in Noncompliance," *The Nonproliferation Review*, 8 no. 1 (Spring 2001), 43–58, https://www.nonproliferation.org/wp-content/uploads/npr/81ali.pdf.

[137] "Chemical Weapons Convention Signatories and States-Parties," Arms Control Association, https://www.armscontrol.org/factsheets/cwcsig.

[138] Bashar al-Assad used chemical weapons against the Syrian population on multiple occasions before joining the CWC in the fall of 2013. Additional use of chemical weapons by al-Assad and Syrian armed forces, however, was reported and confirmed in 2017, proving that Syria had violated the terms of the CWC. Mallory Stewart, "Twenty Years of the Chemical Weapons Convention—Where Do We Go from Here?" The Stimson

Center, July 6, 2017, https://www.stimson.org/2017/twenty-years-chemical-weapons-convention-where-do-we-go-here-0/; “Timeline of Syrian Chemical Weapons Activity, 2012–2022,” Arms Control Association, https://www.armscontrol.org/factsheets/Timeline-of-Syrian-Chemical-Weapons-Activity.

139 Walter Boyne, *The Influence of Air Power Upon History* (Gretna, LA: Pelican, 2003), 99.

140 Brian C. Lavelle, “Zeppelinitis,” (research paper, Air Command and Staff College, March 1997), https://apps.dtic.mil/sti/pdfs/ADA397845.pdf; Laura Walker, “Theatreland Raid by the ‘Baby-Killers,’” Untold Lives blog, October 13, 2015, https://blogs.bl.uk/untoldlives/2015/10/theatreland-raid-by-the-baby-killers--1.html.

141 Christopher Warren, “The First Battle of Britain,” (Air and Space Power Journal, 2018), 10, https://www.airuniversity.af.edu/Portals/10/ASPJ_Spanish/Journals/Volume-30_Issue-3/2018_3_10_warren_s_eng.pdf; “Viewpoint: 10 big myths about World War One debunked,” BBC News, February 25, 2014, https://www.bbc.com/news/magazine-25776836.

142 Lavelle, “Zeppelinitis,” 21.

143 Giulio Douhet, “The Command of the Air*,*” trans. Dino Ferrari (Air Force History and Museums Program, 1998), 58, https://www.airforcemag.com/PDF/MagazineArchive/Documents/2013/April%202013/0413keeperfull.pdf.

144 Wade S. Karren, “Lightning Strikes and Thunder Claps: The Strategic Bomber and Air Superiority,” *Air & Space Power Journal*, 26 no. 6 (November–December 2012), 139, https://www.airuniversity.af.edu/Portals/10/ASPJ/journals/Volume-26_Issue-6/V-Karren.pdf. The phrase “The bomber will always get through” comes from a speech given by Stanley Baldwin to the House of Commons in 1932: Stanley Baldwin, “The Bomber Will Always Get Through,” *Air Force Magazine*, July 1, 2008, https://www.airforcemag.com/article/0708keeperfile/. The full text of Baldwin’s speech is available at https://www.airforcemag.com/PDF/MagazineArchive/Documents/2008/July%202008/0708keeperfull.pdf.

145 Robert A. Pape, *Bombing to Win: Air Power and Coercion in War* (Ithaca, NY: Cornell University Press, 1996), 61, https://www.jstor.org/stable/10.7591/j.ctt1287f6v.

146 “The Hague Rules of Warfare,” Great Was Primary Documents Archive, http://www.gwpda.org/1918p/hagair.html.

147 “The Hague Rules of Warfare.” For analysis of the Hague Rules, see Heinz Marcus Hanke, “The 1923 Hague Rules of Air Warfare—A contribution to the development of international law protecting civilians from air attack,” *International Review of the Red Cross*, 33 no. 292 (March 1993), 20–28, https://international-review.icrc.org/sites/default/files/S0020860400071370a.pdf.

148 “Protection of Civilian Populations Against Bombing From the Air in Case of War, League of Nations, September 30, 1938,” Dannen.com, http://www.dannen.com/decision/int-law.html#E.

149 “Protection of Civilian Populations Against Bombing From the Air in Case of War, League of Nations, September 30, 1938.”

150 “Appeal of President Franklin D. Roosevelt on Aerial Bombardment of Civilian Populations, September 1, 1939,” Dannen.com, http://www.dannen.com/decision/int-law.html#E.

151 “Germany invades Poland,” History.com, https://www.history.com/this-day-in-history/germany-invades-poland.

152 Robert Forczyk, *Case White: The Invasion of Poland 1939* (Oxford: Osprey Publishing, 2019).

153 Larry Holzwarth, “The Bombing Campaign against Hitler’s Third Reich,” History Collection, January 30, 2020, https://historycollection.com/the-bombing-campaign-against-hitlers-third-reich/2/.

[154] Lincoln Riddle, "8 Things You Need to Know About the 1940 Rotterdam Terror Bombing," War History Online, May 14, 2016, https://www.warhistoryonline.com/today-in-history/8-things-need-know-1940-rotterdam-terror-bombing.html?chrome=1.

[155] C. Peter Chen, "Bombing of Hamburg, Dresden, and Other Cities," World War II Database, May 2008, https://ww2db.com/battle_spec.php?battle_id=55.

[156] Derek Dempster and Derek Wood, *The Narrow Margin: The Battle of Britain and the Rise of Air Power 1930–1949,* (Pen and Sword Books,2003), 117.

[157] "Directive No. 17: For the conduct of air and sea warfare against England," World War II Database, http://ww2db.com/doc.php?q=301.

[158] "World War II in Europe: The Blitz," The History Place, https://www.historyplace.com/worldwar2/timeline/about-blitz.htm.

[159] Michael Schmidt-Klingenberg, "Hitlers Bomben auf Europa: 'Wir werden sie ausradieren'" [Hitler's Bomb Terror: "We will eradicate them"], *Der Spiegel*, March 2003, https://magazin.spiegel.de/EpubDelivery/spiegel/pdf/26109882 .

[160] "Why Did the RAF Bomb Cities?" The National Archives of the United Kingdom, https://webarchive.nationalarchives.gov.uk/ukgwa/20131204172054/https:/www.nationalarchives.gov.uk/education/worldwar2/theatres-of-war/western-europe/investigation/hamburg/sources/art/2b/; Max Hastings, *Bomber Command* (Minneapolis: Quarto Publishing Group USA, 2013), 328–333, appendix E.

[161] Douhet, "The Command of the Air*,"* 6–7, 40–41, 57–58.

[162] For example, a river is a natural focal point for two parties looking to divide territory. A military looking to show restraint in how far it was willing to advance on another nation's territory might halt its advance at a river. Indeed, many international borders are rivers. Schelling, *The Strategy of Conflict*, 75.

[163] Thomas C. Schelling, *Arms and Influence*, (New Haven: Yale University Press, 1966), 164.

[164] Schelling, *The Strategy of Conflict*, 75.

[165] Schelling makes this point in *Arms and Influence* in a critique of the McNamara "no cities" doctrine. Schelling, *Arms and Influence,* 165.

[166] Paul Scharre, *Army of None,* (New York: W.W. Norton & Company, 2018), 342.

[167] Albert Wohlstetter, The Delicate Balance of Terror" (RAND Corporation, November 6, 1958), https://www.rand.org/pubs/papers/P1472.html.

[168] Robert McNamara, Commencement address ("No Cities" Speech), (University of Michigan, June 9, 1962), https://www.atomicarchive.com/resources/documents/deterrence/no-cities-speech.html.

[169] Schelling, *Arms and Influence,* 162–166.

[170] William Daugherty, Barbara Levi, and Frank Von Hippel, "Casualties Due to the Blast, Heat, and Radioactive Fallout from Various Hypothetical Nuclear Attacks on the United States," in *The Medical Implications of Nuclear War,* eds. Fredric Solomon and Robert Q. Marston (Washington DC: National Academies Press, 1986), https://www.ncbi.nlm.nih.gov/books/NBK219165/.

[171] Howard Levie, "Submarine Warfare: With Emphasis on the 1936 London Protocol," *International Law Studies,* 70 (1998), 294, https://digital-commons.usnwc.edu/ils/vol70/iss1/12/.

[172] "Peace Conference at the Hague 1899: Instructions to the International (Peace) Conference at the Hague," The Avalon Project, https://avalon.law.yale.edu/19th_century/hag99-03.asp.

[173] "It should be borne in mind that at this point in time most naval experts still considered that the submarine was a weapon to be used for coastal defense, particularly by the smaller and weaker nations which did not have strong navies. Little or no consideration was given to the fact that the submarine might be valuable as a commerce destroyer and on the high seas." Levie, "Submarine Warfare," 295.

[174] "[A] vote on the proposal to ban the submarine was taken in the First Commission and resulted in five votes (Belgium, Bulgaria, Greece, Persia, and Siam) for the prohibition with reservations; five votes (Germany, Great Britain, Italy, Japan, and Rumania) for the prohibition on condition of unanimity; and nine votes (Austria-Hungary, Denmark, France, Netherlands, Portugal, Spain, Sweden and Norway, Turkey, and the United States) in the negative. Russia, Serbia, and Switzerland abstained. That ended all efforts to ban the submarine at the 1899 Hague Peace Conference." Levie, "Submarine Warfare," 295.

[175] Levie, "Submarine Warfare," 296.

[176] "Convention (X) for the Adaptation to Maritime Warfare of the Principles of the Geneva Convention. The Hague, 18 October 1907," International Committee of the Red Cross, https://ihl-databases.icrc.org/ihl/INTRO/225?OpenDocument.

[177] "Declaration concerning the Laws of Naval War. London, 26 February 1909," International Committee of the Red Cross, https://ihl-databases.icrc.org/ihl/INTRO/255.

[178] James Kraska, "Prize Law," Max Planck Encyclopedia of Public International Law, July 5, 2011, https://papers.ssrn.com/sol3/papers.cfm?abstract_id=1876724.

[179] Jon L. Jacobson, "The Law of Submarine Warfare," *International Law Studies,* 64 (1991), 207–208, https://digital-commons.usnwc.edu/cgi/viewcontent.cgi?article=1756&context=ils.

[180] Levie, "Submarine Warfare," 298.

[181] "Germany declares war zone around British Isles," History.com, https://www.history.com/this-day-in-history/germany-declares-war-zone-around-british-isles.

[182] , "The Secretary of State to the Ambassador in Britain [telegram]," U.S. Department of State, Office of the Historian, https://history.state.gov/historicaldocuments/frus1915Supp/d163.

[183] Levie, "Submarine Warfare," 298–299.

[184] Levie, "Submarine Warfare," 303.

[185] Levie, "Submarine Warfare," 303.

[186] Levie, "Submarine Warfare," 303.

[187] "German submarine sinks Lusitania," History.com, https://www.history.com/this-day-in-history/german-submarine-sinks-lusitania.

[188] Levie, "Submarine Warfare," 300.

[189] Levie, "Submarine Warfare," 300.

[190] "Germany agrees to limit its submarine warfare," History.com, http://www.history.com/this-day-in-history/germany-agrees-to-limit-its-submarine-warfare.

[191] Jacobson, "The Law of Submarine Warfare," 209.

[192] Levie, "Submarine Warfare," 305.

[193] "Germany resumes unrestricted submarine warfare," History.com, https://www.history.com/this-day-in-history/germany-resumes-unrestricted-submarine-warfare.

[194] "U.S. Entry into World War I, 1917," U.S. Department of State, Office of the Historian, https://history.state.gov/milestones/1914-1920/wwi.

[195] Levie, "Submarine Warfare," 307.

[196] Treaty relating to the Use of Submarines and Noxious Gases in Warfare. Washington, 6 February 1922," International Committee of the Red Cross, https://ihl-databases.icrc.org/applic/ihl/ihl.nsf/Article.xsp?action=openDocument&documentId=AE48C18B39E087CCC12563CD005182E3; "Treaty for the Limitation and Reduction of Naval Armaments, (Part IV, Art. 22, relating to submarine warfare). London, 22 April 1930," International Committee of the Red Cross, https://ihl-databases.icrc.org/applic/ihl/ihl.nsf/Article.xsp?action=openDocument&documentId=05F68B7BFFB8B984C12563CD00519417; and "Procès-verbal relating to the Rules of Submarine Warfare set forth in Part IV of the Treaty of London of 22 April 1930. London, 6 November 1936," International Committee of the Red Cross, https://ihl-databases.icrc.org/applic/ihl/ihl.nsf/Article.xsp?action=openDocument&documentId=C103186F0C4291EEC12563CD00519832.

[197] Levie, "Submarine Warfare," 315.

[198] Levie, "Submarine Warfare," 316.

[199] Levie, "Submarine Warfare," 317; Jacobson, 210.

[200] Joel Ira Holwitt, "'Execute Against Japan': Freedom-of-the-Seas, the U.S. Navy, Fleet Submarines, and the U.S. Decision to Conduct Unrestricted Warfare, 1919–1941" (PhD diss., Ohio State University, 2005, https://etd.ohiolink.edu/apexprod/rws_etd/send_file/send?accession=osu1127506553&disposition=inline.

[201] "Declaration (IV,3) concerning Expanding Bullets. The Hague, 29 July 1899," International Committee of the Red Cross, https://ihl-databases.icrc.org/applic/ihl/ihl.nsf/385ec082b509e76c41256739003e636d/f1f1fb8410212aebc125641e0036317c

[202] This is different from an open-tip match-grade bullet that is designed to minimize air resistance for high accuracy long-range shooting.

[203] Department of Defense, *Law of War Manual,* 325.

[204] Robin Coupland and Dominique Loye, "The 1899 Hague Declaration concerning Expanding Bullets: A treaty effective for more than 100 years faces complex contemporary issues," *IRRC,* 85 no. 849 (March 2003), https://www.icrc.org/eng/assets/files/other/irrc_849_coupland_et_loye.pdf.

[205] "Practice Relating to Rule 77. Expanding Bullets," IHL Database: Customary IHL, https://ihl-databases.icrc.org/customary-ihl/eng/docs/v2_rul_rule77; and "Rule 77. Expanding Bullets," IHL Database: Customary IHL, https://ihl-databases.icrc.org/customary-ihl/eng/docs/v1_rul_rule77.

[206] Department of Defense, *Law of War Manual,* 323–324.

[207] "The law of war does not prohibit the use of bullets that expand or flatten easily in the human body. Like other weapons, such bullets are only prohibited if they are calculated to cause superfluous injury. The U.S. armed forces have used expanding bullets in various counterterrorism and hostage rescue operations, some of which have been conducted in the context of armed conflict. Department of Defense, *Law of War Manual,* 323.

[208] "The Geneva Naval Conference, 1927," U.S. Department of State, Office of the Historian, https://history.state.gov/milestones/1921-1936/geneva.

[209] "Peace Treaty of Versailles, Articles 159–213: Military, Naval, and Air Clauses," World War I Document Archive, http://net.lib.byu.edu/~rdh7/wwi/versa/versa4.html.

[210] "London Naval Conference," Britannica, https://www.britannica.com/event/London-Naval-Conference.

[211] Also called the Partial Test Ban Treaty. Treaty Banning Nuclear Weapons Tests in the Atmosphere, in Outer Space and Under Water, August 5, 1963, https://2009-2017.state.gov/t/avc/trty/199116.htm#signatory.

[212] Treaty on the Prohibition of the Emplacement of Nuclear Weapons and Other Weapons of Mass Destruction on the Seabed and the Ocean Floor and in the Subsoil Thereof, February 11, 1971, https://2009-2017.state.gov/t/isn/5187.htm.

[213] "The Antarctic Treaty," Secretariat of the Antarctic Treaty, https://www.ats.aq/e/antarctictreaty.html.

[214] Three countries have conducted tests since the CTBT was signed: India (1998), Pakistan (1998), and North Korea (2006, 2009, 2013, and two tests in 2016).

[215] Treaty on Principles Governing the Activities of States in the Exploration and Use of Outer Space, Including the Moon and Other Celestial Bodies,January 27, 1967, https://2009-2017.state.gov/t/isn/5181.htm.

[216] "Seabed Arms Control Treaty," Arms Control Association, https://www.armscontrol.org/treaties/seabed-arms-control-treaty.

[217] Treaty on Principles Governing the Activities of States in the Exploration and Use of Outer Space.

[218] Treaty Between The United States of America and The Union of Soviet Social Republics on The Limitation of Anti-Ballistic Missile Systems (ABM Treaty), May 26, 1972, https://2009-2017.state.gov/t/avc/trty/101888.htm.

[219] Treaty Between The United States Of America And The Union Of Soviet Socialist Republics On The Elimination Of Their Intermediate-Range And Shorter-Range Missiles (INF Treaty), December 8, 1987, https://2009-2017.state.gov/t/avc/trty/102360.htm.

[220] Interim Agreement Between The United State of America and The Union of Soviet Social Republics on Certain Measures with Respect to the Limitation of Strategic Offensive Arms, May 26, 1972, https://fas.org/nuke/control/salt1/text/salt1.htm.

[221] Treaty Between The United States of America and The Union of Soviet Social Republics on the Limitation of Strategic Offensive Arms (SALT II),June 18, 1979, https://2009-2017.state.gov/t/isn/5195.htm.

[222] Treaty Between the United States of America and the Union of Soviet Socialist Republics on Further Reduction and Limitation of Strategic Offensive Arms (START I), July 31, 1991, https://media.nti.org/documents/start_1_treaty.pdf.

[223] Daryl G. Kimball, "U.S. Commits to ASAT Ban," Arms Control Association, May 2022, https://www.armscontrol.org/act/2022-05/news/us-commits-asat-ban; "Russian direct-ascent anti-satellite missile test creates significant, long-lasting space debris," U.S. Space Command, press release, November 15, 2021, https://www.spacecom.mil/Newsroom/News/Article-Display/Article/2842957/russian-direct-ascent-anti-satellite-missile-test-creates-significant-long-last/; and "Fact Sheet: Space Debris from Anti-Satellite Weapons," Union of Concerned Scientists, April 2008, https://www.ucsusa.org/sites/default/files/2019-09/debris-in-brief-factsheet.pdf.

[224] "Fact Sheet: Vice President Harris Advances National Security Norms in Space," The White House, press release, April 18, 2022, https://www.whitehouse.gov/briefing-room/statements-releases/2022/04/18/fact-sheet-vice-president-harris-advances-national-security-norms-in-space/.

[225] In practice, any adjustable dial-a-yield nuclear weapon that can be adjusted to 10 kilotons or less could be used as a neutron bomb.

[226] "Neutron bomb," Britannica, https://www.britannica.com/technology/neutron-bomb; Robert S. Norris and Thomas B. Cochran, "Nuclear weapon," Britannica, https://www.britannica.com/technology/nuclear-weapon; Jonathan Karp, "India Discloses It Is Able To Build a Neutron Bomb," *The Wall Street Journal,* August 17, 1999, https://www.wsj.com/articles/SB934836102919955535; and John T. Correll, "The Neutron Bomb," Air Force Magazine, October 30, 2017, https://www.airforcemag.com/article/the-neutron-bomb/.

[227] "ABM Treaty Fact Sheet," U.S. Department of State, press release, December 13, 2001, https://2001-2009.state.gov/t/ac/rls/fs/2001/6848.htm.

[228] Kingston Reif, "Russia Completes CFE Treaty Suspension," Arms Control Association, April 2015, https://www.armscontrol.org/act/2015-04/news-briefs/russia-completes-cfe-treaty-suspension.

[229] Mark Stokes and Dan Blumenthal, "Can a treaty contain China's missiles?" *The Washington Post,* January 2, 2011, http://www.washingtonpost.com/wp-dyn/content/article/2010/12/31/AR2010123104108.html.

[230] Michael R. Gordon, "U.S. Says Russia Tested Cruise Missile, Violating Treaty," *The New York Times,* July 28, 2014, https://www.nytimes.com/2014/07/29/world/europe/us-says-russia-tested-cruise-missile-in-violation-of-treaty.html?_r=0.

[231] Alec Luhn and Julian Borger, "Moscow may walk out of nuclear treaty after US accusations of breach," *The Guardian*, July 29, 2004, https://www.theguardian.com/world/2014/jul/29/moscow-russia-violated-cold-war-nuclear-treaty-iskander-r500-missile-test-us.

[232] "The Intermediate-Range Nuclear Forces (INF) Treaty at a Glance," Arms Control Association, August 2019, https://www.armscontrol.org/factsheets/INFtreaty.

[233] "U.S. Withdrawal from the INF Treaty on August 2, 2019," U.S. Department of State, press release, August 2, 2019, https://2017-2021.state.gov/u-s-withdrawal-from-the-inf-treaty-on-august-2-2019/index.html.

[234] Treaty Between the United States of America and the Russian Federation on Strategic Offensive Reductions (The Moscow Treaty), May 24, 2002, https://2009-2017.state.gov/t/isn/10527.htm; "New START Treaty," U.S. Department of State, https://www.state.gov/new-start/.

[235] Andrew F. Krepinevich, Jr., "The New Nuclear Age: How China's Growing Nuclear Arsenal Threatens Deterrence," *Foreign Affairs*, 101 no.3 (May/June 2022), https://www.foreignaffairs.com/articles/china/2022-04-19/new-nuclear-age.

[236] Convention on the Prohibition of Military or Any Other Hostile Use of Environmental Modification Techniques, May 18, 1977, https://2009-2017.state.gov/t/isn/4783.htm.

[237] "Convention on the prohibition of military or any hostile use of environmental modification techniques, 10 December 10, 1976," International Committee of the Red Cross, https://ihl-databases.icrc.org/ihl/INTRO/460?OpenDocument.

[238] "The Convention on Certain Conventional Weapons," United Nations, Office for Disarmament Affairs, https://www.un.org/disarmament/the-convention-on-certain-conventional-weapons/.

[239] "Protocol (II) on Prohibitions or Restrictions on the Use of Mines, Booby-Traps and Other Devices. Geneva, 10 October 1980," Article 4, International Committee of the Red Cross, https://ihl-databases.icrc.org/applic/ihl/ihl.nsf/ART/510-820004?OpenDocument.

[240] "Protocol (II) on Prohibitions or Restrictions on the Use of Mines, Booby-Traps and Other Devices, Geneva, 10 october 1980," Article 5, https://ihl-databases.icrc.org/applic/ihl/ihl.nsf/Article.xsp?action=openDocument&documentId=17135717CA8A4938C12563CD0051EE5E.

[241] "Protocol on Prohibitions or Restrictions on the Use of Incendiary Weapons (Protocol III). Geneva, 10 October 1980," International Committee of the Red Cross, https://ihl-databases.icrc.org/applic/ihl/ihl.nsf/Article.xsp?action=openDocument&documentId=14FEADAF9AF35FA9C12563CD0051EF1E.

[242] " U.S. Blinding Laser Weapons" Human Rights Watch Arms Project 12 no. 5 (October 2000), https://www.hrw.org/reports/1995/Us2.htm#P118_16569.

[243] "Protocol on Blinding Laser Weapons (Protocol IV to the 1980 Convention), 13 October 1995," International Committee of the Red Cross https://ihl-

databases.icrc.org/applic/ihl/ihl.nsf/Treaty.xsp?action=openDocument&documentId=70D9427BB965B7CEC12563FB0061CFB2.

[244] Dunlap, "Is it Really Better to be Dead than Blind?"; "Blinding laser weapons: questions and answers," International Committee of the Red Cross, November 16, 1994, https://www.icrc.org/eng/resources/documents/misc/57jmcz.htm; Kelly Geoghegan, "On the Utility of Weapon Bans and Restrictions—Anti-Personnel Mines, Cluster Munitions and Blinding Lasers," Intercross blog, November 5, 2015, http://intercrossblog.icrc.org/blog/on-the-utility-of-weapon-bans-and-restrictions-blinding-lasers-cluster-munitions-and-anti-personnel-mines; Charles Dunlap Jr., "A Better Way to Protect Civilians and Combatants Than Weapons Bans: Strict Adherence to the Core Principles of the Law of War," Intercross blog, December 1, 2015, http://intercrossblog.icrc.org/blog/guest-post-a-better-way-to-protect-civilians-and-combatants-than-weapons-bans-strict-adherence-to-the-core-principles-of-the-law-of-war; and Crootof, "Why the Prohibition on Permanently Blinding Lasers is Poor Precedent for a Ban on Autonomous Weapon Systems."

[245] Dan Drollette Jr has argued that the U.S. military is continuing research into blinding lasers under the "fig leaf" of bettering the understanding of defenses against blinding lasers. His interpretation may or may not be valid, but in any case such research, even if it were intended to better the understanding of how to create a blinding laser, would not be prohibited under the ban. CCW Protocol IV prohibits employing a blinding laser, not research. Developing a weapon with no intent to use it, though, would be a waste of scarce resources. See Dan Drollette Jr., "Blinding them with science: Is development of a banned laser weapon continuing?" *Bulletin of the Atomic Scientists*, September 14, 2014, http://thebulletin.org/blinding-them-science-development-banned-laser-weapon-continuing7598.

[246] Graça Machel, "Impact of Armed Conflict on Children" (United Nations, August 26, 1996), http://www.unicef.org/graca/.

[247] "Myths and Realities About Incendiary Weapons: Memorandum to Delegates of the 2018 Meeting of States Parties to the Convention on Conventional Weapons," Human Rights Watch, November 14, 2018, https://www.hrw.org/news/2018/11/14/myths-and-realities-about-incendiary-weapons.

[248] Mary Wareham, "Dispatches: Incendiary Weapons Pose Civilian Threat in Syria," Human Rights Watch, June 21, 2016, https://www.hrw.org/news/2016/06/21/dispatches-incendiary-weapons-pose-civilian-threat-syria; Mark Hiznay, "Where Is Outrage Over Incendiary Weapons Attacks in Syria?" Human Rights Watch, August 23, 2016, https://www.hrw.org/news/2016/08/23/where-outrage-over-incendiary-weapons-attacks-syria; "Increase in Incendiary Weapon Attacks," Human Rights Watch, December 12, 2016, https://www.hrw.org/news/2016/12/12/increase-incendiary-weapon-attacks; and "Time to Act against Incendiary Weapons: Memorandum to Delegates at the Fifth Review Conference of the Convention on Conventional Weapons," Human Rights Watch, December 12, 2016, https://www.hrw.org/news/2016/12/12/time-act-against-incendiary-weapons.

[249] Marc Garlasco, Fred Abrahams, Bill van Esveld, Fares Akram, and Darryl Li, "Rain of Fire: Israel's Unlawful Use of White Phosphorus in Gaza," Human Rights Watch, March 25, 2009, https://www.hrw.org/report/2009/03/25/rain-fire/israels-unlawful-use-white-phosphorus-gaza.

[250] Daniel P. Mahoney III, "Goalie Without a Mask? The Effect of the Anti-Personnel Landmine Ban on U.S. Army Countermobility Operations" (monograph, School of Advanced Military Studies, United States Army Command and General Staff College, 1996), https://apps.dtic.mil/sti/pdfs/ADA324323.pdf.

[251] "122 States Sign Ottawa Landmine Treaty," Arms Control Association, https://www.armscontrol.org/act/1997-11/arms-control-today/122-states-sign-ottawa-landmine-treaty.

[252] Land mine ban chronology from "Timeline of the International Campaign to Ban Landmines," International Campaign to Ban Landmines, 2012, http://www.icbl.org/media/342067/icb009_chronology_a5_v4-pages.pdf.

[253] "Cluster Munitions a Foreseeable Hazard in Iraq," Human Rights Watch Briefing Paper, March 2003, https://www.hrw.org/reports/031403%20BP%20-%20Cluster%20Munitions%20Hazard%20in%20Iraq%20-%20Formatted.pdf; U.S. General Accounting Office, *Operation Desert Storm: Casualties Caused by Improper Handling of Unexploded U.S. Submunitions,* GAO/NSIAD-93-232 (August 1993), http://archive.gao.gov/t2pbat5/149647.pdf; "'Million bomblets' in S Lebanon," BBC News, September 26, 2006,

http://news.bbc.co.uk/2/hi/middle_east/5382192.stm; and “Kosovo mine expert criticises Nato,” BBC News, May 23, 2000, http://news.bbc.co.uk/2/hi/europe/761092.stm.

[254] Brian Brady, “Nato comes clean on cluster bombs,” *The Independent,* September 15, 2007, http://www.independent.co.uk/news/world/politics/nato-comes-clean-on-cluster-bombs-5329292.html.

[255] Ian MacKinnon, “Forty years on, Laos reaps bitter harvest of the secret war,” *The Guardian*, December 2, 2008, https://www.theguardian.com/world/2008/dec/03/laos-cluster-bombs-uxo-deaths.

[256] “Convention on Cluster Munitions, 30 May 2008,” International Committee of the Red Cross, https://ihl-databases.icrc.org/ihl/INTRO/620?OpenDocument.

[257] “Convention on the Prohibition of the Use, Stockpiling, Production and Transfer of Anti-Personnel Mines and on their Destruction, 18 September 1997,” Article 2, International Committee of the Red Cross, https://ihl-databases.icrc.org/applic/ihl/ihl.nsf/Article.xsp?action=openDocument&documentId=86CACEA6721637D541256585003E3003.

[258] “Yemen: Houthi Landmines Claim Civilian Victims,” Human Rights Watch, September 8, 2016, https://www.hrw.org/news/2016/09/08/yemen-houthi-landmines-claim-civilian-victims.

[259] “Convention Text,” Convention on Cluster Munitions.

[260] “Fitzgibbon wants to keep SMArt cluster shells,”Australian Broadcasting Corporation, May 29, 2008, http://www.abc.net.au/news/2008-05-29/fitzgibbon-wants-to-keep-smart-cluster-shells/2452894.

[261] “Treaty Status,” International Campaign to Ban Landmines, http://www.icbl.org/en-gb/the-treaty/treaty-status.aspx; “Treaty Status,” Cluster Munition Coalition, http://www.stopclustermunitions.org/en-gb/the-treaty/treaty-status.aspx.

[262] For countries outside the treaty, the total stockpile of antipersonnel landmines is estimated at around 50 million. “Why Landmines Are Still a Problem,” International Campaign to Ban Landmines, http://www.icbl.org/en-gb/problem/why-landmines-are-still-a-problem.aspx; “Treaty in Detail,” International Campaign to Ban Landmines, http://www.icbl.org/en-gb/the-treaty/treaty-in-detail/frequently-asked-questions.aspx.

[263] “Landmine treaty meeting concludes with advances in stockpile destruction. 158 countries no longer hold anti-personnel mines,” Anti-Personnel Mine Ban Convention, press release, December 2016, https://www.apminebanconvention.org/uploads/media/15MSP_Closes_advances_in_stockpile_destruction.pdf.

[264] Bryan Bender, “Formerly a leader on land mine ban, Obama now balks,” *Boston Globe*, June 22, 2014, https://www.bostonglobe.com/news/nation/2014/06/21/nearly-two-decades-after-leading-role-remains-sidelines-treaty-banning-landmines/oMeRDQjDB3W2vj4VtuYqWK/story.html.

[265] “Background Briefing on Landmine Use in Ukraine,” Human Rights Watch, June 15, 2022, https://www.hrw.org/news/2022/06/15/background-briefing-landmine-use-ukraine.

[266] “Background Briefing on Landmine Use in Ukraine,”

[267] “U.S. Landmine Policy,” U.S. Department of State, https://2009-2017.state.gov/t/pm/wra/c11735.htm.

[268] “Syria: Cluster Munition Ban Policy,” Landmine and Cluster Munition Monitor, http://www.the-monitor.org/en-gb/reports/2019/syria/cluster-munition-ban-policy.aspx; “Saudi Arabia: Cluster Munition Ban Policy,” Landmine and Cluster Munition Monitor, http://www.the-monitor.org/en-gb/reports/2019/saudi-arabia/cluster-munition-ban-policy.aspx; and Bonnie Docherty, “Russia’s Use of Cluster Munitions and Other Explosive Weapons Shows Need for Stronger Civilian Protections,” Human Rights Watch, March 21, 2022, https://www.hrw.org/news/2022/03/21/russias-use-cluster-munitions-and-other-explosive-weapons-shows-need-stronger#.

[269] “Cluster Munition Monitor 2016” (International Campaign to Ban Landmines–Cluster Munition Coalition, August 2016), http://www.the-monitor.org/media/2394895/Cluster-Munition-Monitor-2016-Web.pdf.

[270] Office of the Secretary of Defense, "Memorandum for Secretaries of the Military Departments, Chairman of Joint Chiefs of Staff, Under Secretary of Defense for Acquisition, Technology and Logistics, Under Secretary of Defense for Policy Commanders of the Combatant Commands, General Counsel of the Department of Defense, (June 19, 2008), https://www.globalsecurity.org/military/library/policy/dod/d20080709cmpolicy.htm.

[271] Thomas Graham Jr., "Avoiding the Tipping Point," Arms Control Association, November 1, 2004, https://www.armscontrol.org/act/2004_11/BookReview.

[272] Treaty on the Non-Proliferation of Nuclear Weapons, July 1, 1968, https://www.un.org/disarmament/wmd/nuclear/npt/text/.

[273] India tested a nuclear device in 1974 and Pakistan in 1998 (following a second Indian test in 1998). North Korea became the only country ever to withdraw from the treaty in 2003, and North Korea has subsequently tested several nuclear weapons. In 2006, North Korea conducted a nuclear test, which was widely believed by experts to be only partially successful. In 2009, North Korea conducted a second, more successful test, followed by an additional nuclear test in 2013, two tests in 2016, and a test in 2017. Israel has never publicly confirmed or denied having nuclear weapons, but it is widely believed to have acquired them in 1966 with the help of France. For more information, see Michael Krepon, "Looking Back: The 1998 Indian and Pakistani Nuclear Tests," Arms Control Association, https://www.armscontrol.org/act/2008-06/looking-back-1998-indian-pakistani-nuclear-tests; "North Korea Nuclear Overview," Nuclear Threat Initiative, October 11, 2018, https://www.nti.org/analysis/articles/north-korea-nuclear/; Avner Cohen, *Israel and the Bomb* (New York: Columbia University Press, 1999); and Avner Cohen, *The Worst-Kept Secret: Israel's Bargain with the Bomb* (New York, Columbia University Press, 2012).

[274] "The Australia Group," https://www.dfat.gov.au/publications/minisite/theaustraliagroupnet/site/en/index.html.

[275] Originally created by the United States, United Kingdom, France, Germany, Italy, and Canada in 1987, the Missile Technology Control Regime (MTCR) now has 35 signatories. For more information, see Missile Technology Control Regime, "Frequently Asked Questions (FAQs)," http://mtcr.info/frequently-asked-questions-faqs/.

[276] "The Wassenaar Agreement at a Glance," Arms Control Association, February 2022, https://www.armscontrol.org/factsheets/wassenaar; "The Wassenaar Arrangement on Export Controls for Conventional Arms and Dual-Use Goods and Technologies," the Wassenaar Arrangement, December 20, 2017, http://www.wassenaar.org/.

[277] "The Hague Code of Conduct against Ballistic Missile Proliferation (HCoC)," Hague Code of Conduct, http://www.hcoc.at/.

[278] "Proliferated Drones," Center for a New American Security, May 2016, http://drones.cnas.org/.